\documentclass[a4paper,12pt]{article}
\usepackage{relsize}
\usepackage{color}
\ifx\TeXtures\undefined
\usepackage{amsfonts}%
\usepackage{amsmath}%
\else
\usepackage{amsmath}%
\usepackage{amsfonts}%
\usepackage{times}
\fi
\usepackage{amsthm} 
\usepackage{epsfig}                

\newcommand{\DATUM}{}              
\newcommand{\change}
{{\marginpar{\#}}}        
\newcommand{\comma}{\: ,}              
\newcommand{\period}{\: .}             
\newcommand{\Om}{\Omega}                

\newcommand{\la}{\langle}                
\newcommand{\ra}{\rangle}

\newcommand{\boH}{{\bf H}}

\newcommand{\boE}{{\bf E}}         

\newcommand{\boW}{{\bf W}}
\newcommand{\boT}{{\bf T}}

\newcommand{\cA}{{\mathcal{A}}}
\newcommand{\cB}{{\mathcal{B}}}
\newcommand{\cD}{{\mathcal{D}}}
\newcommand{\cE}{{\mathcal{E}}}
\newcommand{\cF}{{\mathcal{F}}}
\newcommand{\cG}{{\mathcal{G}}}
\newcommand{\cH}{{\mathcal{H}}}

\newcommand{\cR}{{\mathcal{R}}}
\newcommand{\cS}{{\mathcal{S}}}

\newcommand{\cW}{{\mathcal{W}}}

\newcommand{\RR}{\mathbb{R}}            
\newcommand{\NN}{\mathbb{N}}            
\newcommand{\CC}{\mathbb{C}}            
\newcommand{\ur}{{\underline r}}
\newcommand{\uE}{{\underline E}}

\newcommand{\uW}{{\underline W}}

\newcommand{\ua}{{\underline a}}

\newcommand{\uw}{{\underline w}}
\newcommand{\ue}{{\underline \epsilon}}
\newcommand{\ual}{{\underline \alpha}}
\newcommand{\uv}{{\underline v}}

\newcommand{\dom}{{\rm dom}}

\newcommand{\cirS}{\mathop{\bigcirc\kern -.73em {\scriptstyle{\rm S}}}}

\newcommand{\QED}{\phantom{blablabla}\hfill\qed\newline}  
\renewcommand{\thesection}
{\Roman{section}}                      
\renewcommand{\theequation}
{\thesection.\arabic{equation}}        
\newcommand{\secct}[1]{\section{#1}
\setcounter{equation}{0}}              

\newtheorem{theorem}{Theorem}[section]         
\newtheorem{lemma}[theorem]{Lemma}             
\newtheorem{corollary}[theorem]{Corollary}     
\newtheorem{definition}[theorem]{Definition}   
\newtheorem{remark}[theorem]{Remark}           
\theoremstyle{plain}

\newcommand{\eq}[1]{\begin{align}#1\end{align}}

\def\beq{\begin{equation}}
\def\ene{\end{equation}}

\begin{document}

\bibliographystyle{plain}

\thispagestyle{empty}

\title{Continuous Renormalization Group Analysis of Spectral Problems in Quantum Field Theory}

\author{ 
Volker Bach 
\and
\small{Institut f\"ur Analysis und Algebra; Technische Universit\"at Braunschweig;} \\[-1ex]
\small{ 38092 Braunschweig; Germany (v.bach@tu-bs.de)}
\and
Miguel Ballesteros
\and
\small{Institut f\"ur Analysis und Algebra; Technische Universit\"at Braunschweig;} \\[-1ex]
\small{38092 Braunschweig; Germany (m.ballesteros@tu-bs.de)}
\and 
J\"urg Fr\"ohlich 
$\phantom{dsq}$\\
\small{ Institute for Theoretical Physics; ETH Z\"urich;} \\[-1ex]
\small{ 8093 Z\"urich; Switzerland (juerg.froehlich@itp.phys.ethz.ch)}
}
\date{\DATUM}

\maketitle

\thispagestyle{empty}

\begin{abstract}
The isospectral renormalization group is a powerful method to analyze the spectrum of operators in quantum field theory. It was introduced in 1995 [see \cite{nue}, \cite{BFS98a}] and since then it has been used to prove several results for non-relativistic quantum electrodynamics. After the introduction of the method there have been many works in which extensions, simplifications or clarifications are presented (see \cite{BCFS2003}, \cite{GH2008}, \cite{FGS2009}). In this paper we present a new approach in which we construct a flow of operators parametrized by a continuous variable in the positive real axis. While this is in contrast to the discrete iteration used before, this is more in spirit of the original formulation of the renormalization group introduced in theoretical physics in 1974 \cite{KogutWilson1974}. The renormalization flow that we construct can be expressed in a simple way: it can be viewed as a single application of the Feshbach-Schur map with a clever selection of the spectral parameter. Another advantage of the method is that there exists a flow function for which the renormalization group that we present is the orbit under this flow of an initial Hamiltonian. This opens the possibility to study the problem using different techniques coming from the theory of evolution equations.  
           
\end{abstract}

\setcounter{page}{1}

\secct{Introduction}

\subsection{Historical Context and Description of the Problem}

The processes of emission and absorption of photons by atoms can be rigorously understood in the low-energy limit, if we neglect the creation and annihilation of electrons. The corresponding theory is frequently referred to as \emph{nonrelativistic quantum electrodynamics (NR QED)}. The description of matter and light from the mathematical point of view relies on the study of eigenvalues of operators, which are immersed in the continuum. The study of eigenvalues immersed in the continuum requires sophisticated constructions that do not fall into the realm of regular perturbation theory used to analyze isolated eigenvalues (see \cite{ReedSimonII1980}). There are two methods that have been applied to investigate these questions. The first one, introduced in [\cite{nue}, \cite{BFS98a}], is the spectral renormalization group. Inspired by a construction that Feshbach used in \cite{F1958}, the Feshbach-Schur projection map is defined and developed [\cite{nue}, \cite{BFS98a}]. This method is based on a transformation that permits us to localize the regions of the spectrum that we are interested in. The second method, introduced in \cite{P2003}, produces a sequence of isolated eigenvalues that converges to the desired eigenvalue, that is immersed in the continuum by including ever more momentum shells into the dynamics.

The spectral renormalization group has been extensively used to
analyze spectral problems in non relativistic quantum
electrodynamics. In numerous works this method has been used to prove
several properties of the spectrum of Hamiltonian operators for
different models (see for example \cite{4}-\cite{3} and
\cite{6}-\cite{9}). Although it is a powerful tool to analyze the
spectrum of operators, it yet has the disadvantage of being
technically and conceptually complicated. Further developments of the
original method have been presented in \cite{BCFS2003}, \cite{GH2008},
\cite{FGS2009}.  In these works, new techniques and methods are
presented that simplify the computations and clarify the concepts of
the original procedures in \cite{nue} and \cite{BFS98a}.  In this paper we present a
new approach to the renormalization group described in the following
section. Interestingly, this new approach uses the spatial
  length scale as a flow paramter and is thus closer to the original
  renormalization group introduced by Kogut and Wilson in
  \cite{KogutWilson1974} and subsequently improved by Polchinski in
  1984 \cite{Polchinski1984} and later Wieczerkowski
  \cite{Wieczerkowski1988} and Salmhofer in 1998
  \cite{Salmhofer1998}.

\subsection{Short Description of the Main Results}\label{short}

In this section we describe our main results. We present a short description of the 
method and the most important theorems without giving precise definitions of the operators 
and spaces that we use. The definitions are deferred to later sections.   

Our intention is to present the method, rather than the specific complications of each model of nonrelativistic quantum field theory. For this reason, and for the sake of simplicity, we restrict ourselves to a class of operators that is as simple as possible but that, at the same time, includes all important ingredients. Our method can be easily applied to a bigger class of operators that includes the Pauli-Fierz and the Spin-Boson models.

\vspace{.5cm}

{\bf The Operators}

\vspace{.5cm}

The Hilbert space on which the operators are defined is the symmetric Fock space (see Section \ref{Fock-Space})
\eq{\label{infock}
\cH = \bigoplus_{N = 0}^{\infty} (L^2(\RR^3))^{\otimes_s^N}\period
}
For every $k, \tilde k  \in \RR^3$, we denote by $a^*(k)$ and $a(\tilde k)$ the creation and annihilation operators (see Section \ref{screan}) acting in the Fock space $\cH$. 
The free Hamiltonian is [see \eqref{h.5}]
\eq{\label{int1}
H_f : = \int_{\RR^3} |k| a^{*}(k)a(k)\period 
} 
We fix a positive number $\rho \in (0,1)$ and use the symbol $D_{\rho/2}$ for the disc in the complex place of 
radius $\rho/2$ centered at the origin.

We denote by 
$$ \uw = (w_{m,n})_{m,n  = 0}^\infty $$
a family of measurable functions 
$$
\forall m, n  \geq 0  \: : \: w_{m,n} : D_{\rho/2} \times [0, \infty) \times (\RR^3)^m \times  (\RR^3)^n \to \CC \comma
$$
and, for every $\alpha \geq 0$, by $\chi_\alpha$ the characteristic function in $\RR$ of the set $[0, e^{-\alpha}\rho]$. We identify
$$
\chi_\alpha \equiv \chi_\alpha(H_f)\comma \hspace{1cm} \overline \chi_\alpha \equiv 1 - \chi_\alpha(H_f) \period
$$ 
 We study a class operators of the form
\eq{ \label{orale}
H(z) = T(z) + W(z)\comma
}  
where 
\eq{ \label{sepasa}
T(z) = w_{0,0}(z, H_f) \chi_0
}
and 
\eq{\label{hiint7}
W(z) = \sum_{m + n \geq 1} \chi_0 W_{m,n}(\uw) \chi_0 \comma
}
with
\eq{ \label{hiint6}
W_{m,n}(\uw) = & \int dk^1 \cdots dk^m d \: \tilde k^1 \cdots d\tilde k^n  \:
 a^*(k^1) \cdots a^*(k^m) \\ \notag &
  w_{m,n}(z; H_f; k^1, \cdots, k^m; \tilde k^1, \cdots, \tilde k^n) 
  \: a(\tilde k^1) \cdots a(\tilde k^n) \period
}
The specific properties that sequences of functions $\uw$ satisfy are defined in Section  \ref{secfs}. 

\vspace{.5cm}

{\bf The Feshbach-Schur Map}

\vspace{.5cm}

Our goal is to analyze the spectrum (and eigenvalues) of operators of the form
$$
H = T + W \comma
$$
where $T$ and $W$ do not depend of $z$ and are given by \eqref{sepasa}-\eqref{hiint6}. 
We study in particular the spectral  points in $D_{\rho/2}$ of the operator $H$. For this purpose we define a family of Hamiltonians parametrized by a complex 
number $z \in D_{\rho/2}$ as follows
\eq{\label{param}
 D_{\rho/2} \ni z \longrightarrow H(z) : = H + z.
}
The spectral points of $H$ in $D_{\rho/2}$ are, thus, the complex numbers $-z \in D_{\rho/2}$ such that 
$H(z)$ is not bounded invertible, and the eigenvalues are the complex numbers $- z \in D_{\rho/2}$ such that 
$ H(z) $ is not injective (see Remark \ref{param}).

A mathematical tool that we use is the Feshbach-Schur map 
\eq{\label{pror}
F_\alpha & (H(z)) =  \chi_\alpha H(z) \chi_\alpha  
 -  \chi_\alpha H(z)\overline{\chi_\alpha}(\overline{\chi_\alpha} H(z) \overline{\chi_\alpha}  )^{-1} \overline{\chi_\alpha} H (z) \chi_\alpha \period
}
Eq. \eqref{pror} is well-defined (or exist) if $\overline{\chi_\alpha} H(z) \overline{\chi_\alpha}$ is (bounded) invertible, which we rephrase saying that $H(z)$ belongs to the domain of $F_\alpha$.   

The Feshbach-Schur map has the important property of being isospectral in the sense that 
\begin{itemize}
\item[(a)] $H(z)$ is (bounded) invertible if, and only if, $ F_\alpha(H(z)) $ is (bounded) invertible.

\item[(b)] $H(z)$ is not injective if, and only if, $ F_\alpha(H(z)) $ is not injective.
\end{itemize}

\vspace{.5cm}

{\bf The Rescaled Feshbach-Schur Map}

\vspace{.5cm}

\noindent For every $\alpha \in \RR$,
we denote by $u(\alpha): \mathfrak{h} \to \mathfrak{h}$ 
 the unitary (dilation) operators on $\mathfrak{h}$.
 The unitary operator $u(\alpha)$ is defined by the 
formula 
\beq \label{hid1}
(u(\alpha)\phi)(k) := e^{ -3 \alpha /2}\phi(e^{-\alpha} k), \, 
 \,\forall \phi  \in L^{2}(\RR^3) \period 
\ene
We denote by $\Gamma_\alpha $ the operator that results after lifting $u(\alpha)$ to the 
Fock space. $u(\alpha)$ represents a re-scaling of the photon momenta. 

\noindent We do a slight modification of the Feshbach-Schur map and  define another isospectral map by     
\begin{equation} \label{Hirmn.7}
\widehat{\cR_\alpha}(H(z)) : = e^\alpha \Gamma_\alpha F_\alpha (H(z)) \Gamma_\alpha^*.
\end{equation} 
Re-scaling the Feshbach-Schur map is not necessary, but convenient. It amounts to find the right scale for the photon 
momentum that makes the free photon Hamiltonian $\chi_0 H_f$ invariant, i.e., $\chi_0 H_f$ is a fixed point for $\widehat{\cR_\alpha}$.

The rescaled Feshbach-Schur has the important property of being isospectral in the sense that 
\begin{itemize}
\item[(a)] $H(z)$ is (bounded) invertible if, and only if, $\widehat \cR_\alpha(H(z)) $ is (bounded) invertible.

\item[(b)] $H(z)$ is not injective if, and only if, $\widehat \cR_\alpha(H(z)) $ is not injective.
\end{itemize}
Note that the rescaled Feshbach-Schur map is not defined for all operators $H$ and all $z$, and (a) and (b) are valid only for $H$ and $z$ such that $H(z) \in \dom(\widehat \cR_\alpha)$, i.e., for $H$ and $z$ such that $\overline{\chi_\alpha} H(z) \overline{\chi_\alpha}$ is (bounded) invertible.

Another important property of the rescaled Feshbach-Schur map is that, for $H(z)$ belonging to its domain, there exist operators $T_\alpha(z)$ and $W_{\alpha}(z)$
such that
\eq{\label{ecua}
\widehat \cR_{\alpha}(H(z)) = T_\alpha(z) + W_\alpha(z) \comma
}
where the spectrum of $T_\alpha(z)$ is explicit and $W_\alpha(z)$ decays (in operator norm) exponentially in 
$\alpha$, as $\alpha$ tends to infinity. As $W_\alpha (z)$ decays exponentially in $\alpha$, we obtain increasingly accurate information on the invertibility of $\widehat \cR_{\alpha}(H(z))$- hence on the invertibility of $H(z) = H + z$, thanks to the isospectrality (a), (b)-, as $\alpha$ gets large.

\vspace{.5cm}

{\bf Main Results}

\vspace{.5cm}

One of the main results that we prove in this paper is the following (see Theorem \ref{cpr}):
\begin{theorem}\label{in1}
 There exists a family $\{ \boE_s \}_{s\geq 0}$ of biholomorphic functions 
\eq{\label{s}
\boE_s : D_{\rho/2} \to  \boE_s( D_{\rho/2} ) \subset D_{\rho/2}
}
such that  
\eq{
\forall \zeta \in \boE_s(D_{\rho/2}) \: : \:  H(\zeta) = H + \zeta \in \dom (\widehat \cR_s) \period
}
\end{theorem}   
Using Theorem \ref{in1} we can define a family  $\{ \boH_s(z) \}_{s\geq 0}$ of operators by
\eq{\label{lah}
 \forall z \in D_{\rho/2}\: : \: \boH_s(z)  =  \widehat \cR_s(H(\boE_s(z))), \hspace{1cm}
 \boH_0(z) = H + z = T + z + W  \period
} 
Another result that we prove is the next theorem (see Theorem \ref{cpr})
\begin{theorem}\label{in2}
For every $s \geq 0$, there exist operators $\boT_s(z)$ and $\boW_s(z)$ such that
\eq{\label{hlelo}
\boH_s(z) = \boT_s(z) + \boW_s(z)\comma
}
where the spectrum of $ \boT_s(z) $ can be computed explicitly and there are constants $\mu > 0$ and $C >0$
such that 
\eq{\label{hlelol}
\|  \boW_s(z)  \| \leq C e^{- \mu s/4}\period
}
In the case that $s = 0$ we take $\boT_0(z) = T + z$ and $ \boW_0(z) = W$.

The operator $\boT_s(z)$ is actually a function of $H_f$:
\eq{ \label{compa}
\boT_s(z) = \tau_s(z, H_f), 
}
for a function $\tau : D_{\rho/2}\times [0, \infty) \to \CC $. 
This implies that $\boT_s(z)$ has a simple and well known spectral decomposition.  

\end{theorem}

We call the family of functions $\{\boE_s\}_{s \geq 0}$ the {\it renormalization} flow (or {\it renormalization group}) of the spectral parameter ($z$) and the set of 
operators $\{ \boH_s(z) \}_{s \geq 0}$ the renormalization flow (or renormalization group) of operators with initial condition $\boH_0(z) = H(z)$. The assignment of the name \emph{flow} (or \emph{group}) has actually a mathematical meaning in the sense that 
$\{ \boH_s(z) \}_{s \geq 0}$ is the orbit of $H(z)$ under a flow $\bf \Phi(\cdot, s)$. This is the content of the next theorem (see Theorem \ref{sgp}). 

\begin{theorem}\label{flowsn}
For a suitable space of operators $ H[\cW_\xi]^{(0)}$ there is a flow map 
$$
{\bf \Phi} : H[\cW_\xi]^{(0)}\times [0, \infty) \mapsto H[\cW_\xi]^{(0)} \comma
$$
such that, for every $H(z) \in H[\cW_\xi]^{(0)}$ and every $s \geq 0,$
$$
 {\bf \Phi}(H(z), s) = \boH_s(z) \period
$$ 
The function $\bf \Phi$ satisfies the flow (or semigroup) property
\eq{\label{semg}
{\bf \Phi} (\boH_s(z), t) = {\bf \Phi} (H(z), s+t)\comma \hspace{2cm} 
\forall s,t \geq 0, \, \forall H(z) \in H[\cW_\xi]^{(0)}\period
}

\end{theorem}
   
{\bf The Differential Equation} \\

A natural question to ask is whether the flow $\{ \boH_s(z) \}_{s \geq 0}$ constructed in Theorem \ref{flowsn} is associated to a differential equation.
The answer is affirmative. Actually it is possible to compute (formally) the derivative
\eq{  \label{deri}
\frac{\partial}{\partial t} \boH_t = & \boH_t + [ A, \boH_t] 
+ \Big ( - z + \frac{ \la \boW_t \delta(H_f - 1) \boW_t  \ra_\Omega }{\tau_{t}(z, 1)}\Big )   \frac{\partial}{\partial z} \boH_t  \\ &  - \frac{  \boW_t \delta(H_f - 1) \boW_t   }{\tau_{t}(z, 1)} \comma \notag
} 
where $\delta$ refers to the Dirac delta, $A$ is the generator of the group $( \Gamma_\alpha)_{\alpha \in \RR}$:
$$
\Gamma_\alpha =  e^{\alpha A},
$$
and $\la \cdot \ra_\Omega $ is the vacuum expectation value. Eq. \eqref{deri} is understood in the sense of quadratic forms in $\chi_0 (\cH)$.
 
The quadratic form in the right hand side of \eqref{deri} is not well-defined for the class of operators that we use. 
To make sense of it, it is necessary to assume regularity conditions for the initial Hamiltonian
$H$ and to prove that this regularity is preserved by the renormalization flow. The rigorous justification of 
\eqref{deri} is deferred to a future work, since the inclusion of it here would increase considerably the length of this paper.  

Its is proven in Theorem \ref{itisp} that there is a flow of sequences of functions $\{\uw_s\}_{s \geq 0}$ such that
$$
\forall s \geq 0 \: : \: \boH_s(z) = H(\uw_s(z))  \comma
$$
where $H(\uw_s(z))$ is defined as in \eqref{orale}-\eqref{hiint6} (here we write explicitly the dependence on 
$ \uw_s(z))$. Eq. \eqref{deri} produces additionally a differential equation on the sequences of functions $\{\uw_s\}_{s \geq 0}$, which might actually be more useful that the quadratic-form version of the equation (for the estimates). To make use of the differential equation one should start with an initial condition given by a sequence of functions $\uw$ for which all elements have a finite number of surfaces of discontinuity, outside of them the functions should be  $C^1$ in the photon variables. We expect that the surfaces of discontinuity are preserved by the renormalization flow and that there are no new discontinuities arising.   

To define precisely the differential equation \eqref{deri} on the sequences functions $\{ \uw_s\}_{s \geq 0}$, it is unlikely that the norms involved could include a derivative with respect to the variable $ r \in [0, \infty)$. An intuitive (imprecise) reason is the following: The sequences of functions $\{ \uw_s\}_{s \geq 0}$ have at most one derivative with respect to $r \in [0,\infty)$. If we could compute the derivative in \eqref{deri}  using a norm that already includes one derivative with respect to the parameter $r \in [0, \infty)$, then, as a result, we would have computed two derivatives  with respect to $ r \in [0, \infty)$) (one coming from the norm and the other from the derivative in \eqref{deri}), which is not possible.

\begin{remark}
Although the renormalization semi-group of operators $\{ \boH_s(z) \}_{s \geq 0}$  is an explicit solution to the differential equation \eqref{deri}. It is interesting to construct solutions of \eqref{deri} using standard methods of differential equations. This would generate new alternative techniques to study spectral properties of operators in nonrelativistic quantum electrodynamics.

\end{remark}

\begin{remark}
We could choose $H-z$ instead of $H + z$ to define $H(z)$ in \eqref{param}. This selection would be more natural
from the point of view of spectral theory. Nevertheless, the selection 
$$
H(z) = H + z
$$
is convenient because it simplifies the notation in many places of the paper.
As the complex number $z$ is directly related to the spectrum of $H$, we call it the spectral parameter. 

\end{remark}

\subsection{Comparison with Previous Works}    
The main interest of the previous works (see \cite{BFS98a}  and \cite{BCFS2003}, for example) is
the analysis of the spectrum of operators. Starting with an operator $H(z)$ as in Section \ref{short},
 a sequence of operators $ \{H_{n \alpha_0 }(z)\}_{n \in \mathbb{N}}$ is constructed. The sequence of operators is labeled by a discrete set of numbers $\{ n \alpha_0 \}_{n \in \NN}$ for which it is necessary to take a sufficiently 
 large, fixed, $\alpha_0$. The sequence of operators $  \{ H_{n \alpha_0 }(z)\}_{n \in \mathbb{N}} $ also satisfies \eqref{hlelo} and \eqref{hlelol}, if we take $s = n \alpha_0$. For every $n \in \NN$ there is a biholomorphic function 
\eq{\label{fr}
\widetilde Q_n : D_{\rho/2}  \to \widetilde Q_n(D_{\rho/2}) \subset  D_{\rho/2}
}
that satisfies the following properties: 
\begin{itemize}
\item[(a')] $H_{n\alpha_0}\big( \widetilde Q_n(z)\big)$ is invertible if, and only if, $ H_{(n+1) \alpha_0}(z) $ is invertible. 

\item[(b')] $H_{n\alpha_0}\big( \widetilde Q_n(z) \big)$ is not injective if, and only if, $ H_{(n+1) \alpha_0}(z) $ is not injective. 
\end{itemize}
Items (a') and (b') together with \eqref{hlelol} facilitate the analysis of the spectrum of the original operator by an iterated composition of the functions $ \{\widetilde Q_n \}_{n \in \NN} $. 

The inputs of our new approach are the following:  
\begin{itemize}
 \item We define a continuous family of operators $\{ \boH_s(z)\}_{s \geq 0}$ that is parametrized by a real number $s \in [0, \infty)$.

\item The renormalization group of operators that we construct has a simple and conceptually clear interpretation. 
It is only one single application of the (rescaled) Feshbach-Schur map with a clever selection of the spectral parameter given by the continuous flow of functions $\{ \boE_s\}_{s \geq 0}$ 
[see Theorem \ref{in1}].    

\item In spite of its name, the set of operators $ \{ H_{n \alpha_0}(z) \}_{n \in \NN} $ is not
a (discrete) semigroup (on $\alpha_0 \NN  $) of operators [see Remark \ref{rem}]. It does not satisfy \eqref{semg} for $s, t \in \alpha_0 \NN$. This is because the renormalization of the spectral parameter \eqref{fr} is not properly chosen [see Remark \ref{rem}]. Here we choose instead another function to renormalize the spectral parameter, it is denoted by $Q_\alpha^{-1}$, where $Q_\alpha$ is explicitly written in \eqref{int10}.  It is possible, nevertheless, that the set of operators $ \{ H_{n \alpha_0}(z) \}_{n \in \NN} $  satisfies a groupoid property, instead a group semi-property, [See Remark \ref{remfr}].

\item We can define a flow for which the renormalization group is the orbit of the operator $H(z)$ under this flow [see Theorem \ref{flowsn}].  

\end{itemize}

\noindent In our approach we do not use the smooth Feshbach-Schur map introduced in \cite{BCFS2003} because we don't know how to make it compatible with the semigroup property. Instead, we use the original Feshbach-Schur map projection method 
introduced in \cite{nue} and \cite{BFS98a}, inspired by the procedures of \cite{F1958}. We modify the construction of \cite{BFS98a} in order to fulfill (\ref{semg}). This is an important input of our method. We also use different seminorms to make the computations mathematically precise.

Our results provide a new mathematical structure for the renormalization group. As a semigroup 
$\{\boH_s(z)\}_{s \geq 0}$, can be viewed as the orbit of a flow with initial condition $H(z)$. This opens a new perspective, in which the renormalization group can be regarded as the solution to an (autonomous) evolution equation. This was not possible before because  any solution to an evolution equation has to be a flow (although it might be possible that the previous scheme would give rise to an non-autonomous differential equation, see Remark \ref{remfr}). The study of the renormalization flow, from the perspective of differential equations, is an interesting and challenging mathematical problem.

An important area in evolution equations is the study of stable and unstable manifolds. Although these concepts are already used for the renormalization group (see \cite{BFS98a}, for example), this analysis is done on a sequence of operators which is not proved to have the structure of a (discrete) flow and is not related, therefore, to an (autonomous) evolution equation. Our results give new examples of evolution equations in which the concepts of fixed point, stable and unstable manifolds can be applied to the spectral theory and have a clear interpretation.  Our results provide a nontrivial example of an explicitly solvable problem which has important applications to mathematical physics.

\begin{remark}\label{rem}
To construct a renormalization group of operators the selection (or renormalization) of the spectral parameter is fundamental. The renormalization of the spectral parameter localizes the regions in the complex plane in which the Feshbach-Schur map can be applied and it is, therefore, the ingredient that allows to apply the Feshbach-Schur map iteratively, which in turn define the sequence of operators $ \{ H_{n \alpha_0}(z) \}_{n \in \NN} $. Without the renormalization of the spectral parameter it is possible to apply the Feshbach-Schur map only once and the sequence of Hamiltonians cannot be constructed, the reason being that multiples of the identity define a relevant direction in the space of operators.  

In \cite{BFS98a}, the selection of the spectral parameter is represented by the inverse of the function 
\eq{ \label{rele}
\widetilde Q_n^{-1} = e^{\alpha_0}\la H_{n \alpha_0}(z)  \ra_\Omega \comma }        
where $\la \cdot \ra_\Omega$ is the vacuum expectation value (see Section \ref{S.hamiltonians}). It is this choice which breaks the proof of the semigroup property, despite the fact that the Feshbach-Schur map satisfies this property.    

In contrast with the previous works, we not only define a family of isospectral operators $\{ \boH_s(z) \}_{s \geq 0}$ parametrized by a continuous variable $s$ that satisfies \eqref{lah}, but we prove that the family satisfies a semigroup property (see also Remark \ref{remfr}). This is a consequence of a new definition of the renormalization of the spectral parameter, which produces a new renormalization semigroup of operators. The selection of the spectral parameter is done in such a way that         
\eq{\label{reles}
\la \boH_s(z) \ra_\Omega = z, \hspace{2cm} \forall s \geq 0.
}
Eq.~\eqref{reles} is the key input for the proof of the semi-group property.    

\end{remark}   
   
 \begin{remark} \label{remfr}
Although the set of operators $ \{ H_{n \alpha_0}(z) \}_{n \in \NN} $ does not satisfy a group property, it is possible that it satisfies a similar property (a groupoid property). This means that if we could construct a continuous set of operators $\{ H_s (z)\}_{s \geq 0}$ following the construction of the discrete set $ \{ H_{n \alpha_0}(z) \}_{n \in \NN} $,
then the continuous set of operators might satisfy a non-autonomous differential equation (with initial condition at $0$).  
 
 \end{remark}

\subsection{A Guideline of the Paper}
In this section we give an outline of our paper. First we briefly review the operators that we use (a more detailed description is done in Sections \ref{secfs} and \ref{S.hamiltonians}).



We denote by $\NN_0  = \NN\cup \{ 0\}$ and $B_{\rho}$ the ball of radius $\rho$ and center $0$ in $\RR^3$. We use the symbol $\uw$ to represent a sequence of functions
\eq{\label{int2}
\uw = \big( w_{m,n} \big)_{m,n \in \NN_0}\comma
}
such that, for every $m,n \in \NN_0$, 
$$
w_{m, n} : D_{\rho/2} \times [0, \infty) \times B_\rho^m \times B_\rho^n  \to \mathbb{C}
$$
is a measurable function, that is analytic with respect to the first variable belonging to $D_{\rho/2}$ and symmetric with respect to the variables belonging to $(\RR^{3})^m$  and $(\RR^3)^n$ (see Section \ref{secfs}, the analyticity is understood with respect to the seminorm \eqref{i.5elpex} in the sense of Remark \ref{nes}). 
The functions $w_{m,n}$ can be regarded as functions defined in $ D_{\rho/2} \times [0, \infty) \times \RR^{3m} \times B_\rho^{3n}  $ if we extend them by zero. We do this identification without mentioning it later on. If it is required, we write explicitly the dependence on the spectral parameter $z \in D_{\rho/2}$ as follows 
\eq{\label{zmn}
\uw = \big( w_{m,n} \big)_{m,n \in \NN_0} \equiv \big( w_{m,n}(z) \big)_{m,n \in \NN_0} = \uw(z) \period
}

\noindent In Section \ref{fuspa} we associate to every sequences of functions $\uw$ different seminorms. The seminorms associated to  $w_{m,n}$, 
\eq{\label{int3}
\| w_{m,n} \|^{(\infty)}, \hspace{1cm} \| w_{m,n} \|^{(0)}\comma
}
are defined in \eqref{i.5elpex}-\eqref{i.8el}.
 $ \, $ The most relevant of seminorms 
are introduced in \eqref{i.11.1}-\eqref{i.11}, they are denoted by 
\eq{\label{int4}
\| \uw \|_\xi^{(I)}, \hspace{1cm} \| \uw  \|^{(Z)},  \hspace{1cm}  \| \uw  - \ur\|^{(F)} \comma
} 
where $\ur : = \big( r_{m,n} \big)_{m,n \geq 0}$, with $r_{0, 0}(z, r) = r$, and  $ r_{m, n} = 0 , \: \: m+n \geq 1$.


In \eqref{int4}, $I$ stands for ``interacting", $Z$ refers to the derivative with respect to the spectral parameter $z$ that is considered in the definition of the semi-norm, and $F$ stands for ``free". We denote by $ \cW_\xi $ the space of sequences of functions for which the seminorms \eqref{int4} are finite [see \eqref{i.9}]. 
Here $\xi\in (0,1)$  is a small parameter.  

In Section \ref{basic-a} we define additional restrictions for the sequences of functions $\uw \in \cW_\xi$. 
To this end, we define a set $\uE_\alpha$ consisting of triples of positive numbers    
$$
\ue = (\epsilon_I, \epsilon_Z, \epsilon_F) 
$$
that satisfy certain properties [see Definition \ref{hice}]. \\  
We finally define the set $\widetilde W_\xi $ of sequences of functions $ \uw \in \cW_\xi$ such that 
\eq{\label{int5}
\| \uw \|^{(I)}\leq \epsilon_I, \hspace{1cm} \| \uw  \|^{(Z)}\leq \epsilon_Z,  \hspace{1cm}  \| \uw  - \ur\|^{(F)}
\leq \epsilon_F \comma
} 
for some $\ue \in \uE_\alpha$. With this definition we complete the number of properties that are important concerning 
the functions spaces. The operators that we study are constructed from the sequences of functions as follows (see Section \ref{S.hamiltonians}): For every $\uw \in \cW_\xi$ we define [see \eqref{Wmn}] 
\eq{ \label{int6}
W_{m,n}(\uw) = & \int dk^1 \cdots dk^m d \: \tilde k^1 \cdots d\tilde k^n  \:
 a^*(k^1) \cdots a^*(k^m) \\ \notag &
  w_{m,n}(z; H_f; k^1, \cdots, k^m; \tilde k^1, \cdots, \tilde k^n) 
  \: a(\tilde k^1) \cdots a(\tilde k^n)
}
and by [see \eqref{hw}]
\eq{\label{int7}
H(\uw ) = \sum_{m,n} \chi_0 W_{m,n}(\uw) \chi_0 \comma
}
where $\chi_0$ is a projection introduced in Definition \ref{n}.
The image under $H$ of $\cW_\xi$ is denoted by $H\cW_\xi$. This is the space of operators that we study.  

In Section \ref{Basic-Estimates} we derive some norm estimates on the operators $ W_{m,n}(\uw) $ in terms of the 
seminorms of $\uw \in \cW_\xi$.  

In Section \ref{cf} we address the question of whether a composition of two operators in $H \cW_\xi$ can be written of the form
\eqref{int7}. We state the results that we need and refer to \cite{BFS98a} for the proofs.
Much of the notation that we use throughout the paper is introduced in Section \ref{cf}.  

In Section \ref{S.feshbachmap} we introduce the Feshbach-Schur map [see \eqref{fm.3}]. We prove that if 
$\uw \in \widetilde \cW_\xi$ and $z \leq (\rho/2)e^{-\iota \alpha}$ [see \eqref{rho} and 
\eqref{alphaplus}-\eqref{iota}], then the Feshbach-Schur map
\eq{\label{int.8}
F_\alpha(H(\uw(z)))
}
is well defined (see Lemma \ref{L.fm}). 

In Section \ref{S.RG} we define the renormalization map. It consists on a slight modification of the Feshbach-Schur map, which we call rescaled Feshbach-Schur map [see \eqref{rmn.7}], and a renormalization of the spectral parameter (see Section \ref{rensp}). We denote by 
\eq{\label{int9}
\widehat \cR_\alpha (H(\uw(z)))  
}
the rescaled Feshbach-Schur map applied to $H(\uw(z))$, and by   
\eq{\label{int10}
Q_\alpha(z) = \la \Omega |\; \widehat \cR_\alpha (H(\uw(z)))  \Omega \ra \comma
}
where $\Omega $ is the vacuum.
In Section \ref{rensp} we prove that $Q_\alpha$ is invertible. Then we define the renormalization operator by 
\eq{\label{int11}
 \big(\cR_\alpha (H(\uw))\big)(z) : = \widehat \cR_\alpha (H(\uw(\zeta)))\comma  
}
where $\zeta = Q_\alpha^{-1}(z)$ is the renormalization of the spectral parameter. 

Section \ref{S.contraction} is the biggest part in this paper where most of the laborious computations and estimates are done. There we prove that for every $\uw \in \widetilde \cW_\xi$ there exists a sequence of functions $\widehat \uw = \big( \widehat w_{m,n}\big)_{m,n \in \NN_0} $ such that
\eq{\label{int12}
\cR_\alpha (H(\uw))(z) =   H(\widehat \uw(z))\period
}
We study the seminorms \eqref{int3} for $\widehat \uw$. To this purpose we first construct a sequence of functions 
$\widetilde \uw^{(sym)} = \big( \widetilde w_{m,n}^{(sym)}\big)_{m,n \in \NN_0} $ such that
\eq{\label{int13}
F_\alpha (H(\uw(z))) =\chi_\alpha H(\uw) \chi_\alpha + \chi_\alpha H(\widetilde \uw^{(sym)}(z))\chi_\alpha \period
}
The analysis of the seminorms \eqref{int3} for $\widetilde \uw^{(sym)}$ is done in Section \ref{sano}. This section requires long computations and is divided into many subsections. The title of each subsection indicates which term or which norm is being estimated. The difficult part of Section \ref{S.contraction} is the estimation of the seminorms for $ \widetilde \uw^{(sym)} $. Once this is achieved, the analysis of the seminorms \eqref{int3} for $\widehat \uw$ is straightforward and is done in Section \ref{repite}. This concludes Section \ref{S.contraction}.     

In Section \ref{rf} we construct a series of iterations of the renormalization map as follows:\\
Given a sequence of positive real numbers 
$$
\ual : = \{ \alpha_j\}_{j \in \NN}
$$
and an initial sequence of functions $\uw_\ual^{(0)}\in \widetilde \cW_\xi$, we give conditions on $ \epsilon_I, \epsilon_Z $ and $\epsilon_F$ [see \eqref{int5}]  and on the sequence $\ual$, in order to assure that the iterated renormalization map 
\eq{\label{int13-1}
H(\uw_\ual^{(\ell)}) : =     \Big(\cR_{\alpha_\ell}\circ \cdots\circ \cR_{\alpha_1}\Big)(H(\uw_\ual^{(0)}))
}
is well-defined.

To achieve our purpose, we define a sequence of triples $\{ \ue_\ual^{(\ell)} \}_{\ell \in \NN_0}$ 
\eq{\label{int14}
\ue^{(\ell)}_{\ual} := (\epsilon_{I, \ual}^{(\ell)}, \epsilon_{Z, \ual}^{(\ell)},\epsilon_{F, \ual}^{(\ell)})
}
[see \eqref{i}-\eqref{3}] and prove inductively [see Section \ref{ic}] that for every $\ell \in \NN_0$ a sequence of functions $\uw_\ual^{(\ell)}$ can be constructed, that satisfies \eqref{int13-1} and 
\eq{\label{int15} 
\| \uw^{(\ell)}_\ual \|^{(Z)} \leq \epsilon_{Z, \ual}^{(\ell)} \comma \:
 \| \uw^{(\ell)}_\ual - \ur \|^{(F)} \leq  \epsilon_{F, \ual}^{(\ell)}\comma \:
\| \uw^{(\ell)}_\ual \|^{(I)}_{\xi}\leq \epsilon_{I, \ual}^{(\ell)}\period}
In particular we obtain that the interacting part, that is controlled by $ \epsilon_{I, \ual}^{(\ell)}  $, decreases  exponentially to zero as $\ell$ goes to $\infty$ [see \eqref{3}].  
 
In Section \ref{sacelfua} we prove our main results. We prove Theorems \ref{in1}, \ref{in2} and \ref{flowsn}. We fix an initial sequence of functions $\uw$ satisfying \eqref{int15} for $\ell = 0$ (with $\uw$ instead of $\uw_\ual^{(0)}$). 
For every $s \in [0, \infty)$ we select a sequence $\ual $ and a number 
$ \beta \in [0, \alpha_+]$, see \eqref{alphaplus}, such that 
$$
\sum_{j=1}^\ell \alpha_j  + \beta = s\comma
$$
for some $\ell \in \mathbb{N}_0$ (if $\ell = 0$ we omit the sum).  We define the operator 
$$
\boH_s   = \cR_\beta( H(\uw_\ual^{(\ell)})) \period 
$$
We prove that, for $\ell \geq 1$, there exists $ \uw_\ual^{(\ell, \beta)} \in \widetilde \cW_\xi $ such that
$$
\boH_s  = H(\uw_\ual^{(\ell, \beta)})
$$
(for $\ell = 0$ and $\xi < 1/4$  we can find $ \uw_\ual^{(\ell, \beta)} \in \widetilde \cW_{4\xi} $ satisfying the same equality).\\
We define
$$
\boT_s =(w^{(\ell, \beta)}_\ual)_{0,0}(z, H_f)\comma \hspace{2cm} \boW_s = \boH_s - \boT_s.
$$
Then we have 
$$
\boH_s   =  \boT_s +  \boW_s\period
$$
The operators $\boH_s$ define a family of isospectral operators for which the interacting part $\boW_s$ satisfies \eqref{hlelol} [see (\ref{contaction})]. In Section \ref{sacelfua} we prove that 
the family of operators is well defined, in the sense that it does not depend on $\ual $ and $\beta$. 
The key ingredient is the construction of the (continuous) renormalization of the spectral parameter 
$$
\boE_s : D_{\rho/2} \to D_{\rho_2}\comma
$$
which is an analytic open injective function with analytic inverse. It has the following properties:

\begin{itemize}
\item For every $z \in D_{\rho/2}$, the rescaled Feshbach-Schur map $\widehat \cR_s\Big(H\big(\uw(\boE_s(z))\big)\Big)$ is well defined.

\item The following equation holds true
\eq{\label{acajosn}
\boH_s(z) = \widehat \cR_s\Big(H\big(\uw(\boE_s(z))\big)\Big)\period
}

\end{itemize}  
In Section \ref{renflow} we construct the function $\boE_s$. We prove furthermore \eqref{acajosn} and the exponential decay of the interacting term. \\
In Section \ref{renomfunc} we define a set of sequences functions $\{ \uw_s \}_{s \geq 0}$ such that
\eq{\label{tisn}
\boH_s = H(\uw_s)\period
}
In Section \ref{flowop} we define a space of operators $ H[\cW_\xi]^{(0)} $ and a flow 
$$
{\bf \Phi} :  H[\cW_\xi]^{(0)} \times [0, \infty) \to  H[\cW_\xi]^{(0)}
$$
whose orbits are the sets $\{ \boH_s\}_{s \geq 0}$. In particular this proves that 
the operators $\boH_s$ satisfy a group property. We furthermore define the corresponding flow in the function spaces. 

\vspace{.5cm} 
{\bf Acknowledgments}
\vspace{.5cm}

This project started in collaboration with Israel Michael Sigal. The initial idea was originated in several discussions with him that continued for a long period. We are thankful with Israel Michael Sigal for his contributions to this project. We are grateful with Israel Michael Sigal additionally for having reviewed this manuscript and for his useful comments that improved the presentation of this paper.

\secct{Function Spaces}\label{secfs}

\subsection{Notation}\label{S.notation}

We use the symbol $\NN_0$ for the set 
\eq{ \label{noma1}
\NN_0 : = \mathbb{N}\cup \{ 0 \} \period
}
The ball in $\RR^3$ with radius $r$ and center $0$ is denoted by
\eq{ \label{noma2}
B_r : = \{ x = (x_1, x_2, x_3) \in \mathbb{R}^3 \: :\: |x| \leq r   \}\comma
}
where 
\eq{\label{noma3}
|x| = \big ( x_1^2 + x_2^2 + x_3^2 \big )^{1/2}\period 
}

For every $m \in \NN$, $ B_r^m$ is the Cartesian product of $m$ copies of $B_r$ and 
$B_r^{(m,n)} = B_r^m \times B_r^n$.\\
The disc in the complex plane with radius $r$ and center $0$ is represented by 
\eq{\label{noma4}
D_r : = \{ z \in \mathbb{C} \: : \: |z| \leq r \} \period
}
We remark that \eqref{noma1}-\eqref{noma4} were already partially presented in the introduction. Nevertheless we
repeat them in this notation section to make easier the reader finding them.

We use the symbol   
\eq{\label{st}
\triangle_\rho : = \Big \{ (r, s) \in (0, \infty)\times [0, \infty) \: : \: r + s \leq \rho  \Big \} \period
}

For any set of vectors $ k^1, k^2, \cdots, k^m, \tilde k^1, \cdots \tilde k^n  \in \mathbb{R}^3$ we define 
\eq{\label{noma6}
k^{(m)} : = & (k^1, k^2, \cdots, k^m) \in (\mathbb{R}^3)^{m}\comma 
\\ \notag \tilde k^{(n)} : = & (\tilde k^1, \tilde k^2, \cdots, \tilde k^n) \in (\mathbb{R}^3)^{n}\comma  
\\ \notag k^{(m,n)} := & (k^{(m)}; \tilde k^{(n)})  \in  (\mathbb{R}^3)^{m}\times (\mathbb{R}^3)^{n}  \comma
}
and 
\begin{align}\label{n.1}
 |k^{(m)}| : & = |k^1|\cdot | k^2| \cdots | k^m|, \hspace{2cm} |k^{(m, n)}| :  = |k^{(m)}|\cdot |\tilde k^{(n)}| \comma \\ \notag
 \| k^{(m)} \|_1 : & = \sum_{i= 1}^m |k^i|, \hspace{3.3cm} \| k^{(m, n)} \|_1 :  = \| k^{(m)} \|_1 + \|\tilde k^{(n)} \|_1 \comma \\ \notag 
 dk^{(m)} : & = dk^1\cdot dk^2 \cdots dk^m, \hspace{2cm} dk^{(m, n)} :  = dk^{(m)}\cdot d\tilde k^{(n)},
\end{align}
where $d k^i$ is the Lebesgue measure in $\mathbb{R}^3$. We furthermore identify 
\eq{ \label{sup1}
k^{(m, 0)} : = k^{(m)}\comma \hspace{2cm} k^{(0,n)} : = \tilde k^{(n)}\comma
} 
for $n = 0$ or $m = 0$, respectively, and we omit $k^{(0,0)}$ altogether in case that $m=n=0$. 

\subsubsection{Fixed Parameters}

We select two positive real numbers  $\mu > 0 $ and $\rho > 0$, that are fixed throughout the paper,  
\eq{ \label{rho}
0 < \mu < 1\comma \hspace{2cm}  \rho =  \frac{1}{12^2}   \period
}
Eq.~\eqref{rho} is used several times in our estimations, without mentioning it always. The value or $\rho$ is used to estimate 
$$
4 \pi \rho < 1, \hspace{2cm} 6 \rho^{1/2} \leq \frac{1}{2}
$$
in many proofs throughout the paper.

\subsection{The Function Spaces $\cW(m,n)$}\label{fuspa}

\subsubsection{Definition of the Spaces}

For any pair $ (m, n) \in \mathbb{N}_0^2 $ we denote by $\cW (m, n)$ the space of measurable functions 
$$
w_{m, n} : D_{\rho/2} \times [0, \infty) \times B_\rho^{(m, n)}   \to \mathbb{C}
$$
that are analytic with respect to the variable $z \in D_{\rho/2}$ (see Remark \ref{nes}), symmetric with respect to the variables belonging to $(\RR^{3})^m$  and $(\RR^3)^n$ (separately), and that  satisfy the following properties
 :  

\begin{itemize}
\item[i)] For any $ (m, n) \in \mathbb{N}_0 \times \mathbb{N}_0  $ [recall \eqref{noma1}-\eqref{sup1}],

\begin{align}\label{i.1}
  \sup  \Big \{ & | w_{m, n}( z; r; k^{(m, n)})| \cdot |k^{(m, n)}|^{1/2 - \mu/2}  \\ \notag 
 & + | \frac{\partial}{ \partial z}w_{m, n}( z; r; k^{(m, n)})| \cdot |k^{(m, n)}|^{1/2 - \mu/2}  \\ \notag \: & \Big | \:  z\in D_{\rho/2},\, r \geq 0,\, k^{(m,n)} \in B_\rho^{(m,n)}  \Big \} < \infty \period
\end{align}

\item[ii)] For any $ (m, n) \in \mathbb{N} \times \mathbb{N}_0 \cup \mathbb{N}_0 \times \mathbb{N}, $
\begin{align}\label{i.2}
  \sup  \Big \{  &  \frac{1}{r} \int_{\|k^{(m)}\|_1, \| \tilde k^{(n)} \|_1 \leq \rho } \frac{d k^{(m,n)}}{ |k^{(m,n)}|^{3/2 + \mu/2}}    | w_{m, n}( z; s + r; k^{(m, n)} ) 
    \\ \notag &  - w_{m, n}( z; s; k^{(m, n)})|  
  \Big | \:  z\in D_{\rho/2},\, (r, s) \in \triangle_\rho  \Big \}  < \infty.
\end{align}


\item[iii)] For $m = n  = 0$,
\begin{align} \label{i.4}
  \sup  \Big \{   \tfrac{1}{r} |w_{0, 0}( z; s + r)  - w_{0, 0}( z; s  ) |  
  \Big | \:  z\in D_{\rho/2}, (r, s) \in \triangle_\rho \}< \infty.
\end{align}

\end{itemize}

\begin{remark}\label{nesi}
The elements $  w_{m, n}  \in \cW(m,n) $ are functions $w_{m, n} : D_{\rho/2} \times [0, \infty) \times B_\rho^m \times B_\rho^n  \to \mathbb{C}$. If required, we make the dependence on the variable $z \in D_{\rho/2}$ explicit, 
$$
w_{m,n} \equiv w_{m,n}(z) \period
$$
\end{remark}

\begin{remark}\label{nes}
The analyticity is understood with respect to the seminorm \eqref{i.5elpex} below, in the sense that there exists a function $\frac{\partial}{\partial z} w_{m,n}$ such that
$$
\lim_{h \to 0} \Big \|\frac{w_{m,n}(z + h) - w_{m,n}(z)}{h} - \frac{\partial}{\partial z} w_{m,n}(z) \Big \|_0^{(\infty)} = 0 \period
$$
Note that this implies point-wise analyticity of $ |k^{(m,n)}|^{1/2 - \mu/2} w(\cdot; r; k^{(m,n)})$, as we are 
using the supremum in the definition of $\| \cdot \|_0^{(\infty)}$- not the essential supremum.  
  
\end{remark}

\subsubsection{Semi-norms}

In this section we define some seminorms that are attributed to the space  $\cW (m, n)$.  We list these seminorms below. Let $ w_{m, n}  \in  \cW (m, n)$ and recall \eqref{noma1}-\eqref{sup1}. We define:  
\begin{itemize}
 \item[i)] For any $ (m, n) \in \mathbb{N}_0 \times \mathbb{N}_0  $

\begin{itemize}
\item[a)]
\begin{align} \label{i.5elpex}
 \| w_{m, n}  \|^{(\infty)}_0 : =  \sup  \Big \{ & | w_{m, n}( z; r; k^{(m,n)})| \cdot |k^{(m, n)}|^{1/2 - \mu/2} 
\\ \notag \: & \Big | \:  z\in D_{\rho/2},\, r  \geq 0,\, 
k^{(m, n)} \in B_\rho^{(m,n)}  \Big \}\comma
\end{align}
\item[b)]
\eq{ \label{i.5}
 \| w_{m, n}  \|^{(\infty)} : = \| w_{m, n}  \|^{(\infty)}_0 
 + \| \frac{\partial}{\partial z}w_{m, n}  \|^{(\infty)}_0\period
}

\end{itemize}

\item[ii)] For any $ (m, n) \in \mathbb{N} \times \mathbb{N}_0 \cup \mathbb{N}_0 \times \mathbb{N} $
\begin{align}\label{i.6el}
\| w_{m, n}  \|^{(0)}  : = \sup  \Big \{ &\frac{1}{r} \int_{\|k^{(m)}\|_1, \| \tilde k^{(n)} \|_1 \leq \rho } \frac{d k^{(m,n)}}{ |k^{(m,n)}|^{3/2 + \mu/2}} 
 \\ \notag &  | w_{m, n}( z; s+ r; k^{(m,n)}) - w_{m, n}( z; s; k^{(m,n)})|  
    \\ \notag & \hspace{1cm} \Big | \:  z\in D_{\rho/2}, (r, s) \in \triangle_\rho \Big \}.
\end{align}


\item[iii)] For $m = n  = 0$ 
\begin{align}\label{i.8el}
\| w_{0, 0}  \|^{(0)} : =   \sup  \Big \{\tfrac{1}{r}  |[w_{0, 0}( z; s +  r )  - w_{0, 0}( z; s  ) ] |  
  \Big | \:  z\in D_{\rho/2}, (r, s) \in \triangle_\rho  \Big \}.
\end{align}
\end{itemize} 

\subsubsection{The Spaces of Sequences of Functions $ \cW_\xi $}
We fix a parameter 
\eq{
\xi \in (0, 1).
} 
We use the symbol $\uw$ to denote a general sequence of functions of the form 
\eq{ \label{gen}
\uw : = \big( w_{m, n}  \big)_{m+n \geq 0}\comma
 }
 where $ w_{m, n} \in \cW(m, n), \; \forall m, n \in \mathbb{N}_0^2 $, and 
 $m+n \geq 0$, denotes $(m,n)\in \NN_0^2$. To every such sequence
 $\uw$ we associate the quantities 
\eq{ \label{i.11.1} 
\| \uw \|^{(I)}_{\xi} : = & \frac{2}{(1 - \xi)^2} \sup_{m+n \geq 1 } \xi^{-(m + n)}\Big[ \| w_{m, n}   \|^{(\infty)}   +  \| w_{m, n}   \|^{(0)} \Big]\comma \\
 \label{rmn.14}
\| \uw \|^{(Z)} := & \sup\Big \{ \Big |\frac{\partial }{\partial z} w_{0, 0} (z, r) \Big | : z \in D_{\rho/2}, r \in [0, \rho] \Big \}  \comma
}
with $m+n \geq 1$ denoting $(m,n) \in (\NN_0 \times \NN) \cup (\NN \times \NN_0),$
\eq{ \label{i.11} 
\| \uw - \ur\|^{(F)} : =  \| w_{0,0} - r \|^{(0)}\comma
}
where
\eq{\label{ereel2}
\ur : = \big( r_{m,n} \big)_{m,n \geq 0}
}
with 
\eq{\label{ere1el2}
r_{0, 0}(z, r) = r, \hspace{2cm} r_{m, n} = 0 , \: \: m+n \geq 1\comma
}
and $r$ is identified with the identity mar $r \mapsto r$ on $[0, \rho]$. In our notation $I$ stands for {\it interaction} and $F$ stands for {\it free}. \\
The space $ \cW_\xi $ of sequences of functions is defined by 
\beq \label{i.9}
\cW_\xi : = \Big \{   \uw = \big( w_{m, n}  \big)_{m + n \geq 0 }   \Big |  \| \uw \|^{(I)}_{\xi} + \| \uw \|^{(Z)} + \| \uw - \ur  \|^{(F)} < \infty \Big \} \period 
\ene

\begin{remark}\label{r.uw.z}
The elements $ \underline w =    \big( w_{m, n}  \big)_{m + n \geq 0 } \in \cW_{\xi} $ are sequences of functions $w_{m, n} : D_{\rho/2} \times [0, \infty) \times B_\rho^m \times B_\rho^n  \to \mathbb{C}$ that are analytic with respect to $z \in D_{\rho/2}$. If required, we make  the dependence on the variable $z$ explicit: 
$$
\uw \equiv \uw(z)\comma \hspace{3cm}  w_{m, n}  \equiv w_{m,n}(z)\period
$$
We furthermore use the notation
$$
\frac{\partial}{\partial z} \uw(z) =  \Big( \frac{\partial}{\partial z} w_{m, n}(z)  \Big)_{m + n \geq 0  } \period
$$
\end{remark}

\subsection{Further Definitions and Parameters}\label{basic-a}
\subsubsection{Definitions and Parameters}

\begin{definition}
We introduce three new parameters that we denote by $\alpha_+$, $\alpha$ and $\iota$, they satisfy the following: 
\eq{\label{alphaplus}
\alpha_+  > \frac{6}{\mu} > 6\comma 
}
\eq{\label{ok}
\alpha \in [0, \alpha_+]
}
and
\eq{\label{iota}
\iota = 1 -  \frac{1}{10 \alpha_+} \in \Big(\frac{59}{60}, 1\Big)\period
}
The constants $\alpha_+$ and $\iota$ are fixed. 
\end{definition}

\begin{definition}\label{ele}
For all $\epsilon_I, \epsilon_Z, \epsilon_F > 0$ such that 

\eq{ \label{besto}
(1 - \epsilon_F) -  \frac{1}{2} e^{1/10} - \epsilon_I (\rho e^{-\alpha_+})^{-1}  > \frac{1}{3} \comma
}
and every $\alpha \in [0, \alpha_+]$, we define

\eq { \label{HiE.L.r.2} 
 \cG_\alpha(\epsilon_I, \epsilon_Z, \epsilon_F):= & 
3 \Big(\frac{  \min(\alpha, 1) \epsilon_I^2  }{ e^{- \alpha}\rho} \Big) \Big[ 2 +
  \Big(\frac{3}{ e^{- \alpha}\rho } \Big) 
 \Big(\epsilon_Z  + \epsilon_I \Big)\Big] \cdot
}


\end{definition}

\begin{definition}\label{hice}
For $\alpha \in [0, \alpha_+]$, denote by $\uE_\alpha$ the set of triples of positive numbers 
$$
\ue : = (\epsilon_I, \epsilon_Z, \epsilon_F) \in (\RR^+)^3$$ 
 satisfying \eqref{besto} and the following properties:

\begin{itemize}
\item[(i)]
\beq \label{hirmn.6}
  \epsilon_I  (\rho e^{-\alpha})^{-1}  < \frac{1}{3 \cdot 2^4} \comma 
\ene

\item[(ii)]
\eq{\label{E.L.r.2.1}
\cG_\alpha(\ue) < 1\comma
}
\item[(iii)]
\beq\label{hiE.L.r.2.2}
\frac{ 3 e^{\alpha}\epsilon_I^2  }{(1 - \cG_\alpha(\ue)) e^{- \alpha}\rho }   < \frac{ 1 -\iota}{2}\rho \period  
\ene

\end{itemize}

\end{definition}

\begin{remark}\label{ce}
As $  1   -  \frac{1}{2} e^{1/10} > \frac{1}{3} $, taking $ \epsilon_F $ and $\epsilon_I$ small enough assures 
that \eqref{besto} is satisfied. Eqs.~\eqref{hirmn.6} and \eqref{E.L.r.2.1} are fulfilled for small
$\epsilon_I$. By the fact that $ \cG_\alpha(\ue) $ decreases as $\epsilon_I$ decreases, selecting small $\epsilon_I$
implies \eqref{hiE.L.r.2.2}.

In Definitions \ref{ele} and \ref{hice} we include some simplifications to make the notation shorter. Nevertheless, 
the functions and conditions stated there do not always appear  in that form in the proofs. The next inequalities, that are consequences of Definitions \ref{ele} and \ref{hice}, are useful to follow the arguments 
in some proofs. 

\begin{itemize}
\item[(i)]
\eq { \label{E.L.r.2} 
 \Big(& \frac{  \min(\alpha, 1)  (\epsilon_I\xi)^2  }{ e^{- \alpha}\rho(1 - \epsilon_F) -  \frac{1}{2} e^{- \iota \alpha}\rho - \epsilon_I\xi} \Big)\\ & \hspace{1cm} \cdot \Big[ 2 +
  \Big(\frac{1}{ e^{- \alpha}\rho(1 - \epsilon_F) -  \frac{1}{2} e^{- \iota \alpha}\rho - \epsilon_I\xi } \Big) 
 \Big(\epsilon_Z  + \epsilon_I\xi \Big)\Big] \notag \\ & \hspace{8cm}  \leq 
 \cG_\alpha(\epsilon_I, \epsilon_Z, \epsilon_F) \period \notag
}
 
\item[(ii)] 
\eq{\label{put0}
(1 - \epsilon_F) -  \frac{1}{2} e^{(1 - \iota) \alpha} > 0\comma
}
\item[(iii)]
\beq \label{rmn.6}
 \frac{ \epsilon_I  (\rho e^{-\alpha})^{-1} }{ (1 - \epsilon_F) -  \frac{1}{2} e^{(1 - \iota) \alpha}} < \frac{1}{2^4} \comma 
\ene

\item[(iv)]
\beq\label{E.L.r.2.2}
\frac{(\epsilon_I\xi)^2  }{1 - \cG_\alpha(\ue)}  \Big(\frac{e^{\alpha}}{e^{- \alpha}\rho(1 - \epsilon_F) -  \frac{1}{2} e^{- \iota \alpha}\rho - \epsilon_I\xi } \Big)  < \frac{ 1 -\iota}{2}\rho \period  
\ene

\end{itemize}

\end{remark}

\begin{remark}
The hypotheses stated in Definitions \ref{ele} and \ref{hice} are necessary to define the Feshbach-Schur map (see Definition \ref{D.fmn.1} and Lemma \ref{L.fm}) for $\alpha \in [0,\alpha_+]$. These properties are not used in previous works 
(see \cite{BCFS2003} and \cite{BFS98a}, for example) because there it is required that $\alpha$ is big enough. We need to define the Feshbach-Schur map for $\alpha$ tending to $0$ to be able to construct a continuous renormalization group of operators, in contrast to the discrete one that is used in the other works.

\end{remark}

\subsubsection{The Polydisc $\widetilde \cW_\xi$}

\begin{definition}\label{ass}
We say that $\uw \in \cW_\xi$ belongs to $\widetilde \cW_\xi$  if it satisfies the following properties:
\begin{itemize} 
\item[(a)]
\beq \label{rmn.3}
w_{0, 0}(z, 0) = z \comma
\ene
\item[(b)] 
\beq \label{rmn.5}
\| \uw \|^{(I)}_{\xi} \leq \epsilon_I, \hspace{1cm}  \| \uw \|^{(Z)} \leq \epsilon_Z, \hspace{1cm} 
 \| \uw - \ur \|^{(F)} \leq \epsilon_F \comma
\ene
for some $\ue = (\epsilon_I, \epsilon_Z, \epsilon_F) \in \uE_\alpha .$
\end{itemize}

\end{definition}

\secct{The Space of Operators}\label{S.hamiltonians}

\subsection{Basic Notions}
\subsubsection{The Fock Space} \label{Fock-Space}
We denote by 
\begin{equation} \label{h.1}
\mathfrak{h} : = L^2(\mathbb{R}^3) 
\end{equation}
and by 
\begin{equation} \label{bn.1} 
\cF^{(n)} = \mathfrak{h}^{\otimes_s n } 
\end{equation}
the space of symmetric functions in $ L^2(\mathbb{R}^{3n})  $, for $n \in \NN$. \\
We furthermore denote by 
\eq{\label{bn.2}
\cF^{(n)}_S  : = \cF^{(n)} \cap S(\RR^{3n})\comma
}
where $ S(\RR^{3n}) $ is the Schwartz space.\\
The Bosonic Fock space is given by 
\begin{equation}\label{h.2}
\cF := \bigoplus_{n = 0}^{\infty} \cF^{(n)}\comma 
\end{equation}
where 
$$\cF^{(0)} \equiv \cF_S^{(0)} : =  \CC \Omega$$
and $\Omega$ is the normalized vacuum vector.  

We define additionally
\eq{
\cF_S : = \bigoplus_{n = 0}^{\infty} \cF_S^{(n)}
}
and 
\eq{\label{fin}
\cF_{\mathrm{fin}} : = \bigcup_{n = 0}^\infty \bigoplus_{j = 0}^{n} \cF^{(j)}.
}
the finite particles Fock space. \\
 
\subsubsection{Creation and Annihilation Operators}\label{screan}
For every $k  \in \mathbb{R}^3$, the annihilation operator $a(k)$ takes a function $\psi \in \cF^{(n)}_S$ to the function 
$a(k)\psi \in \cF^{(n-1)}_S$ given by 
\eq{\label{cao.1}
a(k)\psi(k_1, \cdots, k_{n-1}) := \sqrt{n}\psi(k,k_1, \cdots, k_{n-1})\period
}
Using \eqref{cao.1} we define $a(k)$ as an operator from $ \cF_S $ to $\cF_S$, taking additionally
\eq{\label{aomega}
a(k)\Omega : = 0\period
}
Note that $a(k)$ is densely defined on $\cF$, but not closable and hence has no adjoint (as an operator).  
The creation operator $a^*(k)$ is the (formal) adjoint of $a(k)$, it takes a function $\psi \in \cF^{(n)}_S$ to the tempered distribution
$a^*(k)\psi \in (\cF^{(n+1)}_S)'$ given by 
\eq{\label{cao.2}
a^*(k)\psi(k_1, \cdots, k_{n + 1}) : = \frac{1}{\sqrt{n + 1}} \sum_{j =1}^{n+1}\delta(k - k_j)  
\psi(k_1, \cdots,k_{j-1}, k_{j+1}, \cdots  k_{n+1})\period
}  
Let  $k_1, \cdots k_p \in \RR^3$ and $\tilde k_1, \cdots,  \tilde k_q \in \RR^3$, the product 
$$
a(\tilde k_1)\cdots a(\tilde k_q)
$$ 
is a well-defined operator in $\cF_S$, but the product of creation operators 
$$
a^*(k_1)\cdots a^*( k_p)
$$ 
is not. We can, however, define the product 
\eq{\label{well}
a^*(k_1)\cdots a^*( k_p) a(\tilde k_1)\cdots a(\tilde k_q)
}
as a quadratic form, namely, 
\eq{\label{well.1}
\la \psi | a^*(k_1)\cdots a^*( k_p) a(\tilde k_1)\cdots a(\tilde k_q)\phi \ra : =
 \la a(k_1)\cdots a( k_p)\psi |  a(\tilde k_1)\cdots a(\tilde k_q)\phi \ra
}
for any $\psi, \phi \in \cF_S$. But this is only possible if the creation operators are to the left 
of the annihilation operators, i.e., for normal-ordered products.\\ 
The creation and annihilation operators satisfy (formally) the canonical commutation relations 
\begin{equation} \label{h.3}
[ a^{\#}(k), a^{\#}(k')] = 0, \: \: [a(k), a^*(k')] = \delta(k - k'),
\end{equation}
where $a^{\#} = a$ or $a^*$.  We suggest the reader who is not familiar with this operators to review section X.7 in \cite{ReedSimonII1980}. 

\subsection{The Space of Operators $H\cW_\xi$}\label{so}
In this section we define the vector space of operators that we study (see Definition \ref{noooo}). We define a linear
map $H$ that associates to every    
sequence of functions $\uw \in \cW_\xi  $ an operator in $\cF$ (see definition \ref{noo}). The space of operators we are interested in is the image of $\cW_\xi$ under this linear function .   

\begin{definition}[Free Hamiltonian]
The free boson Hamiltonian is the self-adjoint operator 
\begin{equation} \label{h.5}
H_f := \int d^3 k a^*(k)|k|a(k)\period
\end{equation}
It is the operator in $\cF$ that represents the (positive) quadratic form in $\cF_S$ derived from
\eqref{well.1}.  The operator $H_f$ leaves $\cF^{(n)}$ invariant and maps $\psi \in \cF^{(n)}$ to the function 
$\phi = H_f \psi$ given by 
\eq{
\phi(k_1, \cdots, k_n) = (|k_1| + \cdots + |k_n|)\psi(k_1, \cdots, k_n) \comma
}
provided that $\phi \in \cF^{(n)}$.
\end{definition}

\begin{definition}\label{deseo}
For every $k^{(m)} \in \RR^{3m}$ and $\tilde k^{(n)}\in \RR^{3n}$ (see Section \ref{S.notation}), we set
$$
a^*(k^{(m)}) = a^*(k^1)\cdots a^*(k^m),\: \hspace{1cm}  \: a(\tilde k^{(n)}) = a(\tilde k^1)\cdots a(\tilde k^n).
$$
\end{definition}

\begin{definition}\label{n}
We denote by $\chi_0 : \mathbb{R} \to \mathbb{R}$ the characteristic function 
\beq \label{h.5.1}
\chi_0(r) : = \begin{cases}  1, & \text{ if $ r \in [0, \rho] $} \\  0, & \text{otherwise},   \end{cases}
\ene
and by $\chi_\alpha,\;   \overline{\chi_\alpha}$ the functions defined by 
\beq \label{h.5.2}
\chi_{\alpha}(r) = \chi_0(e^{\alpha }r), \: \: \: \: \:  \overline{\chi_\alpha }(r)  := 1 - \chi_\alpha(r) \comma  
\ene
for every $\alpha \in [0, \infty)$. To make the notation shorter, we use the following identifications
\eq{ \label{nesiel2}
\chi_\alpha \equiv \chi_\alpha(H_f), \hspace{2cm} \overline \chi_\alpha \equiv \overline \chi_\alpha(H_f)\period
}
 
\end{definition}

\begin{definition}\label{no}
For every  $  \uw =    \big( w_{m, n}  \big)_{m + n \geq 0 } \in \cW_{\xi} $
we define 
\eq{\label{Wmn}
W_{m, n}(\uw) : =   \int_{B_{\rho}^{m+n}} \: d k^{(m, n)} \:  a^*( k^{(m)}) \: w_{m, n}(z; H_f; k^{(m, n)}) \: 
a( \tilde k^{(n)}) . 
}
$ W_{m, n}(\uw) $ is the operator in $\cF$ representing the quadratic form in $\cF_S$ obtained from
\eqref{well.1}. The existence of this operator is a consequence of Theorem X.44 of \cite{ReedSimonII1980} (see also the proof of Lemma \ref{L.h} below).  
\end{definition}

\begin{definition}[Hamiltonian Operators]\label{noo}
For any  $  \uw =    \big( w_{m, n}  \big)_{m + n \geq 0 } \in \cW_{\xi} $,
we define 
\eq{ \label{hw}
H(\uw) : =  \sum_{m+n \geq 0} \chi_0 W_{m, n}(\uw)\chi_0 \period 
}
The series in \eqref{hw} converges in operator norm and defines, thus, a bounded operator in $\cF$. This is a consequence of Lemma \ref{L.h} below.  
\end{definition}

\begin{remark}\label{quad}
If $ \uw = \big( w_{m, n}  \big)_{m + n \geq 0 } $ is a series of functions such that 
\eq{\label{esano}
\forall m, n \in \NN_0 \: : \: \|  w_{m, n}\|^{(\infty)}_0 < \infty\comma
}
then $W_{m,n}(\uw ) $ defined by \eqref{Wmn} determines an operator in $\cF$, 
and Lemma \ref{L.h} implies that $\chi_0 W_{m,n}(\uw)\chi_0 $ is bounded. In this case, we can also define $ H(\uw) $ as a quadratic form in $\cF_{\mathrm{fin}}$.

\end{remark}

\begin{definition}[Interaction Operators] \label{nooo}
For any  $  \uw =    \big( w_{m, n}  \big)_{m + n \geq 0 } \in \cW_{\xi} $  
we define 
\eq{\label{w}
W(\uw) : = H(\uw) - \chi_0 W_{0,0}(\uw) = \sum_{m+n \geq 1} \chi_0 W_{m,n}(\uw)\chi_0  \period
}
\end{definition}
\begin{definition}\label{noooo}
We denote by $H \cW_\xi$ the vector space of operators of the form
\eq{
H\cW_\xi : = \{ H(\uw) \: | \uw \in \cW_\xi \:  \}\period
}

\end{definition}

\subsection{Basic Estimates}\label{Basic-Estimates}
In this section we study the operators   $W_{m, n}(\uw)$ and give some norm estimates. 
We prove furthermore that the operators $H(\uw)$ and $ W(\uw) $ are bounded and provide some bounds for their norms.

\begin{lemma} \label{L.h}
Suppose that $\uw \in \cW_\xi $, then $\chi_0 W_{m, n}(\uw) \chi_0 $ is a bounded operator and the following estimates are satisfied
\begin{itemize}
\item[] For $m, n \geq 1$,
\begin{align} \label{h.6}
\| \overline{ \chi}_\alpha   
 \chi_0   W_{m, n}(\uw) \overline{ \chi}_\alpha 
 \chi_0   \| 
 \leq  \| w_{m,n} \|^{(\infty)}_0 (4 \pi \rho^{2+ \mu})^{(m+n)/2} \alpha. 
\end{align}
\item[]  For $ n \geq 1$,
\begin{align}  \label{h.7}
\| \chi_0   W_{m, n}(\uw) \overline{ \chi}_\alpha 
 \chi_0  \| 
 \leq   \| w_{m,n} \|^{(\infty)}_0(4 \pi \rho^{2 + \mu})^{(m+n)/2} \alpha^{1/2}\period 
\end{align}
\item[]  For $m  \geq 1$,
\begin{align} \label{h.8}
\| \overline{ \chi}_\alpha   
 \chi_0  W_{m, n}(\uw)  \chi_0 \| 
\leq  \| w_{m,n} \|^{(\infty)}_0 (4 \pi \rho^{2 + \mu})^{(m+n)/2} \alpha^{1/2} \period
\end{align}
\item[] For $m + n \geq 0$ 
\begin{align} \label{h.9}
\|  \chi_0 W_{m, n}(\uw) \chi_0  \|  \leq \| w_{m,n} \|^{(\infty)}_0(4 \pi \rho^{2 + \mu})^{(m+n)/2}  \period
\end{align}
\end{itemize}
\end{lemma}
\noindent {\it Proof:}
We suppose that $ \phi = (\phi_j)_{j \in \mathbb{N}_0}, \psi = (\psi_j)_{j \in \mathbb{N}_0} \in \cF_S$. Suppose furthermore that there exists $ l \in 
\mathbb{N}_0$ such that $ \phi_j = 0 $ for all $j \ne n + l$ and $ \psi_j = 0 $ for all $j \ne m + l$. 
 We estimate the left hand side of Eq.~(\ref{h.6}). We take the operators $ \overline{ \chi}_\alpha  $ and $ \chi_0 $ to the other side of the inner product and compute it integrating with respect to the variable 
 $x^{(\ell)} \in (\RR^3)^\ell$. We get (remember the definition of $H_f$ in (\ref{h.5})):    
\begin{align} \label{h.12}
|\la \psi  |\;  \overline{ \chi}_\alpha    \chi_0 & W_{m, n}(\uw)  \overline{ \chi}_\alpha \chi_0  \phi \ra   | \\ \notag
 \leq &  \int \int  dx^{(\ell)}  d k^{(m, n)}   \Big| \Big (a(k^{(m)}) \overline{ \chi}_\alpha
 \chi_0  \psi\Big )_l(x^{(\ell)}) \Big |  \\ \notag 
&\cdot \Big | \Big (w_{m, n}(z; \|x^{(\ell)}\|_1, k^{(m, n)})   a( \tilde k^{(n)})  
\overline{ \chi}_\alpha  \chi_0 \phi\Big )_{l}(x^{(\ell)})  \Big |\\ \notag \hspace{2cm}
\leq &  \| w_{m,n} \|^{(\infty)}_0  
  \int \int  dx^{(\ell)} \:  d k^{(m, n)}  |k^{(m)}|^{- 1/2 + \mu/2}  \\ \notag 
 & \cdot \Big |\Big (a(k^{(m)}) \overline{ \chi}_\alpha 
 \chi_0  \psi \Big)_l (x^{(\ell)}) \Big |  \\ \notag   
 & \cdot |\tilde k^{(n)}|^{-1/2 + \mu/2}    \Big |\Big (a( \tilde k^{(n)})  
\overline{ \chi}_\alpha  \chi_0 \phi\Big )_l(x^{(\ell)})  \Big| \period
\end{align}
The term $\chi_\alpha \chi_0 $ permits us to restrict the domain of integration and write the right hand side 
of Eq.~(\ref{h.12}) as follows
\begin{align}\label{h.13} 
X : = & |\la \psi  |\;  \overline{ \chi}_\alpha    \chi_0  W_{m, n}(\uw)  \overline{ \chi}_\alpha \chi_0  \phi \ra   | 
\\  \leq & \notag  \| w_{m,n}  \|^{(\infty)}_0  
  \int dx^{(\ell)}  \Big [\int_{  e^{- \alpha}\rho \leq \|k^{(m)}\|_1 + \|x^{(\ell)}\|_1 \leq \rho}    d k^{(m)}  |k^{(m)}|^{ -1/2 + \mu/2 } \\ \notag &    
  \hspace{3.5cm} \cdot |(a(k^{(m)}) \overline{ \chi}_\alpha 
 \chi_0 \psi)_l (x^{(\ell)}) |\Big] \\ \notag & \hspace{3.2cm} \Big [ \int_{  e^{- \alpha} \rho \leq \| \tilde k^{(n)}\|_1  + \|x^{(\ell)}\|_1\leq \rho}  d \tilde k^{(n)}     |\tilde k^{(n)}|^{-1/2 + \mu/2 } 
  \\ \notag & \hspace{4cm} \cdot |  (a( \tilde k^{(n)})  
\overline{ \chi}_\alpha  \chi_0 \phi)_l(x^{(\ell)})  | \Big ] \period
\end{align} 
Using the Cauchy-Schwarz inequality we estimate Eq.~(\ref{h.13}) by 
\begin{align}\label{h.14}
X \leq \notag \| w_{m,n} \|^{(\infty)}_0
  \Big \{ \int & dx^{(\ell)} \Big [\int_{  e^{- \alpha}\rho \leq \|k^{(m)}\|_1 + \|x^{(\ell)}\|_1 \leq \rho}   d k^{(m)}  |k^{(m)}|^{ -1/2 + \mu/2 }  \\ \notag  
 &\cdot |(a(k^{(m)}) \overline{ \chi}_\alpha 
 \chi_0  \psi)_l (x^{(\ell)}) |\Big]^2 \Big \}^{1/2} \\ \notag
  \cdot \Big \{\int & dx^{(\ell)}  \Big [ \int_{  e^{- \alpha} \rho \leq \| \tilde k^{(n)}\| 
   + \|x^{(\ell)}\|_1\leq \rho}  d \tilde k^{(n)}   |\tilde k^{(n)}|^{-1/2 + \mu/2 }  
    \\  &\cdot |   (a( \tilde k^{(n)})  
\overline{ \chi}_\alpha  \chi_0 \phi)_l(x^{(\ell)})  | \Big]^2 \Big \}^{1/2} \comma
\end{align}
which is estimated again using the Cauchy-Swartz inequality by 
\begin{align}\label{h.15}
 \| w_{m,n} \|^{(\infty)}_0  &
  \\ \notag  \cdot \Big \{ \int dx^{(\ell)} & \Big[\int_{  e^{- \alpha} \rho \leq \|k^{(m)}\|_1 + 
  \|x^{(\ell)}\|_1 \leq \rho}   d k^{(m)}      
   |k^{(m)}|\cdot | (a(k^{(m)}) \overline{ \chi}_\alpha 
 \chi_0  \psi)_l (x^{(\ell)}) | ^2 \Big] \\ & 
 \notag \cdot  \Big[\int_{  e^{- \alpha}\rho \leq \|k^{(m)}\|_1 + \|x^{(\ell)}\|_1 \leq \rho}   d k^{(m)}  |k^{(m)}|^{ -2 + \mu }   \Big]  \Big \}^{1/2} \\  \notag
\cdot \Big \{ \int dx^{(\ell)} &
 \Big[\int_{ e^{- \alpha} \rho \leq \|\tilde k^{(n)}\|_1 + \|x^{(\ell)}\|_1 \leq \rho}   d \tilde k^{(n)}    
   | \tilde k^{(n)}| \cdot | (a( \tilde k^{(n)}) \overline{ \chi}_\alpha  
 \chi_0  \phi)_l (x^{(\ell)}) | ^2 \Big] \\ \notag & 
 \cdot \Big[\int_{  e^{- \alpha}\rho \leq \|  \tilde k^{(n)}\|_1 + \|x^{(\ell)}\|_1 \leq \rho}   d \tilde k^{(n)}  |\tilde k^{(n)}|^{ -2 + \mu }  \Big] 
        \Big \}^{1/2} \\ \notag
\leq  \, \| w_{m,n} & \|^{(\infty)}_0   \cdot \sup_{\|x^{(\ell)}\|_1 \leq \rho} \Big[\int_{ e^{- \alpha}\rho \leq \|  k^{(m)}\|_1 + \|x^{(\ell)}\|_1 \leq \rho}   d  k^{(m)}  | k^{(m)}|^{ -2 + \mu}  \Big]^{1/2} 
\\ \notag & \cdot \sup_{\|x^{(\ell)}\|_1 \leq \rho} \Big[\int_{ e^{- \alpha} \rho \leq \| \tilde k^{(n)}\|_1
 + \|x^{(\ell)}\|_1 \leq \rho}   d \tilde k^{(n)}  | \tilde k^{(n)}|^{ -2 + \mu}  \Big]^{1/2} 
\\ \notag 
 &  \cdot \Big[\int   d k^{(m)}     
   | k^{(m)}|\cdot \| (a(  k^{(m)}) \overline{ \chi}_\alpha
 \chi_0  \psi)  \| ^2 \Big]^{1/2} \\ \notag &  \cdot
  \Big[\int   d \tilde k^{(n)}     
   |\tilde k^{(n)}|\cdot \| (a( \tilde k^{(n)}) \overline{ \chi}_\alpha 
 \chi_0  \phi)  \| ^2 \Big]^{1/2}
\end{align}
From (\ref{h.12})-(\ref{h.15}) we obtain
\begin{align} \label{h.16}
|\la \psi  & |\;  \overline{ \chi}_\alpha    \chi_0  W_{m, n}(\uw)  \overline{ \chi}_\alpha \chi_0  \phi \ra   | \\ \notag
\leq & \, \| w_{m,n} \|^{(\infty)}_0 \cdot \| \phi \| \cdot \|\psi \|  \rho^{(n+m)/2}\\ \notag 
& \cdot \sup_{\|x^{(\ell)}\|_1 \leq \rho} \Big[\int_{  e^{- \alpha}\rho \leq \|  k^{(m)}\|_1 + \|x^{(\ell)}\|_1 \leq \rho}   d  k^{(m)}  | k^{(m)}|^{ -2 + \mu }  \Big]^{1/2} 
\\ \notag & \cdot \sup_{\|x^{(\ell)}\|_1 \leq \rho} \Big[\int_{  e^{- \alpha}\rho \leq \| \tilde k^{(n)}\|_1 + \|x^{(\ell)}\|_1 \leq \rho}   d \tilde k^{(n)} 
 | \tilde k^{(n)}|^{ -2 + \mu } \Big]^{1/2}  \period 
\end{align}
To obtain (\ref{h.16}) we estimate the last two lines in (\ref{h.15}) by 
\eq{\label{fino}
 \| (H_f)^{n/2} \chi_0 \phi \| \cdot \| (H_f)^{m/2} \chi_0 \psi \| \leq \| \phi \|\cdot \| \psi \| \rho^{(n+m)/2}} 
 using (III.17)-(III.20) in \cite{BCFS2003}.

Next we estimate the integrals in (\ref{h.16}): 
\begin{align} \label{h.17}
\sup_{\|x^{(\ell)}\|_1 \leq \rho}  \Big[ &\int_{ e^{- \alpha}\rho \leq \|  k^{(m)}\|_1 + \|x^{(\ell)}\|_1 \leq \rho}    d  k^{(m)}  | k^{(m)}|^{ -2 + \mu }  \Big]\\ \notag
\leq & \Big [\int_{ |y| \leq \rho} d^3 y |y|^{-2 + \mu} \Big ]^{m-1} \cdot \sup_{s \in [0,\rho]} \Big [  \int_{ e^{- \alpha}\rho \leq |k_1| + s\leq \rho   } d k_1 \frac{|k_1|^\mu}{|k_1|^2} \Big] 
\\ \notag \leq & \Big (4 \pi \rho^{1 + \mu}\Big )^{m - 1} 4 \pi (1- e^{-\alpha})\rho^{1 + \mu} \leq 
\Big (4 \pi \rho^{1+\mu}\Big )^{m }  \alpha \period 
\end{align}
As $\psi_{m+l} \in \cF_S^{m + l} $ is arbitrary, it follows from  Eqs.~\eqref{h.16} and \eqref{h.17} that   
\eq{\label{elque}
\|    \overline{ \chi}_\alpha    \chi_0   W_{m, n}(\uw)  \overline{ \chi}_\alpha \chi_0  \phi  \|
\leq \| w_{m,n} \|^{(\infty)}_0 \Big (4 \pi \rho^{1+\mu}\Big )^{(m + n)/2 }  \alpha \rho^{(m+n)/2} \| \phi \| \period
}
For a general $\eta = (\eta_j)_{j \in \NN}\in \cF$, we define $ \eta^{(j)} \in \cF $ such that its 
$j-$th component equals $\eta_j$ and the others are zero. Then \eqref{elque} is valid for $\eta^{(j)}$ instead of 
$\phi$. Using the fact that for $ j \ne l $,  $\overline{ \chi}_\alpha    \chi_0  W_{m, n}(\uw)  \overline{ \chi}_\alpha \chi_0 \eta^{(j)}$ and $\overline{ \chi}_\alpha    \chi_0  W_{m, n}(\uw)  \overline{ \chi}_\alpha \chi_0
\eta^{(l)}$ are orthogonal,  we conclude that \eqref{elque} is valid for $\eta$ instead of $\phi$. This implies \eqref{h.6}. Eqs.~\eqref{h.7}-\eqref{h.9} are proved in the same way.
 
\qed

\begin{theorem}\label{eleseguey}
For every $\uw \in \cW_\xi$, $W(\uw)$  and $H(\uw) $  are bounded operators and
\eq{ \label{gue}
\| W(\uw) \| \leq  \xi \|  \uw  \|^{(I)}\comma \hspace{2cm} \| H(\uw) \| 
\leq  \| w_{0, 0} \|^{(\infty)} +  \xi \|  \uw  \|^{(I)}\period  
}

\end{theorem}
\noindent \emph{Proof:} The result follows from \eqref{rho} (which implies that $4 \pi \rho^{2 + \mu} < 1$), \eqref{i.11.1} and Lemma \ref{L.h}.
\QED 

\begin{remark}\label{R.h}
From the proof of Lemma \ref{L.h} we conclude that the operator-valued function 
$$
z \mapsto H(\uw(z))
$$
is analytic; actually the derivative is given by 
$$
\frac{\partial}{\partial{z}} H(\uw(z)) = H(\frac{\partial}{\partial z}\uw(z))\period
$$

\end{remark}
\subsection{Composition Formulae (Wick's Theorem)}\label{cf}
Suppose that $\uw \in \cW_\xi$. A natural question is whether there is a sequence of functions 
$\uw' = \big(  w_{m,n}' \big)_{m + n \geq 0}$ such that
$$
H(\uw)H(\uw) = H(\uw')\period
$$
The answer to this and similar questions is well known. It is actually only an application 
of Wick's Theorem. In this section we state the results that we need concerning the question above and refer 
to \cite{BCFS2003} and \cite{BFS98a} for the proofs.
A big portion of the notation that we use through the paper is introduced here.

\subsubsection{Definitions and Notation}

\begin{definition}\label{D.cprm.2}
Let $ \uw \in \cW_\xi$. We define the operators 
\eq{\label{Wmnpq}
 W_{p, q}^{m, n}(z; r; k^{(m,n)}) : =   \int & d x^{(p)}  d \tilde x^{(q)}  
a^*( x^{(p)}) \\ \notag  & w_{m+p, n+q}\big(z; H_f + r; (k^{(m)}, x^{(p)}); (\tilde k^{(n)}, \tilde x^{(q)})\big) 
a( \tilde x^{(q)}) . 
}

\end{definition}

\begin{definition}\label{D.cprm.3}
For every $L \in \mathbb{N}$ and $m,n \in \NN_0$ we denote by $\cB_L(m, n)$ the set of arrays  $\upsilon = (\bar m, \bar n, \bar p, \bar q) \in \mathbb{N}_0^{4L}$
that satisfy the following properties:
\begin{itemize}
\item The elements  $ \bar m, \bar n, \bar p, \bar q $ belong to $\mathbb{N}_0^{L}$ and their components are denoted by a sub index, for example 
$ \bar m = (m_\ell)_{\ell \in \{ 1, \cdots, L\}} = (m_1, \cdots, m_L)$. 
\item $m_1 + \cdots + m_L = m$ and $\: $ $n_1 + \cdots n_L = n $.
\item For any $\ell \in \{1, \cdots, L  \}$, $ m_\ell + n_\ell + p_\ell + q_\ell \geq 1 $. 
\end{itemize}
For every $ \upsilon = (\bar m, \bar n, \bar p, \bar q) \in \cB_L(m,n)$ we use the following notation:
$$
k^{(\bar m,\bar n)}: = (k_1^{(m_1, n_1)}, \cdots, k_L^{(m_L, n_L)}), \: k^{(m_\ell, n_\ell)} = (k_\ell^{(m_\ell)}; \tilde k_\ell^{(n_\ell)}),
$$ 
$$
r_\ell = r_\ell( k^{(\bar m, \bar n)} ): = \| \tilde k_1^{(n_1)}\|_1 + \cdots + \| \tilde k_{\ell -1}^{(n_{\ell - 1})}\|_1 +  \|  k_{\ell +1}^{(m_{\ell + 1})}\|_1 + \cdots
+ \|  k_{L}^{(m_{L})}\|_1,
$$

$$
\tilde r_\ell = \tilde r_\ell( k^{(\bar m, \bar n)} ) : = r_\ell +  \| \tilde k_{\ell}^{(n_{\ell })}\|_1.
$$
If $m =0 $  or $n = 0$ we omit the corresponding terms. If $m+ n = 0$ then $r_\ell = 0 = \tilde r_\ell$.

Let $ \bar v  = (v_1,\cdots, v_J)\in \NN_0^{J} $,  $ \bar u = (u_J, \cdots, u_L)  \in \NN_0^{L -J + 1} $  for some $J \leq L $, we define
\eq{ \label{cp.1}
k^{(\bar m,\bar n)} \tilde \times \tilde y^{(\bar v)} : = &( (k_1^{(m_1)}; \tilde k_1^{(n_1)}, \tilde y_1^{(v_1)})  , \cdots,  
(k_J^{(m_J)}; \tilde k_J^{(n_J)}, \tilde y_J^{(v_J)}) \\ \notag &, k_{J+1}^{(m_{J+1}, n_{J+ 1})},  \cdots, k_L^{(m_{L}, n_{L})} ) \\ \notag
k^{(\bar m,\bar n)}  \times  y^{(\bar u)} : = & (k_1^{(m_1, n_1)},\cdots, k_{J-1}^{(m_{J-1}, n_{J-1})}, 
 (k_J^{(m_J)},  y_J^{(u_J)}; \tilde k_J^{(n_J)}), \\ & \notag, \cdots,  
(k_L^{(m_L)},  y_L^{(u_L)}; \tilde k_L^{(n_L)})) \period
}

\end{definition}

\begin{definition}\label{D.cprm.4}
Let $\uw \in \cW_\xi$. Suppose that 
\eq{\label{F1} F  : D_{\rho/2} \times \RR \to \CC 
}
is a bounded (measurable) function. We define 
\eq{\label{F2}
F_\ell : = \begin{cases}  F, \: \text{if $\ell \in \{1, \cdots, L-1\}$}\comma \\  1, \: \text{if $\ell = \{0,  L \}$}  \period
\end{cases} 
}

For every $ \upsilon = (\bar m, \bar n, \bar p, \bar q) \in \cB_L(m,n)$ we use the following notation 
$$
\widetilde W_\ell(z;r; k^{(m_\ell, n_\ell)}) :=   W^{m_\ell, n_\ell}_{p_\ell, q_\ell}(z;r;k^{(m_\ell, n_\ell)} ),
$$
and 
\eq{ \label{vupsilon}
V^F_\upsilon (z; r; k^{ (m, n)})  \equiv & V^F_{(\bar m, \bar n, \bar p, \bar q)} (z;r; k^{(m, n)}) \\ \notag : = &
\Big \la \Om   \Big |  \prod_{\ell = 1}^{L} \Big[ \widetilde W_\ell(z; r + r_\ell ( k^{(\bar m, \bar n)} ) ; k^{(m_\ell, n_\ell)}) 
\\ \notag & \hspace{1cm} \cdot F_\ell (H_f + r + \tilde r_\ell( k^{(\bar m, \bar n)} )) \Big]  \Om  \Big  \ra \period 
}
We omit in general the dependence on $ ( k^{(\bar m, \bar n)} ) $ of $r_\ell $ and $\tilde r_\ell$. 

\end{definition}

\begin{definition} \label{D.cprm.4.1}
For every array $ \overline M =(M_1, \cdots, M_N) \in \mathbb{N}_0^{N} $ and any $J \in \{1,\cdots, N  \}$ we denote by 
$$
\overline M_{\leq J} : = (M_1, \cdots  M_J), \: \hspace{2cm}  \overline M_{\geq J} := (M_J, \cdots, M_N).
$$

For every $ \upsilon = (\bar m, \bar n, \bar p, \bar q) \in \cB_L(m,n)$  and any $J \in \{1, \cdots, L \}$ we use the symbols
$$
\upsilon_{\leq J} : = (\bar m_{\leq J}, \bar n_{\leq J}, \bar p_{\leq J}, \bar q_{\leq J}), \:  \upsilon_{\geq J} : = (\bar m_{\geq J}, \bar n_{\geq J}, \bar p_{\geq J}, \bar q_{\geq J}).
$$
For any two arrays $\bar s, \, \bar t \in \NN_0^{N}$ we say that 
$$
\bar s \leq \bar t
$$  
if the inequality holds component-wise. Furthermore we use  the symbol 
$$
|\bar t|_{1} : = t_1 + \cdots + t_N.
$$

\end{definition}

\begin{definition} \label{D.cprm.4.2}
Let $ \upsilon = (\bar m, \bar n, \bar p, \bar q) \in \cB_L(m,n)$ and $\uw \in \cW_\xi $. 
Suppose that $F$ satisfies \eqref{F1} and $F_\ell$ satisfies \eqref{F2}.  
Suppose furthermore that $\bar v = (v_1,\cdots, v_J )  \leq \bar q_{\leq J}$ and that 
 $\bar u = (u_J,\cdots, u_L )  \leq \bar p_{\geq J}$. 
We define 
$$
\widetilde W_{\ell, \bar v}\Big(z;r; k^{(m_\ell)} ; \tilde k^{(n_\ell)}, \tilde y^{( v_\ell)}\Big) :=   W^{m_\ell, n_\ell + v_\ell}_{p_\ell, q_\ell - v_\ell}
\Big(z;r;k^{(m_\ell)}; \tilde k^{(n_\ell)}, \tilde y_j^{(v_\ell)}\Big),
$$
for $\ell \leq J$, and 
$$
\widetilde W_{\ell, \bar u}\Big(z;r; k^{(m_\ell)}, y^{( u_\ell)};\tilde k^{(n_\ell)}\Big) :=   W^{m_\ell + u_\ell, n_\ell }_{p_\ell - u_\ell, q_\ell}
\Big(z;r;k^{(m_\ell)}, y_j^{(u_\ell)}; \tilde k^{(n_\ell)}\Big),
$$
for $  \ell \geq J $. 
We furthermore define
\eq{ \label{vupsilonJmenos} 
V^F_{\upsilon_{\leq J}, \bar v} &\Big(z; r; k^{(m)}; \tilde k^{(n)}, \tilde y^{(|\bar v|_1)}\Big)  \\ \notag & : =  
\Big \la \Om   \Big | \Big[\prod_{\ell = 1}^{J-1}  \widetilde W_{\ell, \bar v} \Big(z; r + r_\ell
 (k^{(\bar m,\bar n)}\tilde \times \tilde y^{(\bar v)})
 ; k^{(m_\ell)};\tilde k^{( n_\ell)}, \tilde y^{(v_\ell)}\Big) \\  \notag
& \hspace{3.8cm} \cdot F_\ell \Big(H_f + r + \tilde r_\ell
(k^{(\bar m,\bar n)}\tilde \times \tilde y^{(\bar v)})
\Big) \Big] \\ \notag &  \hspace{1cm} \cdot
\widetilde W_{J, \bar v} \Big(z; r + r_J
 (k^{(\bar m,\bar n)}\tilde \times \tilde y^{(\bar v)})
 ; k^{(m_J)};\tilde k^{( n_J)}, \tilde y^{(v_J)}\Big) \Om  \Big  \ra, 
}

\eq{ \label{vupsilonJmas}
V^F_{\upsilon_{\geq J}, \bar v} &\Big(z; r; k^{ (m)}, y^{(|\bar u|_1)}; \tilde k^{(n)}\Big) \\ \notag & : =  \Big \la \Om   \Big |  
 \prod_{\ell = J}^{L} \Big[ \widetilde W_{\ell, \bar u}\Big(z; r + r_\ell
 (k^{(\bar m,\bar n)} \times y^{(\bar u)})
 ; k^{(m_\ell)}, y^{(u_\ell)};\tilde k^{( n_\ell)}\Big) \\ \notag & \hspace{3.8cm} \cdot
 F_\ell \Big(H_f + r + \tilde r_\ell
(k^{(\bar m,\bar n)}\times y^{(\bar u)})
\Big) \Big]  \Om  \Big  \ra \period 
}

\end{definition}

\subsubsection{Composition Formulae}

\begin{lemma}\label{L.cprm.1}
Let 
$\uw \in \cW_\xi $. 
Suppose that $F$ satisfies \eqref{F1} and $F_\ell$ satisfies \eqref{F2}.  Then there exists a sequence of functions
$\big(  w_{m,n}^{F, sym}\big)_{m + n \geq 0}$ satisfying \eqref{esano} such that 

\eq{ \label{elmen} 
  W(\uw(z))   
\Big( F(z; H_f) & W(\uw(z)) \Big)^{L-1}    = H\Big (\big( w_{m,n}^{F, sym}\big)_{m + n \geq 0}\Big )
}
in the sense of quadratic forms in $\cF_{\mathrm{fin}}$.\\
$w_{m,n}^{F, sym}$  is the symmetrization with respect to $k^{(m)}$ and $k^{(n)}$ of the function 
\eq{ 
& \widetilde w_{m,n}^F(z; r; k^{(m,n)}) : =    \sum_{\upsilon \in \cB_L(m,n)} 
\Big[ \prod_{\ell =1}^L \binom{m_\ell + p_\ell}{p_\ell} \binom{n_\ell + q_\ell}{q_\ell } \Big]  \notag  V^F_\upsilon (z; r;k^{(m,n)})\period
}

\end{lemma}

\emph{Proof:}
The result follows from Wick's Theorem (see Theorem 3.6 in \cite{BCFS2003}).
\QED

\subsubsection{Recursive Relation}

\begin{lemma}[Wick-Ordering Recursive Relation] \label{recursive-relation}
Let $ \upsilon = (\bar m, \bar n, \bar p, \bar q) \in \cB_L(m,n)$ and $\uw \in \cW_\xi $. 
Suppose that $F$ satisfies \eqref{F1} and $F_\ell$ satisfies \eqref{F2}. For every  $J \in \{ 1, \cdots, L \}$, the following equality holds true
\eq{ \label{E.worr.1} 
V^F_\upsilon  (z; r;k^{(m,n)}) & \notag \\ \notag  = &  \sum_{\bar v \leq \bar q_{\leq (J-1)}} \sum_{\bar u \leq \bar p_{\geq (J+1)}}  \int \Big \{   \prod_{j = 1}^{J -1} d\tilde y^{(v_j)} \binom{q_j}{v_j}  \Big \}   
 \Big \{   \prod_{j = J+1}^{L} d y^{(u_j)} \binom{p_j}{u_j}  \Big \} 
\notag \\ \notag &
 V^F_{\upsilon_{\leq (J-1)}, \bar v}(z; r;k^{(m)}; \tilde k^{(n)}, \tilde y^{(|\bar v|_1)} )
\\ \notag &
\cdot F_{J-1}(r + \tilde r_{J-1}( k^{(m,n)} \tilde \times \tilde y^{(\bar v)}) )F_{J}(r + \tilde r_{J}( k^{(m,n)} \times  y^{(\bar u)}) )
\\ \notag &
\Big \la \prod_{j= 1}^{J-1}a(\tilde y_j^{(v_j)}) \widetilde W_J(z, r  +  r_J(k^{(m,n)}); k^{(m_J, n_J)} )          \prod_{j= J+ 1}^{L}a^*( y_j^{(u_j)}) \Big \ra_\Om
\\ &
\cdot V^F_{\upsilon_{\geq (J+1)}, \bar u}(z; r;k^{(m)}, y^{(|\bar u|_1)}; \tilde k^{(n)}).
}
If, furthermore,  $1 \leq J \leq L-1$, 
\eq{ \label{E.worr.2} 
V^F_\upsilon(z; r;k^{(m,n)}) &  \notag  \\ = & \notag  \sum_{\bar v \leq \bar q_{\leq J}} \sum_{\bar u \leq \bar p_{\geq (J+1)}} \int \Big \{   \prod_{j = 1}^{J} d\tilde y^{(v_j)} \binom{q_j}{v_j}  \Big \}   
\Big \{   \prod_{j = J+1}^{L} d y^{(u_j)} \binom{p_j}{u_j}  \Big \}  
\\ \notag &
V^F_{\upsilon_{\leq J}, \bar v}(z; r;k^{(m)}; \tilde k^{(n)}, \tilde y^{(|\bar v|_1)} )
\\ \notag &
\cdot F_{J}(r + \tilde r_{J}( k^{(m,n)} \times  y^{(\bar u)}) )
\Big \la \prod_{j= 1}^{J}a(\tilde y_j^{(v_j)}) \prod_{j= J + 1}^{L}a^*( y_j^{(u_j)}) \Big \ra_\Om
\\  &
\cdot V^F_{\upsilon_{\geq (J+1)}, \bar u}(z; r;k^{(m)}, y^{(|\bar u|_1)}; \tilde k^{(n)})\comma
}
where 
\eq{\label{omega}
\la (\cdot) \ra_\Omega : = \la \Omega | \; (\cdot) \Omega \ra  \period
}

\end{lemma}

\emph{Proof:}
See the proof of Theorem A.5 of \cite{BFS98}.

\QED

\begin{remark}\label{expl}
The vacuum expectation value appearing in \eqref{E.worr.1} and \eqref{E.worr.2} contains annihilation operators to the left of creation operators (it is not in normal order). The product of creation and annihilation operators that are not normal ordered does not make sense as an operator nor as a quadratic form. In this remark we explain what is the meaning of the vacuum expectation value in \eqref{E.worr.1}. The explanation for \eqref{E.worr.2} is similar and is detailed in \eqref{nn.1}-\eqref{yo-mero}.\\
We notice that the expectation value in Eq.~\eqref{E.worr.1} is different from zero only when $|\bar u|_1 = |\bar v|_1$. To analyze (formally) the product   
$$
 \prod_{j= 1}^{J}a(\tilde y_j^{(v_j)}) \prod_{j= J + 1}^{L}a^*( y_j^{(u_j)}) \comma  
$$    
we use the canonical commutation relations (\ref{h.3}) to take the creation operators to the left. Then we take the vacuum expectation value, which sets to zero all terms containing creation or annihilation operators.  \\ 
We identify 
\eq{ \label{no-sense-0} 
y^{(|\bar u|_1)} & \equiv (y^1, \cdots, y^{|\bar u|_1}) \equiv (y^{(u_{J + 1})}, \cdots, y^{(u_L)})\comma \\ \notag
\tilde y^{(|\bar v|_1)} &  \equiv (\tilde y^1, \cdots, \tilde y^{|\bar v|_1}) \equiv (\tilde y^{(v_{1})}, \cdots, \tilde y^{(v_J)}) \period
}
As a result of the formal computation we obtain 
\eq{ \label{no-sense-1}
\Big \la \prod_{j= 1}^{J}a(\tilde y_j^{(v_j)}) \prod_{j= J + 1}^{L}a^*( y_j^{(u_j)}) \Big \ra_\Omega := \sum_{p \in S_{|\bar v|_1}} \delta\Big( y^{(|\bar{u}|_1)} 
- p (\tilde y^{(|\bar{v}|_1)}) \Big)\comma
} 
where 
\eq{p (\tilde y^{(|\bar{v}|_1)}) : = (\tilde y^{p(1)}, \cdots, \tilde y^{p(|\bar v|_1)}) \period
}
and $  S_{|\bar v|_1}  $ is the set of permutations of the first $|\bar v|_1 $ natural number. We take Eq.~\eqref{no-sense-1}
as a definition. \\
Eq.~(\ref{no-sense-1}) can be taken as a distribution in the variables of $\tilde y^{(|\bar v|_1)}$ if we let $ y^{(|\bar u|_1)} $ fixed.  We understand the integral with respect to $\tilde y^{(|\bar v|_1)}$ in Eq.~(\ref{E.worr.2}) as an application of the distribution (\ref{no-sense-1}) (which makes sense even if the integrand is not a test function).

\end{remark}

\secct{The Feshbach-Schur Map}\label{S.feshbachmap}
In this section we introduce the Feshbach-Schur map. It is a map that takes operators into operators and has the 
important property of being isospectral in the sense that an operator is invertible if, and only if, its Feshbach-Schur map is invertible, and a similarly one can establish a one to one correspondence between eigenvalues and eigenvectors. 
The advantage of studying the Feshbach-Schur map applied to an operator rather that the operator itself is that the Feshbach-Schur map restricts the domain of the operator to a subspace of energies that are close to the spectral region that we are interested on. We can therefore study locally the spectral properties of the operators and neglect the influence of the spectral points that are far away from the region that we want to understand.

\begin{definition}[Feshbach-Schur Map]\label{D.fmn.1}
For every $\uw \in \widetilde \cW_{\xi} $, the Feshbach-Schur map is defined by 
\begin{align} \label{fm.3}
F_\alpha & (H(\uw)) :=  \chi_\alpha H(\uw) \chi_\alpha 
 -  \chi_\alpha H(\uw)\overline{\chi_\alpha}(\overline{\chi_\alpha} H(\uw) \overline{\chi_\alpha}  )^{-1} \overline{\chi_\alpha} H (\uw) \chi_\alpha \comma 
\end{align}
provided that $  \overline{\chi_\alpha} H(\uw) \overline{\chi_\alpha}  $ is bounded invertible. 
\end{definition}

In the next Lemma we establish the invertibility of  $  \overline{\chi_\alpha} H(\uw) \overline{\chi_\alpha}  $ for $ \uw \in \widetilde \cW_\xi $ and $z \leq e^{-\iota \alpha }\rho/2$, which implies that, in this case, the 
Feshbach-Schur map is well defined.    

\begin{lemma} \label{L.fm}

Suppose that $\uw \in \widetilde \cW_{\xi}$, and $|z| \leq e^{-\iota \alpha }\rho/2$.   
Then, for every $ \kappa  \in \big[0 , 1- \frac{e^{(1-\iota)\alpha}}{2(1 - \epsilon_F)}\big),  $  
\begin{align} \label{fm.2.1.1}
\Big \|\frac{ \chi_0\overline{\chi_\alpha}(H_f +\kappa e^{-\alpha}\rho)}{w_{0, 0}(z; H_f ) }  \Big \|  
\leq \frac{[(1 - \kappa)\rho e^{-\alpha}]^{-1}}{(1 - \epsilon_F) -  \frac{1}{2 (1 -\kappa)} e^{(1 - \iota) \alpha}}\comma 
\end{align}
and the operator $ \overline{\chi_\alpha} H(\uw) \overline{\chi_\alpha} $ is invertible. It follows furthermore that
\eq{ \label{fm.2}
 \Big \| \Big(\overline{\chi_\alpha} & H(\uw) \overline{\chi_\alpha}\Big)^{-1}\Big\|\leq \frac{1}{e^{- \alpha}\rho(1 - \epsilon_F) -  \frac{1}{2} e^{- \iota \alpha}\rho - \epsilon_I\xi } \period
}
\end{lemma}

{\noindent \it Proof:}
First we notice that [see (\ref{i.8el}), \eqref{put0}, \eqref{rmn.3} and \eqref{rmn.5}] for $\rho \geq  r \geq e^{- \alpha}\rho - \kappa e^{- \alpha}\rho $
\begin{align} \label{fm.2.1}
|w_{0, 0}(z, r)|  & \geq  r - r\Big |\frac{w_{0, 0} (z, r) - w_{0, 0} (z, 0)}{r} - 1\Big | - | w_{0, 0} (z, 0) | \\ \notag 
& \geq r(1 - \epsilon_F) -  \frac{1}{2(1 - \kappa)} r e^{\alpha} e^{- \iota \alpha} \period
\end{align}
Eq.~(\ref{fm.2.1}) and the functional calculus imply (\ref{fm.2.1.1}); we recall that $W_{0, 0}(\uw)$ commutes with $H_f$, it depends only on $H_f$ and $z$.

We use Neumann series to calculate 
\begin{align}\label{fmn.2}
(\overline{\chi_\alpha} H(\uw) \overline{\chi_\alpha})^{-1}  = & \frac{\overline{\chi_\alpha}\chi_0}{W_{0, 0}(\uw) }\\ \notag
& \cdot \sum_{L = 0}^{\infty} (-1)^L\Big ( W(\uw)\frac{\overline{\chi_\alpha}\chi_0}{W_{0, 0}(\uw) }\Big )^{L} \comma 
\end{align}
which is well-defined whenever
\beq \label{fmn.3}
\begin{array}{l}
\|W(\uw)\frac{\overline{\chi_\alpha}\chi_0}{W_{0, 0}(\uw) } \| < 1 \period
\end{array}
\ene
Eq.~\eqref{rmn.6},  Eq.~\eqref{rmn.5}, Theorem \ref{eleseguey} and, Eq.~(\ref{fm.2.1.1}), ensure that   
inequality (\ref{fmn.3}) is accomplished. Eq.~(\ref{fm.2}) follows from \eqref{rmn.5}, Theorem \ref{eleseguey},  (\ref{fm.2.1.1}) and (\ref{fmn.2}). 


\qed

\begin{corollary}\label{compact-set}[From the proof of Lemma \ref{L.fm}]
Suppose that $\uw \in \widetilde \cW_{\xi}$.
Then $H(\uw(z))$ is invertible for every real number  $z$ in the interval
$$ 
z \in \Big( \frac{1}{4}\rho e^{- \alpha},    \frac{1}{2} \rho\Big). 
$$ 

\end{corollary}
\emph{Proof:}
We notice that \eqref{rmn.3}-\eqref{rmn.5} implies that [see also \eqref{besto}]
$$\Re w_{0,0}(z, r) \geq  z + r - \epsilon_F r \geq z
$$
and therefore $w_{0,0}(z, H_f)$ is invertible. The rest follows from a Neumann series-expansion argument as in the
proof of Lemma \ref{L.fm} and from \eqref{rmn.6}.  
\QED

\begin{remark}\label{R.fmn}
Remark \ref{R.h}, Lemma \ref{L.fm} and Formula (\ref{fm.3}) imply that the operator-valued function 
$$
z \in D_{e^{-\iota \alpha} \rho/2} \mapsto F_\alpha[H(\uw(z))]
$$ 
is analytic, provided that $\uw \in \widetilde \cW_\xi $.     

\end{remark}

\secct{The Renormalization Map $\cR_\alpha$ }\label{S.RG}

\subsection{The Dilation Operator}

For every $\alpha \in \RR$,
we denote by $u(\alpha): \mathfrak{h} \to \mathfrak{h}$ 
 the group of 
 dilation operators on $\mathfrak{h}$.
 The unitary operator $u(\alpha)$ is defined by the 
formula 
\beq \label{d1}
(u(\alpha)\phi)(k) := e^{ -3 \alpha /2}\phi(e^{-\alpha} k), \, 
 \,\forall \phi  \in L^{2}(\RR^3) \period 
\ene
We denote by $\Gamma_\alpha $ the operator that results after lifting $u(\alpha)$ to the 
Fock space.

We have that 
\begin{equation} \label{rmn.1}
\Gamma_\alpha \Omega = \Omega, \: \Gamma_\alpha a^*(k)  \Gamma_\alpha^* = e^{-3\alpha /2}a^*(e^{-\alpha} k), \: 
\Gamma_\alpha a(k)  \Gamma_\alpha^* = e^{-3\alpha /2}a(e^{-\alpha} k).
\end{equation} 
 
\subsection{Definition of the Renormalization Map}
The renormalization map consists on two parts: a rescaled Feshbach-Schur map (see Section \ref{rescal}) and a renormalization of the spectral parameter $(z)$ (see Section \ref{rensp}). Rescaling the Feshbach-Schur map is convenient, but not necessary. The renormalization of the spectral is the fundamental ingredient for the renormalization map. Renormalizing the spectral parameter amounts to find 
the values of $\zeta$ for which the Feshbach-Schur map can be applied not only one but two times. Repeating this procedure we find values of $\zeta$ for which one can apply the Feshbach-Schur map $N-$times for any (fixed) natural number $N$. Each time that we apply the Feshbach-Schur map the interacting term of the resulting operator becomes smaller and we can, therefore, estimate the spectrum more precisely. As the Feshbach-Schur map is isospectral, the more times it can be applied,  
the more accuracy we get on the estimation of the spectrum. The values $\zeta$ for which the Feshbach-Schur map can be applied two times are parametrized by an injective function [see \eqref{rmn.11}]
$$
(Q_\alpha)^{-1} : D_{\rho/2} \mapsto e^{-\iota \alpha}D_{\rho/2}, \hspace{2cm} \zeta = (Q_\alpha)^{-1}(z)\comma  
$$ 
which is what we refer as the renormalization of the spectral parameter $z$. This function is biholomorphic on its range.

\subsubsection{The Rescaled Feshbach-Schur Map}\label{rescal}

The next definition is a slight modification of the Feshbach-Schur map. It introduces a rescaling that is convenient 
(though not necessary) for our analysis.  
\begin{definition}[Rescaled Feshbach-Schur Map]
For every $\uw \in \widetilde \cW_\xi$, we define the following operator, 
\begin{equation} \label{rmn.7}
\widehat{\cR_\alpha}(H(\uw)) : = e^\alpha \Gamma_\alpha F_\alpha (H(\uw)) \Gamma_\alpha^*.
\end{equation} 
\end{definition}

In the next lemma we prove that $  \widehat{\cR_\alpha}(H(\uw))  $ satisfies a semigroup property. 

\begin{lemma}[Semigroup Property] \label{L.rep.1} Let $\alpha, \beta \geq 0$ and
 $\uw \in \cW_\xi$. Let $\cA$ be a simply connected open subset of $\CC$ and $f : \cA \to D_{\rho/2}$ be an analytic function. Suppose that there is a compact infinite subset  $ \mathcal{C} \subset \cA$ such that $ H(\uw(f(z))) $ is invertible 
for all $z \in \mathcal{C}$. \\
Suppose furthermore that $ \overline{\chi}_\alpha H(\uw(f(z))) \overline{\chi}_\alpha $, \\
 $ \overline{\chi}_{\alpha + \beta} H(\uw(f(z))) \overline{\chi}_{\alpha + \beta} $
and
 $ \overline{\chi}_\beta \widehat{\cR_\alpha}\big(H(\uw(f(z)))\big) \overline{\chi}_\beta  $ are invertible for every $z \in \mathcal{C}$, then 
\beq \label{rmn.8}
 \widehat{\cR_\beta} \Big(\widehat{\cR_\alpha}\big(H(\uw(f(z)))\big)\Big) = \widehat{\cR_{\alpha + \beta}}\big(H(\uw(f(z)))\big) \comma
\ene
for all $z \in \cA.$
\end{lemma} 
\noindent \emph{Proof:}
First suppose that $H(\uw(f(z)))$ is invertible. Theorem II.1 in \cite{BFS98a} implies that $  \widehat{\cR_{\alpha + \beta}}\big(H(\uw(f(z)))\big) $, 
$\widehat{\cR_\alpha}\big(H(\uw(f(z)))\big)$ and $ \widehat{\cR_\alpha} \Big(\widehat{\cR_\alpha}\big(H(\uw(f(z)))\big)\Big)  $ are invertible and that 
\begin{align} \label{rmn.9}
\Big(\widehat{\cR_\alpha}\big(H(\uw(f(z)))\big)\Big)^{-1} & =  e^{-\alpha}\Gamma_\alpha \chi_{\alpha} H(\uw(f(z)))^{-1}\chi_{\alpha}  \Gamma_\alpha^*\comma \notag \\ \notag
\Big(\widehat{\cR_{\alpha + \beta}}\big(H(\uw(f(z)))\big)\Big)^{-1} & =  e^{-(\alpha + \beta)}\Gamma_{\alpha + \beta} \chi_{\alpha + \beta} H(\uw(f(z)))^{-1} \\ \notag & \hspace{5cm}\cdot \chi_{\alpha + \beta}  \Gamma_{\alpha + \beta}^*\comma \\ 
\Big[\widehat{\cR_\beta} \Big(\widehat{\cR_\alpha}\big(H(\uw(f(z)))\big)\Big)\Big]^{-1} &  =  e^{-\beta}\Gamma_{\beta}\chi_{\beta}  \Big(\widehat{\cR_\alpha}\big(H(\uw(f(z)))\big)\Big)^{-1} 
\chi_{\beta}  \Gamma_{\beta}^* \comma
\end{align}
which implies that 
\beq \label{rmn.10}
\Big(\widehat{\cR_{\alpha + \beta}}\big(H(\uw(f(z)))\big)\Big)^{-1} = 
\Big[\widehat{\cR_\beta} \Big(\widehat{\cR_\alpha}\big(H(\uw(f(z)))\big)\Big)\Big]^{-1}
\ene
and thus (\ref{rmn.8}) holds true. As equality (\ref{rmn.10}) is valid for $z $ belonging to the infinite compact 
set $\mathcal{C}$ contained in $\cA$  (in which $H(\uw(f(z)))$ is invertible) and both sides of the equality are analytic in $\cA$ [see Remark~\ref{R.h}, Definition~\ref{D.fmn.1} and Remark~\ref{R.fmn}], it follows that (\ref{rmn.10}) holds true for $z \in \cA$.

\QED

\subsubsection{The Renormalization of the Spectral Parameter}\label{rensp}

\begin{definition}
Let $\uw \in \widetilde \cW_\xi$ and $z \in D_{e^{-\iota \alpha} \rho/2}$. We define 
\beq \label{rmn.11}
Q_\alpha (z) : = e^{\alpha} \Big\la  F_\alpha \big( H(\uw(z))\big)  \Big\ra_\Om = \Big \la \widehat \cR_\alpha \big(H(\uw(z))\big)    \Big \ra_\Omega \comma
\ene
where we recall that
\beq \label{rmn.13r}
\la \cdot \ra_\Om = \la \Om \;| \cdot \Om   \ra.
\ene

\end{definition} 

\begin{lemma}\label{L.r.1}
Let $\uw \in \widetilde \cW_\xi$ and $z \in D_{e^{-\iota \alpha} \rho/2}$. The following inequality holds true 

\begin{multline} \label{E.L.r.1} 
\Big| \frac{\partial}{\partial z} \Big\la  H(\uw(z)) \overline{\chi_\alpha}  \Big(\overline{\chi_\alpha} H(\uw(z))
 \overline{\chi_\alpha}\Big)^{-1}
  \overline{\chi_\alpha} H (\uw(z))  \Big\ra_\Omega \Big| \\    \leq \cG_\alpha(\ue), 
\end{multline}
where we recall that $ \cG_\alpha(\ue) $ is defined in \eqref{HiE.L.r.2}.
\end{lemma}

\noindent {\it Proof:}
We compute
\eq{ \notag
 \frac{\partial}{\partial z} & \Big \la   H(\uw(z)) \overline{\chi_\alpha}\Big(\overline{\chi_\alpha} H(\uw(z))
 \overline{\chi_\alpha}  \Big)^{-1}\overline{\chi_\alpha} H (\uw(z))\Big \ra_{\Omega} \\  \notag
 \leq &  \Big|  \Big\la \frac{\partial}{\partial z} W(\uw(z))\overline{\chi_\alpha}\Big(\overline{\chi_\alpha} W(\uw(z))
 \overline{\chi_\alpha}\Big)^{-1}\overline{\chi_\alpha}W(\uw(z))\Big\ra_{\Om} \Big| \\ \notag
& +  \Big|\Big \la W(\uw(z))\overline{\chi_\alpha}\Big(\overline{\chi_\alpha}H(\uw(z))\overline{\chi_\alpha}\Big)^{-1}
  \overline{\chi_\alpha}\frac{\partial}{\partial z} W(\uw(z))   \Big \ra_\Om \Big| \\ \notag 
& + \Big|\Big\la  W(\uw(z))\overline{\chi_\alpha}\Big(\overline{\chi_\alpha} H(\uw(z))
 \overline{\chi_\alpha}\Big)^{-1}\\ \notag & 
 \hspace{2.5cm}\cdot \Big(\frac{\partial}{\partial z}\overline{\chi_\alpha}H(\uw(z))
\overline{\chi_\alpha} \Big)\\ \notag & \hspace{2.5cm} \cdot \Big( \overline{\chi_\alpha} H(\uw(z))
 \overline{\chi_\alpha} \Big)^{-1}
 \overline{\chi_\alpha} W(\uw(z)) \Big\ra_\Om \Big| \\ \notag
\leq & \, 2 \min(\alpha, 1) (\epsilon_I\xi)^2 \Big(\frac{1}{ e^{- \alpha}\rho(1 - \epsilon_F) -  \frac{1}{2} e^{- \iota \alpha}\rho - \epsilon_I\xi }\Big)  \\ \notag
& + \min(\alpha, 1) (\epsilon_I\xi)^2 \Big(\frac{1}{ e^{- \alpha}\rho(1 - \epsilon_F) -  \frac{1}{2} e^{- \iota \alpha}\rho - \epsilon_I\xi }\Big)^2  
\\ \notag & \hspace{8cm} \cdot\Big\| (\frac{\partial}{\partial z}  \overline{\chi_\alpha}  H(\uw(z))
 \overline{\chi_\alpha} )  \Big\| \\ 
 \leq & \,\notag
  2 \min(\alpha, 1)  (\epsilon_I\xi)^2  \Big(\frac{1}{ e^{- \alpha}\rho(1 - \epsilon_F) -  \frac{1}{2} e^{- \iota \alpha}\rho - \epsilon_I\xi  }\Big)
 \\  & +\min(\alpha, 1)  (\epsilon_I\xi)^2  \Big(\frac{1}{ e^{- \alpha}\rho(1 - \epsilon_F) -  \frac{1}{2} e^{- \iota \alpha}\rho - \epsilon_I\xi }\Big)^2  \Big(\epsilon_Z  + \epsilon_I\xi  \Big), \label{unic-azul}
}
where we used the proof of lemma \ref{L.h} and (\ref{fm.2}). Notice that we use the proof of (\ref{h.7}) to estimate the operator $W(\uw)$ to the left in the inner product because the terms containing $W_{m,n}(\uw)$ with $n = 0$ [see (\ref{hw})-(\ref{w})] vanish after taking the inner product. Similarly we use the proof of (\ref{h.8}) to estimate the operator $W(\uw)$ to the right in the inner product. We conclude using \eqref{E.L.r.2} and \eqref{unic-azul}.

\qed

\begin{lemma}\label{L.r.2}
Let $\uw \in \widetilde \cW_\xi$ and $z \in D_{e^{-\iota \alpha} \rho/2}$. Then  $ Q_\alpha  :  Q_\alpha^{- 1} (D_{\rho/2})\cap (e^{-\iota \alpha}D_{\rho/2}) \mapsto D_{\rho/2} $ is a bijection. It follows furthermore that 
\eq{ \label{ddz}
|\frac{d}{dz} Q_\alpha^{-1}| \leq e^{-\alpha}\frac{1}{1 - \cG_\alpha(\ue)}\period
}
The set $  Q_\alpha^{- 1} (D_{\tilde \rho/2}) \cap(e^{-\iota \alpha}D_{\rho/2})  $, for every $\tilde \rho \leq \rho$,  can be estimated in the following way: 
\begin{itemize}
\item[(I)] Every $z \in e^{-\iota \alpha }D_{\rho/2}$ such that 
\eq{\label{estimatez}
e^{\alpha}|z| + e^{\alpha} \cG_\alpha(\ue)< \frac{\tilde \rho}{2}\comma 
}
belongs to $  Q_\alpha^{- 1} (D_{\tilde \rho/2}). $
\item[(II)]
\eq{\label{estimatez1}
| Q_\alpha^{- 1}(\zeta) - e^{-\alpha} \zeta| 
 \leq & \frac{1}{1 - \cG_\alpha(\ue) } (\epsilon_I\xi)^2  \min(\alpha, 1) 
\\ \notag & \cdot 
\Big(\frac{1}{e^{- \alpha}\rho(1 - \epsilon_F) -  \frac{1}{2} e^{- \iota \alpha}\rho - \epsilon_I\xi } \Big)\comma 
}
where $ Q_\alpha^{- 1}(\zeta) $ denotes the unique point in $e^{-\iota \alpha}D_{\rho/2}$ whose image under 
$ Q_\alpha$ is $\zeta$. 

\end{itemize}

\end{lemma} 
 
\noindent{\it Proof:}

We consider the function, 
$$
h(z) = z + e^{- \alpha} \zeta - e^{- \alpha } Q_\alpha(z)\comma
$$
for an arbitrary $\zeta \in D_{\rho/2}$. We search for a fixed point of $ h $. First we notice that as a consequence
\eqref{E.L.r.2.1} and Lemma \ref{L.r.1}, $h$ is a contraction [see \eqref{rmn.5}]:
$$
\Big| \frac{\partial}{\partial z} h(z)\Big| \leq \cG_\alpha(\ue) < 1.  
$$
We define a sequence  $\{ z_n \}_{n \in \mathbb{N}_0}$ recursively in the following way.  
We take  $ z_0  : = e^{- \alpha } \zeta $ and define  
$ z_{n + 1} = h(z_{n}) $ whenever $ z_n \in e^{- \iota \alpha} D_{\rho/2} $. 
We notice that [see \eqref{rmn.3} and (\ref{w})]
\eq{
|z_1 - z_0| & =  |  e^{- \alpha } \zeta  - e^{- \alpha} Q_\alpha( e^{- \alpha} \zeta )|  \\ \notag & =  \Big|
  \Big\la   W(\uw) \overline{\chi_\alpha} \Big(   \overline{\chi_\alpha} H(\uw)
 \overline{\chi_\alpha}  \Big)^{-1}
 \overline{\chi_\alpha} W(\uw)  \Big\ra \Big| \\ \notag
&  \leq  (\epsilon_I\xi)^2 \min(\alpha,1) \Big(\frac{1}{e^{- \alpha}\rho(1 - \epsilon_F) -  \frac{1}{2} e^{- \iota \alpha}\rho - \epsilon_I\xi } \Big)\comma 
}
where we used Lemmata~\ref{L.h} and \ref{L.fm}. Notice that we use (\ref{h.7}) to estimate the operator $W(\uw)$ to the left in the inner product because the terms containing $W_{m,n}(\uw)$ with $n = 0$ [see (\ref{hw})-(\ref{w})] vanish after taking the inner product. Similarly we use (\ref{h.8}) to estimate the operator $W(\uw)$ to the right in the inner product. 

By the contraction property of $h$ we have that 
\eq{\label{ref}
|z_n - z_0 |  \leq & \sum_{j = 0}^{\infty} \cG_\alpha(\ue)^j | z_1 - z_0 | \\ \notag
 \leq & \frac{1}{1 - \cG_\alpha(\ue) } (\epsilon_I\xi)^2 \min(\alpha, 1) 
\\ \notag & \cdot \Big(\frac{1}{e^{- \alpha}\rho(1 - \epsilon_F) -  \frac{1}{2} e^{- \iota \alpha}\rho - \epsilon_I\xi  } \Big). 
}
The desired property $ |z_n| < \frac{1}{2}e^{- \iota \alpha} \rho $ is satisfied if  
\eq{ \notag
\frac{1}{1 - \cG_\alpha(\ue)} (\epsilon_I\xi)^2 \alpha \Big(\frac{1}{e^{- \alpha}\rho(1 - \epsilon_F) -  \frac{1}{2} e^{- \iota \alpha}\rho - \epsilon_I\xi } \Big)    < \frac{1}{2}e^{- \iota \alpha}\rho - e^{- \alpha}\frac{1}{2}\rho, 
}
which is accomplished because \eqref{E.L.r.2.2} holds true.\\
We conclude that 
$$
 \lim_{n \to \infty} z_n =  z \in e^{- \iota \alpha }D_{\rho/2}
$$ 
exists and that $  Q_\alpha(z) = \zeta$. The injectivity of $ Q_\alpha$ restricted to $e^{-\iota \alpha } D_{\rho/2}$ follows from the unicity of the fixed point for the contraction $h$.

Eq.~(\ref{ddz}) follows from (\ref{rmn.3}), (\ref{rmn.11}) and (\ref{E.L.r.1}). Item (I) follows from 
\eqref{rmn.11} and \eqref{E.L.r.1}, since \eqref{estimatez} implies that $| Q_\alpha(z)| < \frac{\tilde \rho}{2}$.  
Item (II) is a consequence of \eqref{ref}.

\qed

\subsubsection{The Renormalization Map} 
\begin{definition} \label{ralpha}
For every $ \uw \in \widetilde \cW_\xi $ we define the renormalization group by 

$$
\cR_\alpha (H (\uw))(z)  : = \widehat \cR_\alpha (H (\uw(\zeta))),
$$
where $ \zeta =  Q_\alpha^{-1} (z) $. 
\end{definition}

\begin{lemma} \label{chito}
For every $ \uw \in \widetilde \cW_\xi $ the following holds true
$$
\la  \cR_\alpha (H(\uw)) \ra_\Om = z. 
$$
\end{lemma}

\emph{Proof:}
The desired result follows from the following computation:
\eq{
\la  \cR_\alpha (H(\uw)) \ra_\Om = \la \widehat \cR_\alpha (H (\uw(\zeta))) \ra_\Om  =  Q_\alpha (\zeta) =  
 Q_\alpha ( (Q_\alpha)^{-1}(z)) = z.
}
\QED

\secct{Analysis of the Renormalization Map $\cR_\alpha$} \label{S.contraction}
In this section we prove that for every $\uw \in \widetilde \cW_\xi $ there is a 
$\widehat \uw =\big( \widehat w_{m,n}\big)_{m + n \geq 0}  $ satisfying \eqref{esano} such that
\eq{\label{noecue}
\cR_\alpha(\uw) = H(\widehat \uw)
}
in the sense of quadratic forms in $  \cF_{\mathrm{fin}}$. We furthermore do some estimations for the 
norms $\| \widehat w_{m,n} \|^{(\infty)}$, $\|\widehat  w_{m,n} \|^{(0)}$, and $\|\widehat  w_{0,0} - r \|^{(0)}$. In the next section (Section \ref{rf}) we give conditions to have 
$ \widehat \uw \in \widetilde \cW_\xi  $ and we prove furthermore that $\| \widehat w \|^{(I)}_\xi$ decreases exponentially in $\alpha$. In this case clearly \eqref{noecue} holds as an operator equality.

\subsection{Analysis of the Feshbach-Schur Map $F_\alpha$} \label{sano}
In this section we prove that for every $\uw \in \widetilde \cW_\xi $ and every $z \in e^{-\iota \alpha} D_{\rho/2}$ there exists
$\widetilde \uw^{(sym)} =\big( \widetilde w_{m,n}^{(sym)}\big)_{m + n \geq 0}  $ satisfying \eqref{esano} such that
\eq{\label{noecuef}
F_\alpha(\uw) = \chi_\alpha H(\uw) \chi_\alpha + \chi_\alpha H(\widetilde \uw^{(sym)})\chi_\alpha
}
in the sense of quadratic forms in $  \cF_{\mathrm{fin}}$. We furthermore do some estimations for the 
norms $\| \widetilde w_{m,n}^{(sym)} \|^{(\infty)}$, $\|\widetilde  w_{m,n}^{(sym)} \|^{(\alpha)}$, 
and $\|\widetilde  w_{0,0}^{(sym)} - r \|^{(\infty)}$.

\subsubsection{Definition of $\widetilde \uw^{(sym)}$}
\paragraph{Notation}$\\ $
\begin{definition}
We fix the function $F$ appearing in \eqref{F1} to be 
\eq{\label{F1ese}
F(z; r) : = \overline \chi_\alpha(r) \chi_0(r)\frac{1}{w_{0,0}(z, r)} 
\comma
}
Notice that term in the denominator is well defined due to \eqref{fm.2.1}. 
The superscripts $(\cdot)^F$  appearing in several objects in Section \ref{cf} are omitted for the rest of the paper, in the understanding that the corresponding function $F$ used is given by \eqref{F1ese}. 
In some proofs we omit writing the dependence on $z$ 
of $w_{0,0}$ and take $ w_{0,0}(z, r) \equiv w_{0,0}( r) $, if keeping track of the variable $z$ is not necessary for the argument. 
 
\end{definition}

\begin{lemma}\label{L.cprm.5}
For every $\uw \in \widetilde \cW_\xi$ and $z \in e^{-\iota \alpha} D_{\rho/2}$ there exists 
$$ \widetilde \uw^{(sym)} = \big( \widetilde w_{m,n}^{(sym)} \big)_{m + n \geq 0} $$ 
satisfying \eqref{esano} such that 
\beq \label{E.L.cprm.5.1}
F_\alpha(H(\uw))(z) = \chi_\alpha  H(\uw)\chi_\alpha +  \chi_\alpha H(\widetilde \uw^{(sym)}(z)) \chi_\alpha .
\ene
$\widetilde w_{m,n}^{(sym)}$  is the symmetrization with respect to $k^{(m)}$ and $k^{(n)}$ of the functions 
\eq{ \label{E.L.cprm.5.2} 
& \widetilde w_{m,n}(z; r; k^{(m,n)}) \\ & : =   \notag  \sum_{L = 2}^{\infty}(-1)^{L - 1}  \sum_{\upsilon \in \cB_L(m,n)} 
\Big[ \prod_{\ell =1}^L \binom{m_\ell + p_\ell}{p_\ell} \binom{n_\ell + q_\ell}{q_\ell } \Big]  \notag  V_\upsilon (z; r;k^{(m,n)})\period
}

\end{lemma}

\noindent \emph{Proof:}
The result follows from the Neumann expansion (\ref{fmn.2}), Definitions \ref{D.cprm.2}-\ref{D.cprm.4} and Lemma \ref{L.cprm.1}. The only missing part is to prove that the functions \eqref{E.L.cprm.5.2} satisfy \eqref{esano}. We defer the proof of this to further sections. It is proved in Theorem \ref{T.C.wmn}.  

\QED

\begin{definition}
For any $ (m, n) \in \mathbb{N} \times \mathbb{N}_0 \cup \mathbb{N}_0 \times \mathbb{N} $ and $\alpha \geq 0$ we define
\begin{align}\label{i.6}
\| w_{m, n}  \|^{(\alpha)}  : = \sup  \Big \{ &\frac{1}{r} \int_{\|k^{(m)}\|_1, \| \tilde k^{(n)} \|_1 \leq e^{-\alpha}\rho } \frac{d k^{(m,n)}}{ |k^{(m,n)}|^{3/2 + \mu/2}} 
 \\ \notag &  | w_{m, n}( z; s+ r; k^{(m,n)}) - w_{m, n}( z; s; k^{(m,n)})|  
    \\ \notag & \hspace{1cm} \Big | \:  z\in D_{\rho/2}, r \in (0, e^{-\alpha}\rho -s], s\in [0,e^{-\alpha}\rho] \Big \}.
\end{align}
In the case that $m = 0$ or $n = 0$ we omit the corresponding variable.


 For $m = n  = 0$ and $\alpha \geq 0$ we denote by 
\begin{align}\label{i.8}
\| w_{0, 0}  \|^{(\alpha)} : =   \sup  \Big \{\tfrac{1}{r}  &|(w_{0, 0}( z; s +  r )  - w_{0, 0}( z; s  ) ) |  
   \\ \notag  &  \Big | \:  z\in D_{\rho/2}, r \in (0, e^{-\alpha}\rho -s], s\in [0,e^{-\alpha}\rho]  \Big \}.
\end{align}

\end{definition}

\subsubsection{Estimates for $\|  \widetilde w_{m,n}\|^{(\infty)}$}

\begin{lemma}\label{L.cprm.6}
Suppose that $L \geq 2$, $m +n \geq 0$ and that $\upsilon \in \cB_L(m, n)$. Suppose furthermore that $\uw \in \widetilde \cW_\xi$ and that $z \in e^{-\iota \alpha}D_{\rho/2}$. Then the following estimate holds true
\eq{ \label{E.L.cprm.6.1} 
\big |V_\upsilon(z;& r; k^{(m,n)})\big | \cdot  |k^{(m)}|^{1/2-\mu/2} \cdot |\tilde k^{(n)}|^{1/2-\mu/2} 
\\ \notag \leq &  \Big(  \frac{(\rho e^{-\alpha})^{-1}}{(1 - \epsilon_F) - \frac{1}{2}e^{(1 -\iota) \alpha}}  \Big)^{L-1}   
   \cdot \rho^{(|\bar p|_1 + |\bar q|_1)/2} \Big[\prod_{\ell =1}^{L} \| w_{m_\ell + p_\ell, n_\ell + q_\ell} \|^{(\infty)}\Big]\period
}
\end{lemma}

\noindent \emph{Proof:}

Eqs.~\eqref{F2}, (\ref{fm.2.1.1}) and (\ref{F1ese}) imply that
$$
\| F_\ell(H_f + r + \tilde r_\ell)\| \leq \frac{(e^{-\alpha}\rho)^{-1}}{(1 - \epsilon_F) - \frac{1}{2}e^{ (1-\iota) \alpha}}, \: \:\ell \in \{ 1, \cdots, L-1 \}.  
$$
The result follows from \eqref{rho} (we estimate $ 4 \pi \rho^{1+\mu} < 1$), the proof of Lemma \ref{L.h} and 
\eq{ \label{P.L.cprm.6.1}\notag 
|V_\upsilon(z;  r; k^{(m,n)})|  \leq &  \| \chi_0   \widetilde W_1 (z; r+   r_1; k^{(m_1, n_1)})  \cdot \overline \chi_\alpha \chi_0(H_f + r + \tilde r_1) \| \\ \notag
 & \cdot \|     F_1(H_f + r + \tilde r_1) \|  \\ & \notag \cdot \Big[\prod_{\ell = 2}^{L-1} \|   \overline \chi_\alpha \chi_0(H_f + r + \tilde r_{\ell-1})   \\ & \hspace{1cm} \notag  \cdot \widetilde W_{\ell} (z; r+   r_\ell; k^{(m_\ell, n_\ell)}) 
  \overline \chi_\alpha \chi_0(H_f + r + \tilde r_\ell)   \| 
\\ \notag & \hspace{6cm} \cdot \|  F_\ell(H_f + r + \tilde r_\ell) \| \Big]
\\ 
&  \cdot \|  \overline \chi_\alpha \chi_0(H_f + r + \tilde r_{L-1}) 
 \cdot \widetilde W_L(z; r+   r_L; k^{(m_L, n_L)})  
 \chi_0\| \period
}

\QED

\begin{lemma}\label{L.cprm.6.z}
Suppose that $L \geq 2$, $m +n \geq 0$ and that $\upsilon \in \cB_L(m, n)$. Suppose furthermore that $\uw \in \widetilde \cW_\xi$ and that $z \in e^{-\iota \alpha}D_{\rho/2}$. Then the following estimate holds true [see \eqref{rmn.5}]
\eq{ \label{E.L.cprm.6.1.z} 
\big |\frac{\partial}{\partial z}&V_\upsilon(z; r; k^{(m,n)})\big | \cdot  |k^{(m)}|^{1/2-\mu/2} \cdot |\tilde k^{(n)}|^{1/2-\mu/2} 
\\ \notag \leq & \notag L (\epsilon_Z+1) \Big(  \frac{(e^{-\alpha}\rho)^{-1}}{(1 - \epsilon_F) - \frac{1}{2}e^{(1 -\iota) \alpha}}  \Big)^{L}   \rho^{(|\bar p|_1 + |\bar q|_1)/2}
 \Big[\prod_{\ell =1}^{L}   \| w_{m_\ell + p_\ell, n_\ell + q_\ell} \|^{(\infty)}\Big] 
\period \notag
}
  
\end{lemma}

\noindent \emph{Proof:}
The proof is similar to the one of Lemma \ref{L.cprm.6}. Here we use the Leibniz rule to compute the derivative. The new ingredient is the term 
$$
\frac{d }{dz} \frac{1}{w_{0,0}} = \Big(\frac{1}{w_{0, 0}}\Big)^{2} \frac{d }{dz} w_{0,0}\comma 
$$ 
which produces the extra
$$
(e^{-\alpha}\rho)^{-1}\epsilon_Z \Big(  \frac{1}{(1 - \epsilon_F) - \frac{1}{2}e^{(1 -\iota) \alpha}}  \Big) 
$$
factor [relative to (\ref{E.L.cprm.6.1})]. 

\QED

\begin{theorem}\label{T.C.wmn}
Suppose that $L \geq 2$, $m +n \geq 0$ and that $\upsilon \in \cB_L(m, n)$. Suppose furthermore that $\uw \in \widetilde \cW_\xi$ and that $z \in e^{-\iota \alpha}D_{\rho/2}$. Then the following estimate holds true
\eq{\label{wmninfty}
\|\widetilde w_{m, n}\|^{(\infty)}\leq \epsilon_I^2 \xi^{m+ n} 4^{m+ n} \frac{(1 - \xi)^2}{2}  A^{(\infty)}(\ue, \alpha)
 \comma
}
where
\eq{ \label{ainfty}
A^{(\infty)}(\ue, \alpha) : = 
\Big(2 +   \epsilon_Z \Big)  
\frac{3 \Big[ \Big(3(e^{-\alpha}\rho)^{-1} \Big)\Big]^2  
 (2^4)^2 }{\Big(1 -  
\Big[\epsilon_I \Big(3(e^{-\alpha}\rho)^{-1} \Big)\Big] 2^4 \Big)^2}\period
}
\end{theorem}
\emph{Proof:}
We use \eqref{i.11.1}, \eqref{rmn.5} and Lemmata \ref{L.cprm.5}, \ref{L.cprm.6} and \ref{L.cprm.6.z} to obtain that 
\eq{ \notag
\| \widetilde w_{m,n} \|^{(\infty)} \leq & \frac{(1 - \xi)^2}{2} \xi^{m+ n} 
\Big(2 +  \epsilon_Z \Big)
 \sum_{L = 2}^{\infty}L\Big[\epsilon_I \Big(\frac{(e^{-\alpha}\rho)^{-1}}{(1 - \epsilon_F) - \frac{1}{2}e^{(1- \iota)\alpha}} \Big)\Big]^{L} \\ &  \cdot \sum_{\upsilon \in \mathcal{B}(m,n)} \Big[\prod_{\ell =1}^L\binom{m_\ell + p_\ell}{p_\ell}\binom{n_\ell + q_\ell}{q_\ell}
   (\xi\rho^{1/2})^{p_\ell + q_\ell} \Big] \period \label{wmninfty.1} 
}
Now we use that $ \binom{M}{N} \leq 2^M$ and (\ref{rho}) to compute 
\eq{ \label{wmninfty.2}
\sum_{\upsilon \in \mathcal{B}(m,n)} & \Big[\prod_{\ell =1}^L\binom{m_\ell + p_\ell}{p_\ell}\binom{n_\ell + q_\ell}{q_\ell}\notag 
   \xi^{p_\ell + q_\ell} \Big(\rho^{1/2}\Big)^{(p_\ell +q_\ell)}\Big] 
\\ \notag & \leq 4^{m+ n}  \sum_{\upsilon \in \mathcal{B}(m,n)}\Big[\prod_{\ell =1}^L ( 2\xi \rho^{1/2})^{p_\ell + q_\ell} 2^{-(m_\ell + n_\ell )}\Big] 
\\ &  \leq \notag
4^{m+ n} \Big[ \sum_{j= 0}^{\infty} (2 \xi \rho^{1/2})^j    \Big]^{2L} 
\Big[ \sum_{j= 0}^{\infty} 2^{-j}    \Big]^{2L} 
\\ & \leq  (2^4)^{L} \cdot  4^{m+ n}  \comma
}
where we used \eqref{rho} to estimate $ 2 \xi \rho^{1/2} < \frac{1}{2} $.
Using (\ref{wmninfty.1}) and (\ref{wmninfty.2}) we obtain that [see also \eqref{besto} 	and \eqref{hirmn.6}]
\eq{\notag \label{wmninfty.3}
\| \widetilde w_{m,n} \|^{(\infty)}\leq &  4^{m+ n} \frac{(1 - \xi)^2}{2} \xi^{m+ n} 
\Big(2 +  \epsilon_Z \Big) \Big[\epsilon_I \Big(3(e^{-\alpha} \rho)^{-1} \Big)\Big] 2^4
\\\notag & \cdot \sum_{L = 2}^{\infty}L\Big[\epsilon_I \Big(3(e^{-\alpha} \rho)^{-1} \Big)\Big]^{L-1} (2^4)^{L-1}  \\ \notag 
 = & \epsilon_I^2  4^{m+ n} \frac{(1 - \xi)^2}{2} \xi^{m+ n} 
\Big(2 +   \epsilon_Z \Big)  \\ &  \cdot
\frac{3 \Big[ \Big(3(e^{-\alpha}\rho)^{-1} \Big)\Big]^2  
 (2^4)^{2} }{\Big(1 -  
\Big[\epsilon_I \Big(3(e^{-\alpha}\rho)^{-1} \Big)\Big] 2^4 \Big)^2}\comma
}
where we used that 
$$ \frac{-x^2 + 2x}{(1 - x)^2} = \Big(\frac{d}{dx} \frac{1}{1-x}\Big) - 1 =\Big(\frac{d}{dx}\sum_{L=0}^\infty x^L\Big) - 1 = \sum_{L=2}^\infty Lx^{L-1}\comma$$
and that for $x \in (0,1)$
$$
\frac{|-x^2 + 2x|}{(1 - x)^2} \leq 3 |x| \frac{1}{(1 - x)^2}\period 
$$
\QED

\subsubsection{Estimates for $\|  \widetilde w_{m,n}\|^{(\alpha)}$}
In this section we analyze the norm  $\|  \widetilde w_{m,n}\|^{(\alpha)}$ and give some estimates for this norm that are similar to \eqref{wmninfty}. The analogous of Theorem \ref{T.C.wmn} is the goal of this section. The proofs are considerably more involved. As in the proof of Theorem \ref{T.C.wmn}, the analysis of $ V_\upsilon(z;  r; k^{(m,n)}) $ 
is essential, but in this case, due to \eqref{i.6} and \eqref{i.8},  we have estimate the difference
\eq{ \label{diff}
\frac{1}{r}|V_\upsilon(z;  r + s; k^{(m,n)}) - V_\upsilon(z;  s; k^{(m,n)})| \comma
}
which is more difficult to handle for the singularity on the factor $\frac{1}{r}$ as $r $ goes to $0$. To overcome this problem we make use of the recursive relation (Lemma \ref{recursive-relation}). \\
 In the first part of this section we study the term \eqref{diff}. This part is very technical. In the second part of this section we conclude   
 using \eqref{E.L.cprm.5.2} as in the proof of Theorem \ref{T.C.wmn}.

\paragraph{Estimates for $\frac{1}{r} |V_\upsilon(z;  r + s; k^{(m,n)}) - V_\upsilon(z;  s; k^{(m,n)})| $} $\\ $

For every $\tilde s \in [0, e^{-\alpha }\rho]$,  $  V_\upsilon(z;  \tilde s; k^{(m,n)}) $ is the expectation value of a product consisting of $L$ factors of the form $\widetilde W_\ell$ and $L-1$ factors of the form $ F_\ell $. We use the Leibniz rule for finite differences to rewrite \eqref{diff} as a sum of $2L -1$ terms: $L $ terms related to the differences of the operators $ \widetilde W_\ell $ and $L-1$ terms containing differences of operators $F_\ell$. This two types of terms are studied differently. Both types of terms are analyzed using the recursive relation  (see Lemma \ref{recursive-relation}). 

The analysis of the terms containing differences of operators $F_\ell$ is done in Lemmata \ref{L.eirh.1} and \ref{L.eirh.3}. In this case, the step from $s$ to $r + s$ [see \eqref{diff}] is divided in smaller steps [see \eqref{ro}-\eqref{ko} and \eqref{ola.1}]. The size of the steps is taken to zero as in a Riemann integral (see the proof of Lemma \ref{L.eirh.3}).         

The analysis of the terms containing  differences of the operators $ \widetilde W_\ell $ is done in Lemma
\ref{L.eirh.2}. 

The main result of this section is Lemma \ref{vn1}, in which we add all estimates from Lemmata \ref{L.eirh.1}-\ref{L.eirh.2}.

\begin{lemma} \label{L.eirh.1}
Suppose that $L \geq 2$, $m +n \geq 0$ and that $\upsilon \in \cB_L(m, n)$. Suppose furthermore that $\uw \in \widetilde \cW_\xi$ and that $z \in e^{-\iota \alpha}D_{\rho/2}$. 
 Let $s \in [0, e^{-\alpha}\rho)$, $r \in (0, e^{-\alpha}\rho - s]$,
and $ t,  \tilde t \geq 0$. \\
Define 
\eq{\label{ro}
r_\varsigma : = e^{-\alpha}\rho \kappa_\varsigma \comma 
}
and
\eq{\label{ko}
\kappa_\varsigma : = \frac{r}{e^{-\alpha}\rho \varsigma}\comma
}
for big enough $\varsigma \in \NN$, such that $ \kappa_\varsigma  \in \big[0 , 1- \frac{e^{(1-\iota)\alpha}}{2(1 - \epsilon_F)}\big)  $ [see \eqref{put0}].

The following estimate holds true
\eq{ \label{E.L.eirh.1.1} \notag
\int & dy^{(|\bar u|_1)} d\tilde y^{(|\bar v|_1)} \int_{\|k^{(m)}\|_1, \| \tilde k^{(n)} \|_1 \leq e^{-\alpha}\rho }  \frac{dk^{(m)}}{|k^{(m)}|^{3/2 + \mu/2}}  \frac{d\tilde k^{(n)}}{|\tilde k^{(n)}|^{3/2 + \mu/2}}  
 \\ & \notag  \Big | V_{\upsilon_{\leq J}, \bar v}(z;t;k^{(m)}; \tilde k^{(n)}, \tilde y^{(|\bar v|_1)} ) 
  \frac{1}{w_{0,0}( br_\varsigma + s  + \tilde r_{J}( k^{(m,n)} \times  y^{(\bar u)}))}\\ &  \notag
\frac{1}{r_\varsigma}\Big[ \overline{\chi_\alpha} \chi_0\Big( (b+1)r_\varsigma + s + \tilde r_{J}( k^{(m,n)} \times  y^{(\bar u)}) \Big)  \\ \notag & \hspace{5cm}
 - \overline{\chi_\alpha} \chi_0\Big( br_\varsigma + s + \tilde r_{J}(k^{(m,n)} \times  y^{(\bar u)}) \Big) \Big]
\\ \notag & \cdot
\Big \la \prod_{j= 1}^{J}a(\tilde y_j^{(v_j)}) \prod_{j= J + 1}^{L}a^*( y_j^{(u_j)}) \Big \ra_\Om  
V_{\upsilon_{\geq (J+1)}, \bar u}(z; \tilde t;k^{(m)}, y^{(|\bar u|_1)};\tilde k^{(n)})\Big | \notag \\ \notag
 \leq & 
\frac{16  \pi}{u!}( 4 \pi \rho )^{n + m + |\bar u|_1 - 1} e^{-(m + n )\alpha} 
\rho^{|\bar u|_1}  (e^{-\alpha}\rho)^{-1}
\\ \notag & \cdot \Big( \frac{(e^{-\alpha}\rho)^{-1}}{(1 - \epsilon_F) - \frac{1}{2}e^{(1-\iota) \alpha}}  \Big)^{L-2}\frac{((1 - \kappa_\varsigma)e^{-\alpha}\rho)^{-1}}{(1 - \epsilon_F) - \frac{1}{2(1 - \kappa_\varsigma)}e^{(1-\iota) \alpha}}
\\ \notag
& \cdot  \prod_{\ell = 1}^J  
\big( \rho \big)^{(p_\ell +  q_\ell - v_\ell)/2}  \| w_{ m_\ell + p_\ell, n_\ell + q_\ell} \|^{(\infty)}
\\ 
& \cdot  \prod_{\ell = J + 1}^L 
\big(  \rho \big)^{(p_\ell +  q_\ell - u_\ell)/2}   \| w_{ m_\ell + p_\ell, n_\ell + q_\ell} \|^{(\infty)} \comma
}
fore every $b \in \{0, \cdots, \varsigma - 1\}$.

\end{lemma}

\noindent \emph{Proof:}
First we notice that we can take $m + n + |\bar u|_1 \geq 1$, because if it is zero 
\eq{\overline{\chi_\alpha} \chi_0\Big((b+1)r_\varsigma + s + \tilde r_{J}( k^{(m,n)} \times  y^{(\bar u)}) \Big) & = 0
\\ \notag &  =
 \overline{\chi_\alpha} \chi_0\Big(br_\varsigma + s + \tilde r_{J}(k^{(m,n)} \times  y^{(\bar u)}) \Big)
}
(notice that $r + s \leq e^{-\alpha}\rho $ and $ (b + 1)r_\xi \leq r $).
We use Remark \ref{expl} to define the vacuum expectation value of 
$$
 \prod_{j= 1}^{J}a(\tilde y_j^{(v_j)}) \prod_{j= J + 1}^{L}a^*( y_j^{(u_j)})  \comma
$$
 which is different from zero only when $|\bar u|_1 = |\bar v|_1$. \\
We identify as in Remark \ref{expl}
\eq{ \label{no-sense-0dos} 
y^{(|\bar u|_1)} & \equiv (y^1, \cdots, y^{|\bar u|_1}) \equiv (y^{(u_{J + 1})}, \cdots, y^{(u_L)})\comma \\ \notag
\tilde y^{(|\bar v|_1)} &  \equiv (\tilde y^1, \cdots, \tilde y^{|\bar v|_1}) \equiv (\tilde y^{(v_{1})}, \cdots, \tilde y^{(u_J)}) \period
}
We have that 
\eq{ \label{no-sense-1dos}
\Big \la \prod_{j= 1}^{J}a(\tilde y_j^{(v_j)}) \prod_{j= J + 1}^{L}a^*( y_j^{(u_j)}) \Big \ra := 
\sum_{p \in S_{|\bar v|_1}} \delta\Big( y^{(|\bar{u}|_1)} 
- p (\tilde y^{(|\bar{v}|_1)}) \Big) 
}  
where 
\eq{p (\tilde y^{(|\bar{v}|_1)}) : = (\tilde y^{p(1)}, \cdots, \tilde y^{p(|\bar v|_1)}) 
}
and $  S_{|\bar v|_1}  $ is the set of permutations of the first $|\bar v|_1 $ integers. 
 
Eq.~(\ref{no-sense-1dos}) can be taken as a distribution in the variables of $\tilde y^{(|\bar v|_1)}$ if we let $ y^{(|\bar u|_1)} $ fixed.  We understand the integral with respect to $\tilde y^{(|\bar v|_1)}$ in Eq.~(\ref{E.L.eirh.1.1}) as an application of the distribution (\ref{no-sense-1dos}).  

 We integrate out the $\tilde y^{(|\bar v|_1)}$ variable in Eq.~(\ref{E.L.eirh.1.1}) and apply (\ref{fm.2.1.1}) and (\ref{E.L.cprm.6.1}) to bound the left hand side of \eqref{E.L.eirh.1.1} by [see also \eqref{rho}, which implies that
 $4 \pi \rho < 1$] 
\eq{ \label{P.L.eirh.1.m1} 
 |\bar u|_1!  \Big(  &\frac{(e^{-\alpha}\rho)^{-1}}{(1 - \epsilon_F) - \frac{1}{2}e^{(1-\iota) \alpha}}  \Big)^{L-2}\frac{((1 - \kappa_\varsigma)e^{-\alpha}\rho)^{-1}}{(1 - \epsilon_F) - \frac{1}{2(1 - \kappa_\varsigma)}e^{(1-\iota) \alpha}}
\notag    
 \\ \notag
& \cdot  \prod_{\ell = 1}^J 
 \big( \rho\big)^{(p_\ell +  q_\ell - v_\ell)/2} 
  \| w_{ m_\ell + p_\ell, n_\ell + q_\ell} \|^{(\infty)}
\\ 
& \cdot  \prod_{\ell = J + 1}^L 
 \big( \rho\big)^{(p_\ell +  q_\ell - u_\ell)/2}   
 \| w_{ m_\ell + p_\ell, n_\ell + q_\ell} \|^{(\infty)}
}
times the integral 
\eq{ \label{P.L.eirh.1.1} 
 \\ \notag &  \int_{ B_{e^{- \alpha}\rho}^{m+ n}}\frac{dk^{(m)}}{|k^{(m)}|^{2 }}   \frac{d\tilde k^{(n)}}{|\tilde k^{(n)}|^{2}}  \int_{\| y^{(|\bar u|_1)} \|_1 \leq \rho}   dy^{(|\bar u|_1)}
 |y^{(|\bar u|_1)}|^{-1 + \mu}    \\ \notag  & \cdot \frac{1}{r_\varsigma}\Big | \overline{\chi_\alpha} \chi_0\big((b+1)r_\varsigma+ s + \tilde r_{J}(k^{(m,n)}\times  y^{(\bar u)}) \big) \\ \notag & \hspace{2cm} - 
 \overline{\chi_\alpha} \chi_0\big( br_\varsigma + s +   \tilde r_{J}(k^{(m,n)} \times  y^{(\bar u)}) \big) \Big |\comma
}
where we used that we can take  $\| y^{(|\bar u|_1)}\|_1 \leq \rho$, otherwise the factors $\chi_0$ in the third and the forth lines of equation (\ref{E.L.eirh.1.1}) would set everything to zero. To obtain \eqref{P.L.eirh.1.m1} we used \eqref{fm.2.1.1}
and that we can substitute
\eq{ & \frac{1}{w_{0,0}( br_\varsigma + s  + \tilde r_{J}( k^{(m,n)} \times  y^{(\bar u)}))}}
by
\eq{
\frac{\chi_0 \Big( br_\varsigma + s + \tilde r_{J}(k^{(m,n)} \times  y^{(\bar u)}) \Big)\overline{\chi_\alpha} \Big( (b+1)r_\varsigma + s + \tilde r_{J}( k^{(m,n)} \times  y^{(\bar u)}) \Big)  }{w_{0,0}( br_\varsigma + s  + \tilde r_{J}( k^{(m,n)} \times  y^{(\bar u)}))}
}
in the second line of \eqref{E.L.eirh.1.1}.

We now estimate the integral in (\ref{P.L.eirh.1.1}). Suppose first that $\bar u = \bar 0$, then $m + n \ne 0$ (see Definition \ref{D.cprm.3}). The integral can be bounded by
\eq{ \label{P.L.eirh.1.2} 
 (4 \pi)^{n + m}\int & d|k^{(m)}|  d|\tilde k^{(n)}| 
   \frac{1}{r_\varsigma}  \Big | \overline{\chi_\alpha} \chi_{0}\big((b+1)r_\varsigma + s + \tilde r_{J}( k^{(m,n)} )\big)  \\ \notag & - \overline{\chi_\alpha} \chi_{0}\big( br_\varsigma + s + \tilde r_{J}( k^{(m,n)}) \big) \Big | \leq 2(4 \pi)^{n + m}(e^{-\alpha}\rho)^{m + n - 1}.
} 
Now we suppose that $\bar u \ne \bar 0$, then the integral in (\ref{P.L.eirh.1.1}) can be bounded by 
\eq{\label{a}
2(4\pi & e^{-\alpha}\rho)^{n + m} (4 \pi)^{|\bar u|_1}   
\int_{ 0\leq r_{1} + \cdots + r_{|\bar u|_1 -1} \leq \rho} r_{1}^{1+\mu}\cdots r_{|\bar u|_1 -1}^{1+\mu} 
\\ & \notag = 2(4\pi e^{-\alpha}\rho)^{n + m} (4 \pi)^{|\bar u|_1} \rho^{(2 + \mu)(|\bar u|_1 - 1)} \frac{\Gamma(2 + \mu)^{|\bar u|_1-1}}{\Gamma((2 + \mu)(|\bar u|_1 -1) + 1)}
\\ \notag & \leq  2(4\pi e^{-\alpha}\rho)^{n + m} (4 \pi)^{|\bar u|_1} \rho^{(2 + \mu)(|\bar u|_1 - 1)}
\frac{\Gamma(2 )^{|\bar u|_1-1}}{\Gamma(2 (|\bar u|_1 -1) + 1)}
\\ \notag & \leq  2(4\pi e^{-\alpha}\rho)^{n + m} (4 \pi)^{|\bar u|_1} \rho^{(2 + \mu)(|\bar u|_1 - 1)} \frac{2}{(|\bar u|_1!)^2}
}
where we used Lemma C.2 in \cite{BFS98a} to compute the integral.\\
Finally Eq.~\eqref{E.L.eirh.1.1} follows from \eqref{P.L.eirh.1.m1}, \eqref{P.L.eirh.1.1},  and 
\eqref{P.L.eirh.1.2}- \eqref{a}.



\QED

\begin{lemma} \label{L.eirh.3}
Suppose that $L \geq 2$, $m +n \geq 0$ and that $\upsilon \in \cB_L(m, n)$. Suppose furthermore that $\uw \in \widetilde \cW_\xi$ and that $z \in e^{-\iota \alpha}D_{\rho/2}$. 
 Let $s \in [0, e^{-\alpha}\rho)$, $r \in (0, e^{-\alpha}\rho - s]$,
and $ t,  \tilde t \geq 0$. The following estimate holds true
\eq{ \label{E.L.eirh.3.1} \notag
\int & dy^{(|\bar u|_1)} d\tilde y^{(|\bar v|_1)} \int_{\| k^{(m)} \|_1, \| \tilde k^{(n)} \|_1 \leq e^{-\alpha}\rho}  \frac{dk^{(m)}}{|k^{(m)}|^{3/2 + \mu/2}}  \frac{d\tilde k^{(n)}}{|\tilde k^{(n)}|^{3/2 + \mu/2}}  
 \\ \notag & \Big | V_{\upsilon_{\leq J}, \bar v}(z;t;k^{(m)}; \tilde k^{(n)}, \tilde y^{(|\bar v|_1)} )  
\frac{1}{r}\Big[  \frac{\overline{\chi_\alpha} \chi_0\Big(s + r + \tilde r_{J}(k^{(m,n)} \times  y^{(\bar u)}) \Big)}{w_{0,0}(s + r + \tilde r_{J}( k^{(m,n)} \times  y^{(\bar u)}))}\\ \notag & \hspace{5cm} -  \frac{\overline{\chi_\alpha} \chi_0\Big(s + \tilde r_{J}(k^{(m,n)} \times  y^{(\bar u)}) \Big)}{w_{0,0}(s + \tilde r_{J}( k^{(m,n)} \times  y^{(\bar u)}))}   \Big]
\\ \notag & \cdot
\Big \la \prod_{j= 1}^{J}a(\tilde y_j^{(v_j)}) \prod_{j= J + 1}^{L}a^*( y_j^{(u_j)}) \Big \ra_\Om \\ & 
\hspace{3cm} \cdot V_{\upsilon_{\geq(J+1)}, \bar u}(z; \tilde t;k^{(m)}, y^{(|\bar u|_1)};\tilde k^{(n)})\Big | \notag \\ \notag
 \leq &
 \frac{32  \pi}{u!}( 4 \pi \rho )^{n + m + |\bar u|_1 - 1} e^{-(m + n)\alpha} 
\rho^{|\bar u|_1}  
\\ \notag &  \Big( \frac{(e^{-\alpha}\rho)^{-1}}{(1 - \epsilon_F) - \frac{1}{2}e^{(1-\iota) \alpha}}  \Big)^{L}   \Big( \| \uw  - \ur \|^{(F)} + 1\Big) 
\\ \notag
& \cdot  \prod_{\ell = 1}^J  
\big(  \rho \big)^{(p_\ell +  q_\ell - v_\ell)/2}  \| w_{ m_\ell + p_\ell, n_\ell + q_\ell} \|^{(\infty)}
\\ 
& \cdot  \prod_{\ell = J + 1}^L 
\big(  \rho \big)^{(p_\ell +  q_\ell - u_\ell)/2}   \| w_{ m_\ell + p_\ell, n_\ell + q_\ell} \|^{(\infty)} \period
}

\end{lemma}
\emph{Proof:}
Define as in Lemma \ref{L.eirh.1} 
$$
r_\varsigma  = e^{-\alpha}\rho \kappa_\varsigma, \comma 
$$
and
$$
\kappa_\varsigma : = \frac{r}{e^{-\alpha}\rho \varsigma}\comma
$$
for big enough $\varsigma \in \NN$, such that $ \kappa_\varsigma  \in \big[0 , 1- \frac{e^{(1-\iota)\alpha}}{2(1 - \epsilon_F)}\big)  $, see \eqref{put0}.

We adopt some notation for the proof of this Lemma. We denote by $F_1(s,r)$ the left hand side of Eq.~(\ref{E.L.eirh.3.1}), by $F_2(br_\varsigma + s, r_\varsigma)$ and $R_1(\kappa_\varsigma)$ the left and the right hand side of
 Eq.~(\ref{E.L.eirh.1.1}), respectively. We furthermore denote by $F_3(s,r)$ the following  
\eq{ \label{E.L.eirh.3.2} \notag
F_3(s,r) := \int & dy^{(|\bar u|_1)} d\tilde y^{(|\bar v|_1)}  \int_{\|\tilde k^{(n)}  \|_1, \| k^{(m)} \|\leq e^{-\alpha }\rho} \frac{dk^{(m)}}{|k^{(m)}|^{3/2 + \mu/2}}  \frac{d\tilde k^{(n)}}{|\tilde k^{(n)}|^{3/2 + \mu/2}}  
  \\ \notag & \Big | V_{\upsilon_{\leq J}, \bar v}(z;t;k^{(m)}; \tilde k^{(n)}, \tilde y^{(|\bar v|_1)} )  
 \overline{\chi_\alpha} \chi_0\Big(r + s + \tilde r_{J}(k^{(m,n)} \times  y^{(\bar u)}) \Big) \\ \notag &
\frac{1}{r}\Big[  \frac{1}{w_{0,0}(r + s + \tilde r_{J}( k^{(m,n)} \times  y^{(\bar u)}))}\\ \notag & \hspace{5cm} -  \frac{1}{w_{0,0}(s + \tilde r_{J}( k^{(m,n)} \times  y^{(\bar u)}))}   \Big]
\\ \notag & \cdot
\Big \la \prod_{j= 1}^{J}a(\tilde y_j^{(v_j)}) \prod_{j= J + 1}^{L}a^*( y_j^{(u_j)}) \Big \ra_\Om \\ & 
\hspace{3cm} \cdot V_{\upsilon_{\geq (J+1)}, \bar u}(z; \tilde t;k^{(m)}, y^{(|\bar u|_1)};\tilde k^{(n)})\Big |  \period
}
Then we have that 
\eq{ \label{ola.1}
F_1(s,r) \leq & \frac{r_\varsigma}{r} \sum_{ b = 0}^{\varsigma-1}  F_1(br_\varsigma + s, r_\varsigma)  \\ \notag   \leq & \frac{1}{\varsigma}\sum_{ b = 0}^{\varsigma-1}  F_2(br_\varsigma + s, r_\varsigma)  
+
\frac{1}{\varsigma} \sum_{ b = 0}^{\varsigma-1}  F_3(br_\varsigma + s, r_\varsigma)   \period
}
Now we do the following estimation using (\ref{i.4}), \eqref{i.11} and (\ref{fm.2.1.1}) 
\eq{ \label{ola.2} & \Big |\overline{\chi_\alpha}\chi_0 \Big((b +1)r_\varsigma + s + \tilde r_{J}(k^{(m,n)} \times  y^{(\bar u)}) \Big) \\ \notag & 
\hspace{3cm}  \cdot  \frac{1}{r_\varsigma}\Big[  \frac{1}{w_{0,0}((b +1)r_\varsigma + s + \tilde r_{J}( k^{(m,n)} \times  y^{(\bar u)}))} \\ \notag & 
\hspace{5cm} -  \frac{1}{w_{0,0}(br_\varsigma + s + \tilde r_{J}( k^{(m,n)} \times  y^{(\bar u)}))}  \Big] \Big |\\  \notag
\leq \Big | &
\frac{\overline{\chi_\alpha} \chi_0 \Big((b +1)r_\varsigma + s + \tilde r_{J}(k^{(m,n)} \times  y^{(\bar u)}) \Big)}{w_{0,0}(br_\varsigma + s + \tilde r_{J}( k^{(m,n)} \times  y^{(\bar u)}))} \Big |\\ \notag &  \cdot\Big |
\frac{\overline{\chi_\alpha}\chi_0 \Big( (b +1)r_\varsigma + s + \tilde r_{J}(k^{(m,n)} \times  y^{(\bar u)}) \Big)}{w_{0,0}(
 (b +1)r_\varsigma + s + \tilde r_{J}( k^{(m,n)} \times  y^{(\bar u)}))}
\Big | \\ & \notag   \cdot
\frac{1}{r_\varsigma}\Big | \Big[ w_{0,0}((b +1)r_\varsigma + s + \tilde r_{J}( k^{(m,n)} \times  y^{(\bar u)})) \\ \notag
& \hspace{5cm} -  w_{0,0}(b r_\varsigma +s + \tilde r_{J}( k^{(m,n)} \times  y^{(\bar u)}))  \Big] \Big | \\ & \notag \leq
\Big(\frac{(\rho e^{-\alpha})^{-1}}{(1 - \epsilon_F) - \frac{1}{2}e^{(1 - \iota)\alpha}}\Big)
\Big(\frac{((1 - \kappa_\varsigma)\rho e^{-\alpha})^{-1}}{(1 - \epsilon_F) - \frac{1}{2(1 - \kappa_\varsigma)}e^{(1 - \iota)\alpha}}\Big)
 \Big( \| \uw - \ur\|^{(F)} + 1\Big) \period
}
Following the proof of Lemma \ref{L.eirh.1} and using (\ref{ola.2}) we obtain that
\eq{ \label{ola.3}
F_3(s,r) \leq \Big( \| \uw - \ur\|^{(F)} + 1\Big)  \Big(  \frac{ 1 }{(1 - \epsilon_F) - \frac{1}{2}e^{(1-\iota) \alpha}}  \Big)R_1(\kappa_\varsigma)\period
}
Taking the limit $\varsigma \to \infty$ in \eqref{ola.1} we obtain, using Lemma \ref{L.eirh.1} and \eqref{ola.3}, 
$$
F_1(s,r)\leq 2 \Big( \| \uw - \ur\|^{(F)} + 1\Big)  \Big(  \frac{ 1 }{(1 - \epsilon_F) - \frac{1}{2}e^{(1-\iota) \alpha}}  \Big)R_1(0)\comma
$$
which is (\ref{E.L.eirh.3.1}).
\QED

\begin{lemma} \label{L.eirh.2}
Suppose that $L \geq 2$, $m +n \geq 0$ and that $\upsilon \in \cB_L(m, n)$. Suppose furthermore that $\uw \in \widetilde \cW_\xi$ and that $z \in e^{-\iota \alpha}D_{\rho/2}$. 
 Let $s \in [0, e^{-\alpha}\rho)$ and $r \in (0, e^{-\alpha}\rho - s]$. The following estimate holds true
\eq{  \label{E.L.eirh.2.1}
 \int & dy^{(|\bar u|_1)}  d\tilde y^{(|\bar v|_1)} \int_{\| k^{(n)}\|, \| \tilde k^{(n)} \|\leq e^{-\alpha}\rho}   \frac{dk^{(m)}}{|k^{(m)}|^{3/2 + \mu/2}}  \frac{d\tilde k^{(n)}}{|\tilde k^{(n)}|^{3/2 + \mu/2}} 
\\ \notag & \Big | V_{\upsilon_{\leq (J-1)}, \bar v}(z; r + s; k^{(m)}; \tilde k^{(n)}, \tilde y^{(|\bar v|_1)} )
\\ \notag &
\cdot F_{J-1}(r+s +  \tilde r_{J-1}(k^{(m,n)} \tilde \times \tilde y^{(\bar v)}) )F_{J}(r + \tilde r_{J}(k^{(m,n)} \times  y^{(\bar u)}))
\\ \notag &
\cdot \frac{1}{r}\Big[ \Big \la \prod_{j= 1}^{J-1}a(\tilde y_j^{(v_j)}) \widetilde W_J(z;(r + s)   +  r_J(k^{(m,n)});k^{(m_J, n_J)} ) 
\\ \notag &
\hspace{.7cm}- \widetilde W_J(z;s  + r_J(k^{(m,n)});k^{(m_J, n_J)} ) \prod_{j= J+ 1}^{L}a^*( y_j^{(u_j)}) \Big \ra_\Om
\Big]
\\ \notag &
\hspace{5cm} \cdot V_{\upsilon_{\geq (J+1)}, \bar u}(z; s;k^{(m)}, y^{(|\bar u|_1)}; \tilde k^{(n)})\Big |.
\\ \notag 
\leq &  \Big(  \frac{(e^{-\alpha}\rho)^{-1}}{(1 - \epsilon_F) - \frac{1}{2}e^{(1-\iota) \alpha}}  \Big)^{L-1}
(4\pi e^{-\alpha}\rho)^{m + n - m_J - n_J }  
\\ \notag &
  \frac{|\bar{u}|_1!}{(|\bar{u}|_1 - q_J)!}\frac{|\bar{v}|_1!}{(|\bar{v}|_1 - p_J)!} 
\frac{\rho^{(1+ \mu)p_J}}{p_J^{(1 + \mu)p_J} } \frac{\rho^{(1+ \mu)q_J}}{q_J^{(1 + \mu)q_J} }
\\ \notag &  \Big[ \prod_{\ell \in \NN_L\setminus \{ J \} }   \Big( \rho\Big)^{(p_l + q_l)/2}  \cdot \| w_{m_\ell + p_\ell, n_\ell + q_\ell} \|^{(\infty)}   
\Big] 
 \\ \notag & \cdot  \frac{(4\pi)^{(|\bar v|_1 - p_J)} \rho^{(2+ \mu)(|\bar v|_1 - p_J)} }{(|\bar v|_1 - p_J)!} \cdot  \|  w_{m_J + p_J, n_J + p_J}  \|^{(0)}\period
}
\end{lemma}
\emph{Proof:}  
First we notice that if $m + n = 0$, we can take   $|\bar u|_1, |\bar v|_1 \geq 1$, because if one of them is zero then either 
\eq{ F_{J-1}(s + r +  \tilde r_{J-1}(k^{(m,n)} \tilde \times \tilde y^{(\bar v)}) )= 0 
}
or 
\eq{F_{J}(r + \tilde r_{J}(k^{(m,n)} \times  y^{(\bar u)})) = 0\period
}

For every natural number $N$, we define 
\beq \label{nn}
\NN_N : = \{ 1, \cdots N \},
\ene 
and 
$$
S_{M,N}
$$
the set of injective functions from $ \NN_M \to \NN_N $. For any two sets $A$, $B$, we denote by 
$S(A, B)$ the set of injective functions from $A $ to $B$.

We use the notations (\ref{no-sense-0}) for $y^{(|\bar {u}|_1)}$ and for $\tilde y^{(|\bar v|_1)}$ (with $J - 1$ instead of $J$). We furthermore write 
\beq \label{nn.1}
x^{(p_J)} =  (x^1, \cdots, x^{p_J}), \hspace{3cm} \tilde x^{(q_J)} =  (\tilde x^1, \cdots, \tilde x^{q_J}).
\ene

We recall Eq.~(\ref{Wmnpq}) and Definition \ref{D.cprm.4} to remark that the vacuum expectation value appearing in 
(\ref{E.L.eirh.2.1}) gives rise to a term of the form  
\eq{ \label{yo-mero} \notag
& \Big \la \Big(\prod_{j= 1}^{J-1}a(\tilde y_j^{(v_j)})\Big)  a^*(x^{(p_J)}) \Big[ w_{m_J + p_J, n_J + q_J}\Big(z; (r + s) + H_f  +  r_J( k^{(m,n)})\\ \notag & \hspace{8cm}; k^{(m_J)}, x^{(p_J)}; \tilde k^{(n_J)}, \tilde x^{(q_J)}\Big) \\ \notag
& \: \: \: -  w_{m_J + p_J, n_J + q_J}\Big(z;s + H_f  +  r_J(k^{(m,n)});k^{(m_J)}, x^{(p_J)} 
;\tilde k^{(n_J)}, \tilde x^{(q_J)}\Big)\Big]
\\  & \: \: \: \cdot a(\tilde x^{(q_J)}) \Big(\prod_{j= J+ 1}^{L}a^*( y_j^{(u_j)})\Big)
\Big \ra_\Om \period
}

Eq.~(\ref{yo-mero}) is not well defined as a quadratic form, because it contains creation operators to the right of annihilation operators. To define it properly we first notice that it is different from zero only if 
$|\bar v|_1 \geq p_J$, $ |\bar u|_1 \geq q_J $ and $|\bar v|_1 - p_J = |\bar u|_1 - q_J$.  A formal computation using the canonical commutation relations (\ref{h.3}) and the push forward formulas
\beq \label{push-forward}
H_f a^*(k) = a^*(k) (H_f + |k| ), \hspace{2cm} (H_f + |k|) a(k) = a(k) H_f
\ene 
permit us to write (\ref{yo-mero}) in the following way (that we take as a definition)
\eq{ \label{yo-mero.1} \notag
& \sum_{\pi_1 \in S_{p_J, |\bar v|_1}} 
  \sum_{\pi_2 \in S_{q_J, |\bar u|_1}} \sum_{\pi_3 \in 
 S(\NN_{|\bar v|_1}\setminus \pi_1(\NN_{p_J}), \NN_{|\bar u|_1}\setminus \pi_2(\NN_{q_J}))} 
\Big[ \prod_{j_1 = 1}^{p_J}\prod_{j_2 = 1}^{q_J}\prod_{j_3 \in \NN_{|\bar v|_1}\setminus \pi_1(\NN_{p_J})}  \notag
 \\ \notag & \: \: \: \: \: \: \delta( \tilde y^{\pi_1(j_1)} - x^{j_1}) 
 \delta(\tilde x^{j_2} -  y^{\pi_2(j_2)})\delta(\tilde y^{j_3} -  y^{\pi_3(j_3)}) \Big]
 \\ & \: \: \: \cdot  \notag \Big[ w_{m_J + p_J, n_J + q_J}\Big(z;(r + s) + \sum_{j_4 \in \NN_{|\bar v|_1}\setminus \pi_1(\NN_{p_J})}
 |\tilde y^{j_4}|  +  r_J(k^{(m,n)})\\ \notag & \hspace{7cm}; k^{(m_J)}, x^{(p_J)};\tilde k^{(n_J)}, \tilde x^{(q_J)}\Big) \\ \notag
& \: \: \: -  w_{m_J + p_J, n_J + q_J}\Big(z; s + \sum_{j_4 \in \NN_{|\bar v|_1}\setminus \pi_1(\NN_{p_J})}
 |\tilde y^{j_4}| +  r_J(k^{(m,n)})
 \\  &   \hspace{7cm}; k^{(m_J)}, x^{(p_J)} 
;\tilde k^{(n_J)}, \tilde x^{(q_J)}\Big)\Big]\period
}
The product of terms of the form $ \delta( \tilde y^{\pi_1(j_1)} - x^{j_1})  $ together with the operator $F_{J-1}$ in the third line of equation \eqref{E.L.eirh.2.1} imply that 
\eq{\label{fix.1}
\| k^{(m_J)} \|_1 + \| x^{(p_J)} \|_1 \leq \rho
}
and similarly we obtain 
\eq{\label{fix.2}
\| \tilde k^{(n_J)} \|_1 + \| \tilde x^{(q_J)} \|_1 \leq \rho \comma
}
and 
\eq{\label{fix.3}
(r + s) + \|\tilde y^{\bar v} \|_1 +  r_J(k^{(m,n)}) \leq \rho\period
}

Eq.~(\ref{yo-mero.1}) defines a distribution. The integral with respect to the variables $\tilde y^{(|\bar v|_1)}$ and $y^{(|\bar u|_1)}$ denotes the application of this distribution to the integrand, which is well defined even though the integrand is not a test function.  

Using Lemma \ref{L.cprm.6} and (\ref{nn})-(\ref{fix.2}) we bound the l.h.s of Eq.~(\ref{E.L.eirh.2.1}) by  
\eq{ \notag \label{yo-mero.2} & \Big(  \frac{(e^{-\alpha}\rho)^{-1}}{(1 - \epsilon_F) - \frac{1}{2}e^{(1-\iota) \alpha}}  \Big)^{L-1}  
\frac{|\bar{u}|_1!}{(|\bar{u}|_1 - q_J)!}\frac{|\bar{v}|_1!}{(|\bar{v}|_1 - p_J)!}(|\bar v|_1 - p_J)!
 \\ \cdot \notag &
\Big[ \prod_{\ell \in \NN_L\setminus \{ J \} }   \Big( \rho \Big)^{(p_l + q_l)/2}  \| w_{m_\ell + p_\ell, n_\ell + q_\ell} \|^{(\infty)}   
 \int_{B_{e^{-\alpha}\rho}^{m_\ell + n_\ell}}\frac{dk^{(m_\ell)}}{|k^{(m_\ell)}|^{2}}  \frac{d\tilde k^{(n_\ell)}}{|\tilde k^{(n_\ell)}|^{2}}  
 \Big] \\ \notag & \cdot \hspace{.1cm} \int_{\|\tilde y^{(|\bar v|_1 - p_J)}\|_1 \leq \rho} \frac{d 
 \tilde y^{(|\bar v|_1 - p_J)}}{|  \tilde y^{(|\bar v|_1 - p_J)} |^{1 - \mu}} \\ & \notag
\int_{ \|x^{(p_J)}\|_1 + \| k^{(m_J)}\|_1 , \|\tilde x^{(q_J)}\|_1 + \|\tilde k^{(n_J)}\|_1
 \leq \rho }\frac{dk^{(m_J)}}{|k^{(m_J)}|^{3/2 + \mu/2}}  \frac{d\tilde k^{(n_J)}}{|\tilde k^{(n_J)}|^{3/2 + \mu/2}}
 \\ \notag &  \cdot  \frac{dx^{(p_J)}}{|x^{(p_J)}|^{1/2 - \mu/2}} \frac{d\tilde x^{(q_J)}}{|x^{(q_J)}|^{1/2 - \mu/2}}  
 \\ & \hspace{1.5cm} \cdot  \notag \frac{1}{r}\Big| \Big[ w_{m_J + p_J, n_J + q_J}\Big(z;(r + s) +
  \|\tilde y^{(|\bar v|_1 - p_J)} \|_1 +  r_J(k^{(m,n)})\\ \notag & \hspace{7cm}; k^{(m_J)}, x^{(p_J)};\tilde k^{(n_J)}, \tilde x^{(q_J)}\Big) \\ \notag
& \hspace{1cm} -  w_{m_J + p_J, n_J + q_J}\Big(z;s + 
 \|\tilde y^{(|\bar v|_1 - p_J)} \|_1 +  r_J(k^{(m,n)})
 \\  &   \hspace{7cm}; k^{(m_J)}, x^{(p_J)} 
; \tilde k^{(n_J)}, \tilde x^{(q_J)}\Big)\Big]\Big| \period
}

Lemma C.3 in \cite{BFS98a} implies that given $g : \RR^{p_J + q_J} \to \CC$ and $t \geq 0$
\eq{ \label{he}
& \int_{ \|x^{(p_J)}\|_1, \|\tilde x^{(q_J)}\|_1 \leq \rho - t }\frac{dx^{(p_J)}}{|x^{(p_J)}|^{1/2 - \mu/2}} \frac{d\tilde x^{(q_J)}}{|x^{(q_J)}|^{1/2 - \mu/2}} g(x^{(p_J)}, \tilde x^{q_J}) \\ \notag &
\leq  \frac{\rho^{(1 + \mu)p_J}}{p_J^{(1 + \mu)p_J} } \frac{\rho^{(1 + \mu)q_J}}{q_J^{(1 + \mu)q_J} }\int_{ \|x^{(p_J)}\|_1, \|\tilde x^{(q_J)}\|_1 \leq \rho - t }\frac{dx^{(p_J)}}{|x^{(p_J)}|^{3/2 + \mu/2}} \frac{d\tilde x^{(q_J)}}{|x^{(q_J)}|^{3/2 + \mu/2}} g(x^{(p_J)}, \tilde x^{q_J})\period
} 
Using Eqs.~(\ref{i.6}), \eqref{fix.3} and (\ref{he}), we bound (\ref{yo-mero.2}) by   
\eq{ \label{yo-mero.3} & \Big(  \frac{(e^{-\alpha}\rho)^{-1}}{(1 - \epsilon_F) - \frac{1}{2}e^{(1-\iota) \alpha}}  \Big)^{L-1}
(4\pi e^{-\alpha}\rho)^{m + n - m_J - n_J }  
\\ \notag &
  \frac{|\bar{u}|_1!}{(|\bar{u}|_1 - q_J)!}\frac{|\bar{v}|_1!}{(|\bar{v}|_1 - p_J)!}(|\bar v|_1 - p_J)!  
\frac{\rho^{(1+\mu)p_J}}{p_J^{(1 + \mu)p_J} } \frac{\rho^{(1+\mu)q_J}}{q_J^{(1 + \mu)q_J} }
\\ \notag &  \Big[ \prod_{\ell \in \NN_L\setminus \{ J \} }   \Big(\rho \Big)^{(p_l + q_l)/2}  \cdot \| w_{m_\ell + p_\ell, n_\ell + q_\ell} \|^{(\infty)}    
\Big] 
 \\ \notag & \cdot \hspace{.1cm} \int_{\|\tilde y^{(|\bar v|_1 - p_J)}\|_1 \leq \rho} \frac{d \tilde y^{(|\bar v|_1 - p_J)}}{|  \tilde y^{(|\bar v|_1 - p_J)} |^{1 - \mu}} \|  w_{m_J + p_J, n_J + p_J}  \|^{(0)}\period
} 
 
We conclude (\ref{E.L.eirh.2.1}) using 
\eq{\label{mm}
\int_{\|\tilde y^{(|\bar v|_1 - p_J)}\|_1 \leq \rho} \frac{d \tilde y^{(|\bar v|_1 - p_J)}}{|  \tilde y^{(|\bar v|_1 - p_J)} |^{1 - \mu}} \leq 
\frac{(4\pi)^{(|\bar v|_1 - p_J)} \rho^{(2+ \mu)(|\bar v|_1 - p_J)} }{(|\bar v|_1 - p_J)!^2}\comma
}
which is proved as in (\ref{a}).
\QED

\begin{lemma} \label{vn1}
Suppose that $L \geq 2$, $m +n \geq 0$ and that $\upsilon \in \cB_L(m, n)$. Suppose furthermore that $\uw \in \widetilde \cW_\xi$ and that $z \in e^{-\iota \alpha}D_{\rho/2}$. 
 Let $s \in [0, e^{-\alpha}\rho)$ and $r \in (0, e^{-\alpha}\rho - s]$. 
 The following estimate holds true
\eq{\label{vn1.1} \notag
\int & _{  \| k^{(m)} \|_1, \| \tilde k^{(n)} \|_1 \leq e^{-\alpha}\rho}      \frac{dk^{(m)}}{|k^{(m)}|^{3/2 + \mu/2}}  \frac{d\tilde k^{(n)}}{|\tilde k^{(n)}|^{3/2 + \mu/2}} \\ \notag & \hspace{3cm} 
\cdot \frac{1}{r} \Big | V_{\upsilon}(z;s+r;k^{(m)}; \tilde k^{(n)} )  - V_{\upsilon}(z;s;k^{(m)}; \tilde k^{(n)} )  \Big |\\ \notag
  \leq &
64  \pi L \rho^{(|\bar q|_1 + |\bar q|_1)/2}  3^{|\bar p|_1 + |\bar q|_1} \Big(\frac{(e^{-\alpha}\rho)^{-1}}{(1 - \epsilon_F) - \frac{1}{2} e^{(1-\iota)\alpha}}\Big)^{L} \\ & \cdot \Big[ \prod_{\ell = 1}^{L} \| w_{m_\ell + p_\ell, n_\ell + q_\ell} \|^{(\infty)} +   \| w_{m_\ell+ p_\ell, n_\ell + q_\ell} \|^{(0)}  \Big] \period
}
\end{lemma}

\noindent \emph{Proof:} 
We use the Leibniz formula and (\ref{vupsilon}) to compute 
\eq{ \label{lf.1}
\Big| & \frac{1}{r}  V_\upsilon (z; r + s; k^{ (m, n)}) -  V_{\upsilon} (z;s; k^{(m, n)})\Big| \\ \notag & \leq 
\Big |\Big \la \Om   \Big |  \frac{1}{r} \Big[\widetilde W_1(z; r + s + r_1 ( k^{(\bar m, \bar n)} ) ; k^{(m_1, n_1)}) \\ \notag & \hspace{5cm}- \widetilde W_1(z; s + r_1 ( k^{(\bar m, \bar n)} ) ; k^{(m_1, n_1)})\Big] 
\\ \notag & \hspace{.5cm} \cdot F_1 (H_f + r + s + \tilde r_1( k^{(\bar m, \bar n)} )) \cdots \widetilde W_L(z; r + s + r_L ( k^{(\bar m, \bar n)} ) ; k^{(m_L, n_L)})  \Om  \Big  \ra \Big| 
\\ & \notag \hspace{.5cm}+
\Big |\Big \la \Om \Big |  \widetilde W_1(z; s + r_1 ( k^{(\bar m, \bar n)} ) ; k^{(m_1, n_1)})
\\ \notag & \hspace{1.5cm} \cdot\frac{1}{r} \Big[ F_1 (H_f + r + s + \tilde r_1( k^{(\bar m, \bar n)} )) - F_1 (H_f + s + \tilde r_1( k^{(\bar m, \bar n)} ))  \Big] \\ \notag & \hspace{5cm} \cdots \widetilde W_L(z; r + s + r_L ( k^{(\bar m, \bar n)} ) ; k^{(m_L, n_L)})  \Om  \Big  \ra\Big| \\ \notag & 
\hspace{5cm} + \cdots  \\ & \notag 
 \hspace{.5cm} +\Big |\Big \la \Om \Big | \Big[ \widetilde W_1(z; s + r_1 ( k^{(\bar m, \bar n)} ) ; k^{(m_1, n_1)}) \\ \notag & \hspace{1.5cm}
\cdots F_{L-1}(H_f + s + \tilde r_{L-1}( k^{(\bar m, \bar n)} ))
\\ \notag & \hspace{1.5cm} \cdot  \frac{1}{r} \Big[\widetilde W_L(z; r + s + r_L ( k^{(\bar m, \bar n)} ) ; k^{(m_L, n_L)}) \\ \notag & \hspace{5cm}- \widetilde W_L(z; s + r_L ( k^{(\bar m, \bar n)} ) ; k^{(m_L, n_L)})\Big]   \Om  \Big  \ra\Big| 
}
Next we apply Lemma \ref{recursive-relation} several times. We apply (\ref{E.worr.1}) with 
\eq{ \label{mmmta.1}& \frac{1}{r}\Big[\widetilde W_J(z; r + s + r_J ( k^{(\bar m, \bar n)} ) ; k^{(m_J, n_J)}) \\ \notag & \hspace{5cm}- \widetilde W_J(z; s + r_J ( k^{(\bar m, \bar n)} ) ; k^{(m_J, n_J)})\Big] 
}
instead of $ \widetilde W_J(z; r + r_J ( k^{(\bar m, \bar n)} ) ; k^{(m_J, n_J)}) $, and we apply (\ref{E.worr.2}) with 
\eq{\label{mmmta.2} \frac{1}{r}\Big[ F_J (H_f + r + s + \tilde r_J( k^{(\bar m, \bar n)} )) - F_J (H_f + s + \tilde r_J( k^{(\bar m, \bar n)} ))  \Big]
}
instead of $ F_J (H_f + r + \tilde r_J( k^{(\bar m, \bar n)} )) $.\\
We estimate the terms of the form (\ref{mmmta.1}) using Lemma \ref{L.eirh.2} and the terms of the form (\ref{mmmta.2}) using Lemma \ref{L.eirh.3}. 

We estimate the right hand side of Eq.~(\ref{E.L.eirh.3.1}) in Lemma \ref{L.eirh.3}
using that $\| \uw - \ur \|^{(F)}\leq 1$ and (\ref{rho}), which implies that $4 \pi \rho < 1$, by 
[see \eqref{put0} and \eqref{rmn.5}]     
\eq{ \label{mmmta.3}
& 64 \pi \rho^{(|\bar p|_1 + |\bar q|_1)/2} 
\Big(\frac{(\rho e^{-\alpha})^{-1}}{(1 - \epsilon_F) - \frac{1}{2} e^{(1-\iota)\alpha}}\Big)^{L}\Big[ \prod_{\ell = 1}^{L}\| w_{m_\ell + p_\ell, n_\ell + q_\ell} \|^{(\infty)}\Big] 
}
(notice that  $\bar u \leq \bar p$, $\bar v \leq \bar q$ and $|\bar u|_1 = |\bar v|_1$).\\
In Eq.~(\ref{E.L.eirh.2.1}) in Lemma \ref{L.eirh.2} we bound $\frac{|\bar u|_1!}{(|\bar u|_1 - q_J)! q_J^{(1 + \mu)q_J}}$ by $2^{|\bar u|_1}$ and  
$\frac{|\bar v|_1!}{(|\bar v|_1 - p_J)! p_J^{(1 + \mu )p_J}}$ by $2^{|\bar v|_1}$. We use that $|\bar v |_1 - p_J = |\bar u|_1 - q_J$ (otherwise the left hand side of the equation is zero) and \eqref{rho} to bound the left hand side of (\ref{E.L.eirh.2.1}) by 
\eq{ \label{mmmta.4}
&\rho^{(|\bar p|_1 + |\bar q|_1)/2} 2^{|\bar v|_1 + |\bar u|_1} \Big(\frac{(e^{-\alpha}\rho)^{-1}}{(1 - \epsilon_F) - \frac{1}{2} e^{(1-\iota)\alpha}}\Big)^{L-1} \\ & \notag \hspace{3cm} \cdot \Big[ \prod_{\ell = 1, \ell \ne J}^{L}\| w_{m_\ell + p_\ell, n_\ell + q_\ell} \|^{(\infty)}\Big]  \| w_{m_J+ p_J, n_J + q_J} \|^{(0)}  \period
}
Next we bound the terms in (\ref{lf.1}) containing (\ref{mmmta.2}) using Lemmata \ref{recursive-relation} 
and \ref{L.eirh.3}, (\ref{mmmta.3}) and the fact that 
$\sum_{n=0}^N \binom{N}{n} = 2^{N}$ by 
\eq{ \label{mmmta.5}
& 64 \pi 2^{|\bar p|_1 + |\bar q|_1}  \rho^{(|\bar p|_1 + |\bar q|_1)/2} 
\Big(\frac{(\rho e^{-\alpha})^{-1}}{(1 - \epsilon_F) - \frac{1}{2} e^{(1-\iota)\alpha}}\Big)^{L}\Big[ \prod_{\ell = 1}^{L}\| w_{m_\ell + p_\ell, n_\ell + q_\ell} \|^{(\infty)}\Big] \period
} 
and we bound the terms  in (\ref{lf.1}) containing (\ref{mmmta.1}) using Lemmata \ref{recursive-relation} and  \ref{L.eirh.2}, (\ref{mmmta.4}) and the fact that 
$\sum_{n=0}^N \binom{N}{n}2^{n} = 3^{N}$ by 
\eq{ \label{mmmta.6} 
& \rho^{(|\bar p|_1 + |\bar q|_1)/2}   3^{|\bar p|_1 + |\bar q|_1}\Big(\frac{(e^{-\alpha}\rho)^{-1}}{(1 - \epsilon_F) - \frac{1}{2} e^{(1-\iota)\alpha}}\Big)^{L-1}  \\ \notag & \hspace{3cm} \cdot \Big[ \prod_{\ell = 1, \ell \ne J}^{L}\| w_{m_\ell + p_\ell, n_\ell + q_\ell} \|^{(\infty)}\Big]  \| w_{m_J+ p_J, n_J + q_J} \|^{(0)}  \period
}
We obtain finally (\ref{vn1.1}) from (\ref{lf.1}) and (\ref{mmmta.5})-(\ref{mmmta.6}).

\QED

\paragraph{Estimates for $\|  \widetilde w_{m,n}\|^{(\alpha)}$} $\\ $

\begin{theorem}\label{T.wmn1}
For $m + n \geq 0$ the following estimate holds true
\eq{\label{wmn1}
\|\widetilde w_{m, n}\|^{(\alpha)}\leq \epsilon_I^2 \xi^{m+n} 4^{m+n}\frac{(1- \xi)^2}{2}  A^{(0)}(\ue, \alpha) \comma}
where
\eq{\label{aalpha} A^{(0)}(\ue, \alpha) : = 128  \pi 
\frac{3 \Big[ \Big(3(e^{-\alpha}\rho)^{-1} \Big)\Big]^2  
 (2^4)^{2} }{\Big(1 -  
\Big[\epsilon_I \Big(3(e^{-\alpha}\rho)^{-1} \Big)\Big] 2^4 \Big)^2} \period
}
\end{theorem}
\emph{Proof:}
We use Lemmata \ref{L.cprm.5} and \ref{vn1} [see also \eqref{i.11.1}, \eqref{rmn.5} and 
\eqref{E.L.cprm.5.2}-\eqref{i.8}] to get the following 
\eq{ \label{wmn1.1}
\|\widetilde w_{m,n} \|^{(\alpha)} 
 \leq & 64 \pi\xi^{m+n}  (1 - \xi)^2 \sum_{L = 2}^{\infty}L \Big(\frac{\epsilon_I (e^{-\alpha}\rho)^{-1}}{(1 - \epsilon_F) - \frac{1}{2} e^{(1-\iota)\alpha}}\Big)^{L}  \\ & \notag \cdot \sum_{\upsilon \in \mathcal{B}(m,n)}\rho^{(|\bar q|_1 + |\bar q|_1)/2}  3^{|\bar p|_1 + |\bar q|_1} \Big[\prod_{\ell =1}^L\binom{m_\ell + p_\ell}{p_\ell}\binom{n_\ell + q_\ell}{q_\ell}\notag \cdot  \xi^{p_\ell + q_\ell}  \Big]\period 
}
We compute as in (\ref{wmninfty.2}) using \eqref{rho} (which implies that $6 \rho^{1/2} \leq 1/2$) and 
\eqref{hirmn.6} to obtain
\eq{\label{wmn1.2}
\sum_{\upsilon \in \mathcal{B}(m,n)} & \rho^{(|\bar q|_1 + |\bar q|_1)/2}  3^{|\bar p|_1 + |\bar q|_1} \Big[\prod_{\ell =1}^L\binom{m_\ell + p_\ell}{p_\ell}\binom{n_\ell + q_\ell}{q_\ell}\notag \cdot  \xi^{p_\ell + q_\ell}  \Big]
   \\  \leq &
  4^{m+n} \sum_{\upsilon \in \mathcal{B}(m,n)}  \rho^{(|\bar q|_1 + |\bar q|_1)/2}  6^{|\bar p|_1 + |\bar q|_1} 
   \xi^{|\bar p|_1+ |\bar q|_1}
   2^{-m-n}
   \\ \notag   \leq & 4^{m+n} \Big( \sum_{j=0}^{\infty}(6\rho^{1/2}\xi)^j)\Big)^{2L}
\Big( \sum_{j=0}^{\infty} 2^{-j}\Big)^{2L}   
   \leq  
 4^{m+n} (2^4)^L\period
 } 
As in (\ref{wmninfty.3}) we conclude using (\ref{wmn1.1})-(\ref{wmn1.2}) that
\eq{\label{wmn1.3}
\|\widetilde w_{m,n} \|^{(\alpha)} 
 \leq & 64 \pi\xi^{m+n} 4^{m+n} (1 - \xi)^2 \sum_{L = 2}^{\infty}L 
 \Big(3\epsilon_I (e^{-\alpha}\rho)^{-1} \Big)^{L}(2^4)^L \\ \notag
\leq & \epsilon_I^2 \xi^{m+n} 4^{m+n}(1- \xi)^2  64  \pi 
\frac{3 \Big[ \Big(3(e^{-\alpha}\rho)^{-1} \Big)\Big]^2  
 (2^4)^{2} }{\Big(1 -  
\Big[\epsilon_I \Big( 3 (e^{-\alpha}\rho)^{-1} \Big)\Big] 2^4 \Big)^2} \period
}
\QED

\subsection{Analysis of the Renormalization Map $\cR_\alpha$: Definition and Properties  of $\widehat \uw$ }\label{repite}

In this section we prove that for every $\uw \in \widetilde \cW_\xi $ there is a 
$\widehat \uw =\big( \widehat w_{m,n}\big)_{m + n \geq 0}  $ satisfying \eqref{esano} such that
\eq{\label{noelcue}
\cR_\alpha(\uw) = H(\widehat \uw)
}
in the sense of quadratic forms in $  \cF_{\mathrm{fin}}$. We furthermore derive some estimations for the 
norms $\| \widehat w_{m,n} \|^{(\infty)}$, $\|\widehat  w_{m,n} \|^{(0)}$, $\|\widehat  w_{0,0} - r \|^{(0)}$, 
and $\| \widehat \uw \|^{(Z)}$.\\

As we already constructed  $  \widetilde \uw^{(sym)} $ in Lemma \ref{L.cprm.5} satisfying
\beq 
F_\alpha(H(\uw))(z) = \chi_\alpha H(\uw)\chi_\alpha + \chi_\alpha H(\widetilde \uw^{(sym)}(z) \chi_\alpha \comma
\ene
and established some important properties in Theorems \ref{T.C.wmn} and \ref{T.wmn1}, the definition and analysis
of $\widehat \uw$ is straightforward and follows from \eqref{rmn.7} and Definition \ref{ralpha}.

\begin{theorem}
Let $\uw \in \widetilde \cW_\xi$. Then there is a sequence of functions 
$\widehat \uw =\big( \widehat w_{m,n}\big)_{m + n \geq 0}  $ satisfying \eqref{esano} such that
\eq{\label{elguero}
\cR_\alpha(\uw) = H(\widehat \uw)\period
}
The functions $\widehat w_{m,n}$ are given by 
\eq{\label{noguero}
\widehat w_{m,n}(z, r, k^{(m,n)}) = & e^{- (3/2)(m+n)\alpha} e^{\alpha} w_{m,n}(Q_\alpha^{-1}(z); e^{-\alpha }r; e^{-\alpha} k^{(m,n)}) \\ \notag & +
e^{- (3/2) (m+n)\alpha} e^{\alpha}\widetilde w^{(sym)}_{m,n}(Q_\alpha^{-1}(z); e^{-\alpha }r; e^{-\alpha} k^{(m,n)})\period
}
\noindent \emph{Proof:}
The result follows from \eqref{hw}, \eqref{rmn.7}, Definition \ref{ralpha} and \eqref{E.L.cprm.5.1}.

\QED

\end{theorem}

\begin{definition}\label{renker}
Let $\uw \in \widetilde \cW_\xi$. We define
\eq{\label{kern}
\cR_\alpha(\uw) : = \widehat \uw \period
}
\end{definition}

\begin{theorem}\label{T.h.wmn}
Let $\uw \in \widetilde \cW_\xi$. For every $m + n \geq 0$ the following estimate holds true
\eq{\label{hwmninfty}
\|\widehat w_{m, n}\|^{(\infty)}\leq & \xi^{m+ n}\epsilon_I \frac{(1 - \xi)^2}{2}e^{-(1 + \mu/2)(m+n)\alpha}e^{\alpha}\\ \notag & \cdot \Big[ 1 +   \epsilon_I 4^{m+ n}  A^{(\infty)}(\ue, \alpha) \Big]
\Big(1 +  \frac{e^{-\alpha}}{1- \mathcal{G}_\alpha(\ue)}\Big)
 \period
}

\end{theorem}
\emph{Proof:}
It follows from  (\ref{i.5}), (\ref{ddz}) and (\ref{noguero}) that
\eq{\label{hwmn.1} \notag
\|\widehat w_{m,n} \|^{(\infty)}  \leq & e^{-(1 + \mu/2)(m + n)\alpha}e^{\alpha} \Big[ \| w_{m,n} \|^{(\infty)} +
\|\widetilde w_{m,n}^{(sym)} \|^{(\infty)}\Big]\Big(1 +  \frac{e^{-\alpha}}{1- \mathcal{G}_\alpha(\ue)}\Big) \\ \notag \leq & \xi^{m+ n}\epsilon_I \frac{(1 - \xi)^2}{2}e^{-(1 + \mu/2)(m+n)\alpha}e^{\alpha}\\  & \cdot \Big[ 1 +   \epsilon_I 4^{m+ n}  A^{(\infty)}(\ue, \alpha) \Big]
\Big(1 +  \frac{e^{-\alpha}}{1- \mathcal{G}_\alpha(\ue)}\Big)\comma
}
where we used Theorem~\ref{T.C.wmn} [see also \eqref{i.11.1}].

\QED

\begin{theorem}\label{T.hwmn1}
Let $\uw \in \widetilde \cW_\xi$. For every $m + n \geq 1$ the following estimate holds true
\eq{\label{hwmn1}
\|\widehat w_{m, n}\|^{(0)}\leq \xi^{m+n}\epsilon_I \frac{(1 - \xi)^2}{2}e^{ -\mu (m + n)\alpha/2} 
\Big[1 + 4^{m+n}   \epsilon_I A^{(0)}(\ue, \alpha) \Big]
    \period}

\end{theorem}
\emph{Proof:}
It follows from (\ref{i.6}) and (\ref{noguero}) that
\eq{\label{hwmn0.1}
\|\widehat w_{m,n} \|^{(0)}  \leq & e^{ -\mu (m + n)\alpha/2}   \Big[ \| w_{m,n} \|^{(\alpha)} +
\|\widetilde w_{m,n}^{(sym)} \|^{(\alpha)}\Big] \\ \notag \leq &
\frac{(1 - \xi)^2}{2}\Big[e^{ -\mu (m + n)\alpha/2}  + 4^{m+n} e^{ -\mu (m + n)\alpha/2}  \epsilon_I 
A^{(0)}(\ue, \alpha) \Big]
 \epsilon_I \xi^{m+n} \comma}
where we used Theorem~\ref{T.wmn1} [see also \eqref{i.11.1}].

\QED

\begin{theorem}\label{w00alpha}
Let $\uw \in \widetilde \cW_\xi$. The following estimate holds true
\eq{\label{w00alpha.1}
\|\widehat w_{0, 0} - r\|^{(0)}\leq \|w_{0, 0}- r\|^{(0)} + \epsilon_I^2 \frac{(1 - \xi)^2}{2}
A^{(0)}(\ue, \alpha) \period
}
\end{theorem}
\emph{Proof:}
The result follows from Theorem~\ref{T.wmn1} and (\ref{noguero})  as follows
\eq{  \notag \| \widehat w_{0, 0} - r \|^{(0)} \leq & \sup \Big\{  \tfrac{1}{r} \Big|e^{\alpha}\big(w_{0,0}(Q_\alpha^{-1}(z), e^{-\alpha}(r+ s))
-  w_{0,0}(Q_\alpha^{-1}(z), e^{-\alpha} s)\big) - r\Big| \\ \notag &
\hspace{.8cm} +  \tfrac{1}{r} \Big|e^{\alpha}\big( \widetilde w_{0,0}(Q_\alpha^{-1}(z), e^{-\alpha}(r+ s))
-  \widetilde w_{0,0}(Q_\alpha^{-1}(z), e^{-\alpha}s)\big) \Big| \\ 
\notag & \hspace{4.7cm} \Big | z \in D_{\rho/2}, \,s \in [0,\rho], r \in (0,\rho - s] \Big\}  
 \\  \leq & \|w_{0, 0} - r\|^{(0)} +  \| \widetilde w_{0,0} \|^{(\alpha)} \notag \\  \leq &
  \|w_{0, 0} - r\|^{(0)}  + \epsilon_I^2 \frac{(1 - \xi)^2}{2}
A^{(0)}(\ue, \alpha) \period
} 
\QED


\begin{theorem}\label{tauw}
Let $\uw \in \widetilde \cW_\xi$. 
We recall that [see (\ref{rmn.14})]
\beq \label{hatrmn.14}
\| \underline{\widehat w} \|^{(Z)} = \sup\Big \{ \Big |\frac{\partial }{\partial z} \widehat w_{0, 0} (z, r) \Big | : z \in D_{\rho/2}, r \in [0, \rho] \Big \}\period 
\ene
 It follows that 
\eq{ \label{tauw.1}
\| \underline{\widehat w} \|^{(Z)} \leq \Big[\| \underline{w} \|^{(Z)}   + \epsilon_I^2
\frac{(1 - \xi)^2}{2} A^{(\infty)}(\ue, \alpha) \Big]\frac{1}{1 - \cG_\alpha(\ue)}\period
}

\end{theorem}
\emph{Proof:}
The result follows from (\ref{ddz}), (\ref{wmninfty}), and (\ref{noguero}).

\QED

\secct{Iterated Applications of the Renormalization Map}\label{rf}

Given a sequence of positive real numbers 
$$
\ual : = \{ \alpha_j\}_{j \in \NN}
$$
and an initial sequence of functions $\uw_\alpha^{(0)}\in \widetilde \cW_\xi$, we give conditions on $ \epsilon_I, \epsilon_Z $ and $\epsilon_F$ [see \eqref{rmn.5}]  and on the sequence $\ual$, in order to assure that the iterated renormalization map 
\eq{\label{fua}
\uw_\alpha^{(\ell)} : =     \cR_{\alpha_\ell}\circ \cdots\circ \cR_{\alpha_1}(\uw^{(0)})
}
is well defined.

To achieve our purpose, we define a sequence of triples $\{ \ue_\ual^{(\ell)} \}_{\ell \in \NN_0}$:
\eq{\label{fues}
\ue^{(\ell)}_{\ual} := (\epsilon_{I, \ual}^{(\ell)}, \epsilon_{Z, \ual}^{(\ell)},\epsilon_{F, \ual}^{(\ell)})\period
}
In Section \ref{aundp} we prove that for every $\beta \in [0, \alpha_+]$ [see \eqref{alphaplus}], 
$\ue_\ual^{(\ell)}\in \underline{E}_\beta$ (see Definition \ref{hice}). We analyze additionally the numbers $\cG_\beta(\ue_\ual^{(\ell)})$, $A^{(\infty)}(\ue_\ual^{(\ell)}, \beta)$, and $A^{(0)}(\ue_\ual^{(\ell)}, \beta) $.  
Section \ref{aundp} consists on a series of definitions and numerical computations that are used in Section \ref{ic}. 
  
In Section \ref{ic} we construct the sequence of functions $\uw_\ual^{(\ell)}$ satisfying \eqref{fua}. We proceed inductively applying 
the numerical computations obtained in section \ref{aundp} to Theorems \ref{T.h.wmn}, \ref{T.hwmn1}, \ref{w00alpha} and \ref{tauw}. We prove furthermore that 
\eq{\label{IS.1hy} \| \uw^{(\ell)}_\ual \|^{(Z)} \leq \epsilon_{Z, \ual}^{(\ell)} \comma
 \| \uw^{(\ell)}_\ual - \ur \|^{(F)} \leq  \epsilon_{F, \ual}^{(\ell)}\comma
\| \uw^{(\ell)}_\ual \|^{(I)}_{\xi}\leq \epsilon_{I, \ual}^{(\ell)}\period}
In particular we obtain that the interacting part, that is controlled by $ \epsilon_{I, \ual}^{(\ell)}  $, decreases  exponentially to zero as $\ell$ goes to $\infty$ [see \eqref{3}].

\subsection{Assumptions and Analysis of the Parameters}\label{aundp}

\begin{definition}
We fix a real number $\alpha_{-}\geq 
\frac{6}{\mu}$ and  suppose that $ \alpha_{+} > \alpha_{-} $ [see \eqref{alphaplus}]. We denote by $\cS(\alpha_{-}, \alpha_{+})$ the set of sequences 
$$\ual : = \{\alpha_j\}_{j \in \NN}$$ 
such that 
\eq{ \label{ual}
\forall j \in \NN \: : \:   \alpha_{-}\leq \alpha_j \leq \alpha_{+}.
}
For every $\ual \in \cS(\alpha_-, \alpha_+)$ and every $j \in \NN$ we define 
\eq{ \label{ualj}
|\ual|_j : = \mu(\alpha_1 + \cdots + \alpha_j) \period
}

\end{definition}

\subsubsection{The Sequence $\{ \ue_\ual^{(\ell)}\}_{\ell \in \NN_0}$}

\begin{definition}
We assume that  
$ \epsilon_{Z, \ual}^{(0)}$, $ \epsilon_{F, \ual}^{(0)}$ and $ \epsilon_{I,\ual}^{(0)}$ are positive numbers that
satisfy the following properties 
\begin{itemize}
\item[(i)]
\eq{\label{i} \epsilon_{Z, \ual}^{(0)} = 1 \period}
\item[(ii)]
\eq{\label{ii}\epsilon_{F, \ual}^{(0)}   \leq \frac{1}{10}\Big( 1 - \frac{1}{2}e^{\frac{1}{10}}\Big)
\period
}
\item[(iii)] 
\eq{\label{iii}\epsilon_{I, \ual}^{(0)} \leq \frac{1}{2}\cdot  \frac{1}{10^7} e^{-2 \alpha_+} \rho^2  
\period}
\end{itemize}

For every $\ell \geq 1$, 
we define
\begin{itemize}
\item[(1)]
\eq{\label{1}\epsilon_{Z, \ual}^{(\ell)} 
: = &\prod_{j= 0}^{\ell-1} \frac{1}{1 - 10^{-12} e^{-|\ual|_j /2 }}  
\\ \notag & + 
\frac{1}{10^7} \sum_{j = 0}^{\ell-1} e^{-|\ual|_j/2}  \prod_{l= j}^{\ell-1} \frac{1}{1 - 10^{-12} e^{-|\ual|_l /2 }}    \period} 
\item[(2)]
\eq{\label{2}\epsilon_{F, \ual}^{(\ell)} : = \epsilon_{F, \ual}^{(0)} + \frac{1}{10^7}\sum_{j = 0}^{\ell-1}   
e^{-|\ual|_j/2} \period
}
\item[(3)] 
\eq{\label{3}\epsilon_{I, \ual}^{(\ell)} : =  \epsilon_I^{(0)} e^{-|\ual|_\ell /4}\period
}
\end{itemize}

We denote 
\eq{\label{fue}
\ue^{(\ell)}_{\ual} := (\epsilon_{I, \ual}^{(\ell)}, \epsilon_{Z, \ual}^{(\ell)},\epsilon_{F, \ual}^{(\ell)})\period
}

\end{definition}

\begin{theorem}\label{T.parameters}
Let $\ual \in \cS(\alpha_-, \alpha_+)$.
For every $\ell \in \NN_0$ and every $\beta \in [0, \alpha_+]$, 
\eq{\label{parapapa}
\ue^{(\ell)}_{\ual}  \in \uE_\beta
}
(see Definition \ref{hice}).\\
It follows furthermore that
\eq{\label{T.parameters.1}
\cG_\beta(\ue_\ual^{(\ell)}) \leq & e^{-|\alpha|_\ell/2}\frac{(e^{-\alpha_+}\rho)^2 }{10^{12}}\comma
\\ \notag
A^{(\infty)}(\ue_\ual^{(\ell)}, \beta) \leq & 3\cdot 10^4 (\rho e^{-\beta})^{-2}\comma
\\ \notag
A^{(0)}(\ue_\ual^{(\ell)}, \beta)  \leq & 3\cdot 10^6 (\rho e^{-\beta})^{-2} \period
}  

\end{theorem}
\emph{Proof:}
It follows from (\ref{ii}), \eqref{iii}, (\ref{2}) and \eqref{3} that
\eq{\label{P.1}
1 -\epsilon_{F, \ual}^{(\ell)} - \frac{1}{2}e^{1/10}- 
\epsilon_{I, \ual}^{(\ell)}\xi (\rho e^{-\alpha_+})^{-1}   \geq \frac{9}{10} (1 - \frac{1}{2}e^{1/10}) - \frac{1}{100}
> \frac{1}{3}\comma
}
which implies (\ref{besto}). Eq. (\ref{hirmn.6}) follows from  \eqref{iii} and (\ref{3}). 
Eq.~(\ref{1}) implies that 
\eq{\label{epsilonz}\epsilon_{Z, \ual}^{\ell} \leq 2\comma}
for all $\ell \in \mathbb{N}_0$,  which together with (\ref{iii}) and (\ref{3}) implies that [see (\ref{HiE.L.r.2})]
\eq{\label{gw}
\cG_\beta(\ue^{(\ell)}) \leq e^{-  |\alpha|_\ell /2}\frac{(e^{-\alpha_+}\rho)^2 }{10^{12}} < 1\period
}
Similarly one verifies that (\ref{hiE.L.r.2.2}) is satisfied for all $\ell$.  
Using (\ref{ainfty}), (\ref{iii}), (\ref{3}), (\ref{P.1}) and (\ref{epsilonz}) we get
\eq{\label{Ainfty}
A^{(\infty)}(\ue^{(\ell)}_\ual, \beta) \leq 3\cdot 10^4(\rho e^{-\beta})^{-2}\period
 }
In the same way we obtain using additionally (\ref{aalpha}) that 
\eq{\label{Aalpha}
A^{(0)}(\ue^{(\ell)}_\ual, \beta) \leq  3\cdot 10^6(\rho e^{-\beta})^{-2}\period
 } 

\QED

\subsection{Inductive Construction}\label{ic}

\subsubsection{Induction Basis}

In this section we assume that $\uw^{(0)}_{\alpha} \in  \cW_{\xi} $ satisfies
\begin{itemize}

\item[($a_0$)]
\eq{\label{a0} \| \uw^{(0)}_\alpha \|^{(Z)} \leq \epsilon_{Z, \ual}^{(0)} \period}
\item[($b_0$)]
\eq{ \label{b0}\| \uw^{(0)}_\alpha - \ur \|^{(F)} \leq  \epsilon_{F, \ual}^{(0)} \period
}
\item[($c_0$)] 
\eq{\label{c0}\| \uw^{(0)}_\alpha \|^{(I)}_{\xi}\leq \epsilon_{I, \ual}^{(0)} \period} 
\end{itemize}
\begin{remark}
By Theorem \ref{T.parameters}, 
\eq{\label{parapapapone}
\ue^{(0)}_{\ual}  \in \uE_\beta
}
and therefore $ \uw^{(0)}_\alpha   \in \widetilde \cW_\xi$ [see \eqref{a0}-\eqref{c0}], which implies that  
$\cR_\beta(\uw^{(0)}_{\ual})$  
is well defined for every $\beta \in [0, \alpha_+]$.
\end{remark}

\subsubsection{Induction Step}
\begin{theorem}\label{is}
Let $\uw^{(0)}_\ual \in  \cW_{\xi} $. Suppose that for every $j \in \{ 0, \cdots, \ell \}$ exists 
$\uw^{(j)}_\ual \in  \cW_{\xi} $ satisfying 
\begin{itemize}

\item[(a)]
\eq{\label{IS.1} \| \uw^{(j)}_\ual \|^{(Z)} \leq \epsilon_{Z, \ual}^{(j)} \comma}
\item[(b)]
\eq{\label{IS.2} \| \uw^{(j)}_\ual- \ur \|^{(F)} \leq  \epsilon_{F, \ual}^{(j)\comma}
}
\item[(c)] 
\eq{\label{IS.3}\| \uw^{(j)}_\ual \|^{(I)}_{\xi}\leq \epsilon_{I, \ual}^{(j)}\period} 
\end{itemize}
Suppose furthermore that for $j \in \{0, \cdots, \ell-1\}$
\eq{\label{jmas1}\cR_{\alpha_{j+1}}(\uw^{(j)}_\ual) = \uw^{(j+ 1)}_\ual
\period}

If follows that 
\eq{ \label{jmas1.1}
\uw^{(\ell + 1)}_\ual : = \cR_{\alpha_{\ell + 1}}(\uw^{(\ell)}_\ual)
}
is well defined and satisfies  (\ref{IS.1})-(\ref{IS.3}) with $\ell + 1$ instead of $j$. \\
If we additionally suppose that $\xi < \frac{1}{4}$, $\beta \in [0,\alpha_+]  $ and we define 
$$
\uw^{(\ell, \beta)}_\ual: = \cR_\beta(\uw^{(\ell)}_\ual),
$$
it follows that $\uw^{(\ell, \beta)}_\ual \in \cW_{4\xi}$ and 
\eq{\label{beta-in}
 \| \uw^{(\ell, \beta)}_\ual \|^{(Z)} \leq &\epsilon_{Z, \ual}^{(\ell + 1)} \comma \\ \notag
 \| \uw^{(\ell, \beta)}_\ual - \ur \|^{(F)} \leq & \epsilon_{F, \ual}^{(\ell + 1)}\comma \\ \notag
\| \uw^{(\ell, \beta)}_\ual \|^{(I)}_{4\xi}\leq & \epsilon_{I, \ual}^{(\ell)} \frac{(1 - \xi)^2}{(1 - 4\xi)^2}   \period
}

\end{theorem}
\emph{Proof:}
By Theorem \ref{T.parameters}, 
\eq{\label{parapapaponele}
\ue^{(j)}_{\ual}  \in \uE_\beta
}
and therefore $\uw^{(j)}_{\ual} \in \widetilde \cW_\xi$ [see \eqref{a0}-\eqref{c0}], which implies that  
$\cR_\beta(\uw^{(j)}_{\ual})$  
is well defined for every $j \in \{ 0, \cdots, \ell\}$ and every $\beta \in [0, \alpha_+]$.\\
Theorems \ref{T.h.wmn} and \ref{T.parameters} and Eq.~(\ref{epsilonz}) imply that for
 $m + n \geq 1  $
\eq{ \label{1.1}
\| (w^{(\ell + 1)}_\ual)_{m,n} \|^{(\infty)}\leq  & \notag   \xi^{m+n}  \epsilon_{I, \ual}^{(\ell)} \frac{(1 - \xi)^2}{2}  \Big(1 + \frac{e^{-\alpha_{\ell + 1}}}{1 - 10^{-12}}\Big) \\ \notag &
\Big[ e^{-\mu\alpha_{\ell +1}/2}+  4^{m+n}e^{-(m+n)\mu\alpha_{\ell + 1}/2}
\frac{3 e^{-|\ual|_\ell/4}}{1000}\Big] 
\\  \leq &\frac{1}{2}  \xi^{m+n} e^{-\mu\alpha_{\ell + 1}/4} e^{- |\ual|_\ell /4}\epsilon_I^{(0)} \frac{(1 - \xi)^2}{2} \comma
}
where we used that $4^{m + n} e^{-(m + n) \mu \alpha_{\ell + 1}/4} \leq 1$ and that 
$ e^{-\mu \alpha_{\ell + 1}/2 } \leq e^{- 3/2}e^{-\mu \alpha_{\ell + 1}/4 } $ (here we use that $\mu \alpha_- \geq 6$).

Theorems \ref{T.hwmn1} and \ref{T.parameters} imply that
\eq{\label{1.2} \notag
\| (w^{(\ell + 1)}_\ual)_{m,n} \|^{(0)} \leq & \xi^{m+n}\epsilon_{I, \ual}^{(\ell)}\frac{(1 - \xi)^2}{2} 
\Big[ e^{-\mu\alpha_{\ell + 1}/2}  +  4^{m+n}e^{-(m+n)\mu\alpha_{\ell + 1}/2}\frac{3 e^{-|\ual|_\ell/4}}{20}\Big]   
\\  \leq & \frac{1}{2} e^{-\mu\alpha_{\ell + 1}/4}\frac{(1 - \xi)^2}{2} \xi^{m+n} e^{- |\ual|_\ell /4}\epsilon_{I, \ual}^{(0)} \period
}
where we used that $4^{m + n} e^{-(m + n) \mu \alpha_{\ell + 1}/4} \leq 1$ and that 
$ e^{-\mu \alpha_{\ell + 1}/2 } \leq e^{- 3/2}e^{-\mu \alpha_{\ell + 1}/4 } $ (here we use that $\mu \alpha_- \geq 6$).

Theorems \ref{w00alpha} and \ref{T.parameters} imply that
\eq{\label{1.3}
\| \uw^{(\ell+1)}_\ual - \ur \|^{(F)}\leq & \epsilon_{F, \ual}^{(\ell)} + 
(\epsilon_{I, \ual}^{(\ell)})^2 \frac{(1 - \xi)^2}{2} A^{(0)}(\ue_\ual^{(\ell)}, \alpha_{\ell + 1}) 
\\ \notag \leq & 
 \epsilon_{F, \ual}^{(\ell)} + \frac{1}{10^7} (e^{-\alpha_{+}}\rho)^2 e^{- |\alpha|_\ell /2}\frac{(1 - \xi)^2}{10}
\period
}
Theorems \ref{tauw} and \ref{T.parameters} imply that
\eq{\label{1.4}
\| \uw_\ual^{(\ell + 1)} \|^{(Z)} \leq & \Big[\| \uw_\ual^{(\ell)} \|^{(Z)}  + (\epsilon_{I, \ual}^{(\ell)})^2\frac{(1 - \xi)^2}{2} A^{(\infty)}(\ue_\ual^{(\ell)}, \alpha_{\ell + 1}) \Big] 
\frac{1}{1 - \cG_\alpha(\ue_\ual^{(\ell)})}
\notag
\\ \leq &  \Big[ \epsilon_Z^{(\ell)} + \frac{1}{10^7} (e^{-\alpha_+}\rho)^2 e^{- |\alpha|_\ell /2}\frac{(1 - \xi)^2}{10}\Big] \frac{1}{1 - e^{-12}e^{- |\alpha|_\ell /2}}  \period
}
Eqs.~(\ref{IS.1})-(\ref{IS.3}) with $\ell + 1$ instead of $j$ follows from (\ref{i.11.1})-(\ref{i.11}), (\ref{1})-(\ref{3}) and (\ref{1.1})-(\ref{1.4}). Eq.~\eqref{beta-in} is proved similarly. 

\QED

\secct{The Renormalization Flow} \label{sacelfua}

We fix an initial sequence of functions $\uw$ satisfying \eqref{a0}-\eqref{c0} (with $\uw$ instead of $\uw_\ual^{(0)}$). 

For every $s \in [0, \infty)$ we select a sequence $\ual \in \cS(\alpha_-, \alpha_+)$ and a number 
$ \beta \in [0, \alpha_+]$ such that 
$$
\sum_{j=1}^\ell \alpha_j  + \beta = s\comma
$$
for some $\ell \in \mathbb{N}_0$ (if $\ell = 0$ we omit the sum).  We define the operator 
$$
\boH_s   = \cR_\beta( H(\uw_\ual^{(\ell)}))\comma
$$
where we take $  \uw_\alpha^{(0)} = \uw$ to construct $ \uw_\ual^{(\ell)} $ (see Theorem \ref{is}).\\
The operators $\boH_s$ define a family of isospectral operators for which the interacting part decays exponentially in $s$ as goes to infinity [see \ref{contaction}]. In this section we prove that 
the family of operators is well defined, in the sense that it does not depend on $\ual $ and $\beta$. 
The key ingredient is the construction of the (continuous) renormalization of the spectral parameter 
$$
\boE_s : D_{\rho/2} \to D_{\rho/2}\comma
$$
which is an analytic open injective function with analytic inverse. It has the following properties:

\begin{itemize}
\item For very $z \in D_{\rho/2}$, the rescaled Feshbach-Schur map $\widehat \cR_s\Big(H\big(\uw(\boE_s(z))\big)\Big)$ is well defined [see \eqref{rmn.7}].

\item The following equation holds true
\eq{\label{acajo}
\boH_s(z) = \widehat \cR_s\Big(H\big(\uw(\boE_s(z))\big)\Big)\period
}

\end{itemize}  

Eq.~\eqref{acajo} and the isospectrality  of the operator $\widehat \cR_s$ imply that the analysis of the spectrum of the original operator $ H(\uw(\zeta))  $ is equivalent to the analysis for the operators 
$ \boH_s(z) $ for $ \zeta = \boE_s(z)$. As the interacting part of the operator $\boH_s$ goes exponentially to zero
as $s $ goes to infinity, then $ \boH_s(z) $ is easier to analyze than the original Hamiltonian. If we take 
$s \to \infty$, the spectrum of the resulting Hamlitonian is explicit.    \\
In Section \ref{renflow} we construct the function $\boE_s$. We prove furthermore \eqref{acajo} and the exponential decay of the interacting term. \\
In Section \ref{renomfunc} we define a set of sequences functions $\{ \uw_s \}_{s \geq 0}$ such that
\eq{\label{ti}
\boH_s = H(\uw_s)\period
}
In section \ref{flowop} we define a space of operators $ H[\cW_\xi]^{(0)} $ and a flow 
$$
{\bf \Phi} :  H[\cW_\xi]^{(0)} \times [0, \infty) \to  H[\cW_\xi]^{(0)}
$$
whose orbits associated are the sets $\{ \boH_s\}_{s \geq 0}$. In particular this proves that 
the operators $\boH_s$ satisfy a group property. We define furthermore the corresponding flow in the function spaces. There we have to construct an equivalence class of function spaces, due to fact that the mapping
$$
\uw \to H(\uw) 
$$
is not injective. The use of equivalence classes of functions is also applied to \eqref{ti}.

\subsection{The Renormalization Flow of Operators}\label{renflow}

\begin{lemma}\label{lero-lero}
Let $O$ be a bounded operator defined in a Hilbert space $\cH$ and let $P_1$  and $P_2$ be commuting projections 
such that $P_1 P_2 = P_2$. Suppose furthermore that 
$$
\overline P_1 O\overline P_1, 
$$ 
is invertible (with bounded inverse), where 
$$
\overline P_i = 1 - P_i, \hspace{3cm} i = 1, 2\period 
$$ 
The Feshbach-Schur map is defined as in \eqref{fm.3}: 
\eq{\label{feshO}
F_{P_1}(O) = P_1O P_1 - P_1 O (\overline P_1 O \overline P_1 )^{-1} O P_1.
}
It follows that if $ \overline P_2 F_{P_1}(O)\overline P_2 $ is invertible with bounded inverse, then also 
$ \overline P_2 O \overline P_2  $ is invertible with bounded inverse.

\end{lemma}
\noindent \emph{Proof:}
As $\overline P_1 \overline P_2 = \overline P_1$, we can define the Feshbach-Schur map as in 
(\ref{feshO}) with $\overline P_2 O \overline P_2$ instead of $O$.
\eq{\label{feshO.1}
F_{P_1}(\overline P_2 O \overline P_2) = P_1 \overline P_2 O \overline P_2 P_1 - P_1 \overline P_2 O (\overline P_1 O \overline P_1 )^{-1}  O \overline P_2 P_1  =  \overline P_2 F_{P_1}(O) \overline P_2.
}
As by assumption $  \overline P_2 F_{P_1}(O) \overline P_2 $ is invertible with bounded inverse it follows by the basic properties of the Feshbach-Schur map (see \cite{BFS98a}) that $\overline P_2 O \overline P_2 $ is invertible with bounded inverse. 
\QED

\begin{theorem}\label{elpex}
Assume the hypotheses and notations of Theorem \ref{is}. We define for every $\beta \in [0, \alpha_+]$ [see \eqref{rmn.11}] 
\eq{\label{Qell}
 Q_\beta^{(\ell, \ual)}(z) : = & \Big \la \widehat \cR_\beta \big(H(\uw_\ual^{\ell}(z))\big)  \Big \ra_\Omega \comma \\
\notag \cE^{(\ell, \ual, \beta)} : = & (Q_{\alpha_1}^{(0, \ual)})^{-1}\circ \cdots \circ (Q_{\alpha_\ell}^{(\ell - 1, \ual)})^{-1} \circ (Q_\beta^{(\ell, \ual)})^{-1}  \period  
}
Let $z \in D_{\rho/2}$ and take $\zeta = \cE^{(\ell, \ual, \beta)}(z)$. Then

\eq{
\chi_{ |\ual|_\ell/\mu + \beta} H(\uw^{0}(\zeta))\chi_{ |\ual|_\ell/\mu + \beta}
}
is invertible (with bounded inverse) and 
\eq{ \label{chido}
H(\uw_\ual^{(\ell, \beta)}(z)) = \widehat \cR_{|\ual|_\ell/\mu + \beta} (H(\uw_\ual^{(0)}(\zeta)))\period
}

\end{theorem}
\noindent \emph{Proof:}
For simplicity we prove the assertion only for $\beta = 0$. If $\beta \ne 0$ the same argument works but the notation is more complicated.    
We use induction in $\ell$. For $\ell = 1$ the result follows from the definition of $\cR_{\alpha_1}$ (see Definition \ref{ralpha}). We suppose that \eqref{chido} is valid for $\ell - 1$ and we prove it for $\ell $.  \\
Let  $z' = (Q^{(\ell- 1, \ual)}_{\alpha_{\ell}})^{-1}(z) \in e^{-\iota \alpha_\ell } D_{\rho/2}$ (see Lemma \ref{L.r.2}). 

By hypothesis 
\eq{ \label{hyp}
H(\uw_\ual^{(\ell-1)}(z')) = \widehat \cR_{|\ual|_{\ell - 1}/\mu } (H(\uw_\ual^{(0)}(\zeta)))\period
}
Lemma \ref{L.fm} and Theorem \ref{is} imply that [see \eqref{rmn.7}] 
\eq{
\overline \chi_{\alpha_\ell }H(\uw_\ual^{(\ell-1)}(z'))\overline \chi_{\alpha_\ell} = &
\overline \chi_{\alpha_\ell }  \widehat \cR_{|\ual|_{(\ell -1)}/\mu } (H(\uw_\ual^{(0)}(\zeta))) 
\overline \chi_{\alpha_\ell}
  \notag \\ = & e^{|\alpha|_{(\ell-1)} /\mu} \Gamma_{| \ual|_{(\ell-1)} /\mu}
\overline \chi_{ |\ual|_\ell/\mu} \notag \\ & F_{|\ual|_{(\ell - 1)/\mu}} (H(\uw_\ual^{(0)}(\zeta)))\overline 
\chi_{ |\ual|_\ell/\mu}
\Gamma_{ |\ual|_{(\ell-1)}/\mu}^*}
is invertible. Lemma \ref{lero-lero} implies that  
$ \overline \chi_{|\ual|_{\ell/\mu}} H(\uw_\ual^{(0)}(\zeta)) \overline \chi_{ |\ual|_{\ual/\mu}} $ is invertible. 
We apply Lemma \ref{L.rep.1} with $f = \cE^{(\ell,\ual ,0)}$, using Corrolary \ref{compact-set} and 
Theorem \ref{T.parameters} to prove the existence of the set $\mathcal{C}$, to conclude that 
\eq{\label{buena}
\widehat \cR_{\alpha_\ell } (\widehat \cR_{|\ual|_{(\ell - 1)/\mu} } (H(\uw_\ual^{(0)}(\zeta)))) =
 \widehat \cR_{|\ual|_\ell/\mu } (H(\uw_\ual^{(0)}(\zeta))) \comma
}
which together with Definition \ref{ralpha} and \eqref{hyp} accomplish the induction step.

\QED

\begin{lemma} \label{bueno}
Let $\ual, \tilde \ual \in \cS(\alpha_-, \alpha_+)$. Suppose that there exist $\ell, \tilde \ell \in \NN_0 $ 
such that 
\eq{
\sum_{j=1}^\ell \alpha_j = \sum_{j=1}^{\tilde \ell}\tilde \alpha_j 
}
and 
\eq{
\cE^{(\ell, \ual, 0)} = \cE^{(\tilde \ell, \tilde \ual, 0)} \period
}

Suppose furthermore that there exist $\beta , \tilde \beta \in [0, \alpha_+]$
and  natural numbers $\ell'> \ell$ and  $ \tilde \ell'> \tilde \ell$ such that
\eq{ \label{la-condicion}
\sum_{j = \ell + 1}^{\ell'} \alpha_j + \beta = \sum_{j = \tilde \ell + 1}^{\tilde \ell'} \tilde \alpha_j + \tilde \beta = b_0
< 2 \alpha_+  \period
}
Then 
\eq{\label{then}
\cE^{(\ell', \ual, \beta)} = \cE^{(\tilde \ell', \tilde \ual, \tilde \beta)} \period
}

\end{lemma}

\noindent \emph{Proof:}
First we notice that Theorem \ref{elpex} implies that 
\eq{ \label{la-condicion-f}
H(\uw_\ual^\ell(z) ) = H(\uw_{\tilde \ual}^{\tilde \ell}(z)) : = H_{in}(z)  \period
}
We denote 
$$
\cE_{in} : =  (Q_{\alpha_{\ell + 1}}^{(\ell, \ual)})^{-1}\circ \cdots \circ (Q_{\alpha_{\ell'}}^{(\ell'-1, \ual)})^{-1} \circ (Q_\beta^{(\ell', \ual)})^{-1}
$$
and 
$$
\tilde  \cE_{in} : = (Q_{\tilde \alpha_{\ell + 1}}^{(\tilde \ell, \ual)})^{-1}\circ \cdots \circ (Q_{ \tilde \alpha_{\tilde \ell'}}^{(\tilde \ell'-1, \tilde \ual)})^{-1} \circ (Q_{\tilde \beta}^{(\tilde \ell', \tilde \ual)})^{-1} \period
$$
Now we iterate \eqref{estimatez} to obtain that $z$ belongs to $\cE_{in}(D_{\rho/2})$ whenever 
\eq{\label{chido.3} 
\frac{\rho}{2} >\,  & e^{\alpha_{\ell + 1} + \cdots + \alpha_{\ell'} + \beta}|z| + e^{\alpha_{\ell + 1} + \cdots + \alpha_{\ell'} + \beta}\cG_{\alpha_{\ell +1}}(\ue_\ual^{(\ell)}) 
  \notag \\ &  + e^{\alpha_{\ell + 2} + \cdots + \alpha_{\ell'} + \beta}\cG_{\alpha_{\ell +2}}(\ue_\ual^{(\ell + 1)}) 
+ \cdots + e^{ \beta}\cG_{\beta}(\ue_\ual^{(\ell')})  \period
}
 Eqs.~\eqref{T.parameters.1} and \eqref{la-condicion} imply that if $|z|$ is sufficiently small, it belongs to $\cE_{in}(D_{\rho/2})$. Using similar arguments we conclude that  
  $\cE_{in}(D_{\rho/2})  \cap \tilde \cE_{in}(D_{\rho/2})$ contains an open (not empty) ball. Now we define the functions 
\eq{\label{g} 
& g(\zeta) =\la \widehat R_{ b_0}(H_{in}(\zeta))\ra_\Omega, \hspace{3cm} \zeta \in  \cE_{in}(D_{\rho/2})\comma 
\\ & \notag
 \tilde g(\zeta) = \la \widehat R_{ b_0}(H_{in}(\zeta))\ra_\Omega, \hspace{3cm} \zeta \in \tilde \cE_{in}(D_{\rho/2}) \period 
}  
That the functions $g$ and $\tilde g$ exist is a consequence of the proof of Theorem \ref{elpex}. 
Lemma \ref{chito} and Theorem \ref{elpex} imply that for every 
$z \in D_{\rho/2} $
\eq{ \label{eq1}
g \circ \cE_{in}(z)  =  \tilde g \circ \tilde \cE_{in}(z) = z\period
}
As $ \cE_{in} $ and $\tilde \cE_{in}$ are bijective over their images, Eq.~\eqref{eq1} implies that they coincide 
in $ g( \cE_{in}(D_{\rho/2})  \cap \tilde \cE_{in}(D_{\rho/2}) ) $ (actually $g$ restricted to 
$\cE_{in}(D_{\rho/2})  \cap \tilde \cE_{in}(D_{\rho/2})$ is the (analytic) bijective inverse of $\cE_{in}$ and 
$\tilde \cE_{in}$). As $ \cE_{in} $ and $\tilde \cE_{in}$ are analytic, they coincide
in $D_{\rho/2}$, which implies \eqref{then}.

\QED

\begin{theorem}\label{T}
Suppose that 
 $ \ual , \tilde \ual \in \cS(\alpha_-, \alpha_+) $, with $\alpha_+ \geq 2 \alpha_-$, and that $\uw^{(0)}_\ual = \uw^{(0)}_{\tilde \ual}$. Suppose furthermore that there are 
 $\ell, \tilde \ell \in \NN_0$ and $\beta, \tilde \beta  \in [0, \alpha_+ ]$ such that
\eq{ \label{sum}
\sum_{j = 1}^\ell \alpha_j + \beta = \sum_{j = 1}^{\tilde \ell} \tilde \alpha_j + \tilde \beta = a \comma
}
 then 
\eq{ \cE^{(\ell, \ual, \beta)} = \cE^{(\tilde \ell, \tilde \ual, \tilde \beta)}\period
 }

\end{theorem}
\noindent \emph{Proof:}
In the proof of this lemma we construct four sequences $\ual, \tilde \ual, \ual^{(1)}, \tilde \ual^{(1)}$ belonging to
$S(\alpha_-, \alpha_+)$. To simplify the formulas we define the following notation that is used only for this proof
[see \eqref{ualj}]:
\eq{ \label{shrothand}
\forall \ell \in \NN : a_\ell = \frac{1}{\mu}|\ual|_\ell, \hspace{.5cm}
\tilde a_\ell = \frac{1}{\mu}|\tilde \ual|_\ell, 
\hspace{.5cm} a_\ell^{(1)} = \frac{1}{\mu}|\ual^{(1)}|_\ell, \hspace{.5cm} \tilde a_\ell^{(1)} 
= \frac{1}{\mu}|\tilde \ual^{(1)}|_\ell \period
}

Define  $\epsilon_0 : = \frac{1}{2}(\alpha_+- \alpha_-) $.
We use induction on $a$. For $a \leq \epsilon_0$ the result follows from Lemma \ref{bueno}. We suppose that the result is valid for $a \leq L \epsilon_0$  and we prove it for $a \leq (L+1)\epsilon_0$.  Suppose that \eqref{sum} holds for
$a \leq (L+1)\epsilon_0$. Let $J $ be the minimal $j$ such that $a - 2 \alpha_+ \leq  a_j$ and 
$\tilde J$ the minimal $j$ such that $a - 2 \alpha_+\leq  \tilde a_j$. Suppose, without loss of generality, that 
$a_J \leq \tilde a_{\tilde J}$. It follows that $a - \tilde a_{\tilde J}  > \alpha_+$. Let $\tilde \ual^{(1)} \in \cS(\alpha_-, \alpha_+)$ be a sequence that coincides with $\tilde \ual$ in the first $\tilde J$ entries and such that $\tilde \alpha^{(1)}_{j}  = \alpha_-$ for $j > \tilde J $.
We denote by $\tilde \ell^{(1)}$ the maximal $j$ such that $\tilde a^{(1)}_j \leq a  $ and by 
$\tilde \beta^{(1)} = a - a^{(1)}_{\tilde \ell^{(1)}}.$ It follows that 
\eq{ \label{zero}
a - \tilde a_{\tilde J + 1}^{(1)} > \epsilon_0\comma \hspace{1cm} 
 \tilde a_{\tilde J + 1}^{(1)} - a_J \geq \alpha_-\period
}
The fact that $\alpha_+ \geq 2 \alpha_-$ implies that there is a sequence $\ual^{(1)}\in \cS(\alpha_-, \alpha_+)$ that coincides with $\ual$ in the first $J$ entries and such that 
$ a^{(1)}_{J^{(1)}} = \tilde a_{\tilde J + 1}^{(1)}  $ for some $J^{(1)} \in \NN$.  
We denote by $ \ell^{(1)}$ the maximal $j$ such that $ a^{(1)}_j \leq a  $ and by 
$ \beta^{(1)} = a - a^{(1)}_{ \ell^{(1)}}.$  \\
Lemma \ref{bueno} (and the induction hypothesis) implies that 
\eq{\label{an}
\cE^{(\tilde \ell, \tilde \ual, \tilde \beta)} = &
 \cE^{(\tilde \ell^{(1)}, \tilde \ual^{(1)}, \tilde \beta^{(1)})}\comma 
 \\ \notag 
\cE^{( \ell,  \ual,  \beta)} = &
 \cE^{( \ell^{(1)}, \ual^{(1)},  \beta^{(1)})}\comma \\ \notag 
  \cE^{( \ell^{(1)},  \ual^{(1)},  \beta^{(1)})} = &
 \cE^{(\tilde \ell^{(1)}, \tilde \ual^{(1)}, \tilde \beta^{(1)})} 
\period
}
\QED
Theorems \ref{elpex} and \ref{T} imply the following:
 
\begin{theorem} \label{T1}
Suppose that 
 $ \ual , \tilde \ual \in \cS(\alpha_-, \alpha_+) $, with $\alpha_+ \geq 2 \alpha_-$, and that $\uw^{(0)}_\ual = \uw^{(0)}_{\tilde \ua}$. Suppose furthermore that there are 
 $\ell, \tilde \ell \in \NN_0$ and $\beta, \tilde \beta  \in [0, \alpha_+]$ such that
\eq{ \label{sum-o}
\sum_{j = 1}^\ell \alpha_j + \beta = \sum_{j = 1}^{\tilde \ell} \tilde \alpha_j + \tilde \beta = s \comma
}
 then 
\eq{ \label{H}
H(\uw_\ual^{(\ell, \beta)}) = H(\uw_{\tilde \ual}^{(\tilde \ell, \tilde \beta)})\period
}

\end{theorem}

\begin{definition}\label{esta}

Let $ \uw^{(0)} \in \widetilde \cW_{\xi}  $ satisfy \eqref{a0}-\eqref{c0}. 
Suppose that $\alpha_+ \geq 2 \alpha_-$. For every  $s \geq 0$
we denote by 
$$
\boH_s\equiv \boH_s(z) : =  H(\uw_{ \ual}^{(\ell, \beta)})\comma 
$$
where $ \uw_{ \ual}^{(\ell, \beta)} $ is the sequence of functions constructed in Theorem \ref{is} 
such that $ \uw_{ \ual}^{(0)} = \uw^{(0)}$, $\ual$ is an element of $S(\alpha_-, \alpha_+)$ and
\eq{ \label{sump}
\sum_{j = 1}^\ell \alpha_j + \beta = s \period
}
We define additionally
\eq{ \label{Ep}
\boE_s : =  \cE^{(\ell, \ual, \beta)}\comma
}
and
\eq{ \label{Hp}
 \boT_s \equiv &\boT_s(z)  : =  (\uw_{ \ual}^{(\ell, \beta)})_{0,0}(H_f)\comma 
\\ \notag \boW_s\equiv & \boW_s(z) : =  \boH_s - \boT_s\period 
}

\begin{remark}
The fact that  $\alpha_+ \geq 2 \alpha_-$ implies that if $s \geq \alpha_-$ we can take $\beta = 0$ in \eqref{sump}.
\end{remark}

\end{definition} 

\begin{theorem}[Contraction Property] \label{cpr}

For every $s \geq 0$ and every $ \uw^{(0)} \in \widetilde \cW_{\xi}  $, 
\eq{\label{aca}
\boH_s = \widehat \cR_s(\boH_0(\boE_s(z)))\comma
}
and for $s \geq \alpha_-$ 
\eq{\label{contaction}
\|\boW_s \|\leq \frac{1}{10^7} e^{-2 \alpha_+} \rho^2 e^{- \mu s /4}\period
}

\end{theorem}
\emph{Proof:}
The result follows from Theorems \ref{is}, \ref{elpex}, \ref{T} and \ref{T1}.
\QED

\subsection{Renormalization Flow on the Function Spaces}\label{renomfunc}

\begin{definition}\label{classes}
We say that two elements 
$$\uw = \big( w_{m,n}\big)_{m + n \geq 0}, \uw' = \big( w'_{m,n}\big)_{m + n \geq 0} \in \cW_\xi $$ 
are equivalent if for every $m, n \in \NN_0$
\eq{\label{equiv}
\chi_0(r + \| k^{(m)} \|_1) & \chi_0(r + \| \tilde k^{(m)} \|_1)w_{m,n}(z;r; k^{(m,n)})\\  & = \notag
\chi_0(r + \| k^{(m)} \|_1) \chi_0(r + \| \tilde k^{(m)} \|_1)w'_{m,n}(z;r; k^{(m,n)})\period
}
We denote by $[\cW_\xi]$ the quotient space of $\cW_\xi$ induced by the equivalence relation \eqref{equiv}.
The elements of $[\cW_\xi]$ are denoted by $[\uw] = \big( [w_{m,n}]\big)_{m + n \geq 0}  $, where
$   [\uw] $ is the equivalent class of $\uw \in \cW_\xi$. The quantities \eqref{i.11.1}-\eqref{i.11} are defined
for $ [\uw] \in [\uW_\xi]$ as usual for the quotient spaces, taking the infimum over the elements in the equivalence class.    
\end{definition}

\begin{remark}\label{r-equiv}
Suppose that $\uw$ and $\uw'$ are equivalent. Then $\cR_\alpha(\uw)$ and $\cR_\alpha(\uw')$ are equivalent.
\end{remark}
\noindent \emph{Proof}
First we notice that [see \eqref{Wmnpq} and Definition \ref{D.cprm.4}]  
\eq{\label{r-e-equiv1}
& \chi_0(H_f + e^{-\alpha}r + e^{-\alpha}\tilde r_{\ell-1}(k^{(\bar m, \bar n)})) \widetilde W_\ell(z; r + e^{-\alpha} r_{\ell}(k^{(\bar m, \bar n)}); 
e^{-\alpha}k^{(m_\ell, n_\ell)}) \notag \\ & \notag  \hspace{7cm} \cdot\chi_0(H_f + e^{-\alpha} r + e^{-\alpha}\tilde r_\ell(k^{(\bar m, \bar n)})) \\ \notag
 & =   \int  d x^{(p_\ell)}  d \tilde x^{(q_\ell)}a^*( x^{(p_\ell)}) \chi_0(H_f + e^{-\alpha}r + e^{-\alpha}\tilde r_{\ell-1}(k^{(\bar m, \bar n)}) + \| x^{p_\ell} \|_1) 
 \\  & \hspace{1cm} \cdot   w_{m_\ell+p_\ell, n_\ell+q_\ell}\big(z; H_f + e^{-\alpha}r + e^{-\alpha} r_{\ell}(k^{(\bar m, \bar n)}); (e^{-\alpha}k^{(m_\ell)}, x^{(p_\ell)})\notag  \\ \notag &  \hspace{9cm}; (e^{-\alpha}\tilde k^{(n_\ell)}, \tilde x^{(q_\ell)})\big) \notag \\  & \hspace{3.5cm} \cdot
\chi_0(H_f + e^{-\alpha}r + e^{-\alpha}\tilde r_{\ell}(k^{(\bar m, \bar n)}) + \| \tilde x^{q_\ell} \|_1) 
a( \tilde x^{(q_\ell)}) 
}
does not depend of the values of $w_{w_{m_\ell+p_\ell, n_\ell+q_\ell}}(z; t; y^{(m_\ell + p_\ell, n_\ell + q_\ell)}  )$
with $t  + \| y^{(m_\ell + p_\ell)} \|_1 > \rho$ or  $t + \| \tilde y^{(n_\ell + q_\ell)} \|_1 > \rho$ and, therefore, if we change $\uw$ for $\uw'$ it does not change. This  (and similar arguments) implies that [see \eqref{vupsilon}] 
\eq{\label{vu-equiv}
\chi_0(r + \| k^{(m)} \|_1) \chi_0(r + \| \tilde k^{(m)} \|_1)
V_\upsilon(\zeta, e^{-\alpha}r, e^{-\alpha} k^{(m, n)})
}
does not change if we substitute $\uw$ for $\uw'$ (notice that we can take $p_1 = 0 = q_L $, otherwise everything is zero).  
 
The desired result is an easy consequence of (\ref{E.L.cprm.5.1}), (\ref{elguero}), (\ref{noguero}), (\ref{vu-equiv}) and similar computations.  

\QED

Remark \ref{r-equiv} permits us define the following
\begin{definition}\label{ren-equiv}
For every $ [\uw] \in [\cW_\xi] $  we define the operator 
$$
H([\uw]) : = H(\uw)
$$ 
and the renormalization map
$$
\cR_\alpha([\uw]): = [\cR_\alpha(\uw)]\period 
$$
\end{definition}

\begin{definition}\label{wchica}
It follows from \cite{BCFS2003} that the map 
\eq{
 [\uw] \in  [\cW_\xi] \mapsto H([\uw])
}
is an injection. 
We denote by 
$$
 [\uw_s] \equiv [\uw_s(z)]
$$
the sequence of functions that satisfy
$$
\boH_s = H( [\uw_s])\period
$$

\end{definition}

\begin{theorem}
Let $ \xi < \frac{1}{4} $.
For every $s \geq \alpha_-$
\eq{ \label{www}
\|  [\uw_s] \|_\xi^{(I)} \leq & \frac{1}{10^7}e^{-2 \alpha_+}\rho^2e^{- \mu s /4}\comma \\  \notag
\|  [\uw_s - \ur] \|^{(F)} \leq & \frac{1}{10}(1-\frac{1}{2}e^{1/10}) + \frac{2}{10^7} \comma \\  \notag
\|  [\uw_s] \|^{(Z)} \leq & 2\period 
}
For every $s < \alpha_-$.
\eq{ \label{www1}
\|   [\uw_s] \|_{4\xi}^{(I)} \leq & \frac{1}{10^7}e^{-2 \alpha_+}\rho^2 \frac{(1 - \xi)^2}{(1 - 4\xi)^2}\comma \\  \notag
\|  [\uw_s -\ur] \|^{(F)} \leq & \frac{1}{10}(1-\frac{1}{2}e^{1/10}) + \frac{2}{10^7}\comma \\  \notag
\|   [\uw_s] \|^{(Z)} \leq & 2\period
}

\end{theorem}
\emph{Proof:}
The result is a direct consequence of Theorem \ref{is}.
\QED

\subsection{The Flow Operator and the Semigroup Property}\label{flowop}

\subsubsection{Definitions and Notation}

\begin{definition}
We denote by $E^{(\infty)}$ the set [see \eqref{fue}] 
\eq{ \label{epsiloninfty}
E^{(\infty)} : = \Big \{ \ue^{(\ell)}_\ual \: \Big | \: \ual \in \cS(\alpha_-, \alpha_+), \alpha_+ > 2\alpha_-,\ell \geq 0  \Big \} \comma
}
and by 
\eq{\label{winfinity}
\cW_\xi^{(\infty)} 
}
the set of sequences $\uw \in \cW_\xi$ satisfying 
\eqref{IS.1}-\eqref{IS.3} (with $ \uw$ instead of $\uw^{(j)}_\ual$), for some $\ue^{(j)} \in E^{(\infty)}$.

\end{definition}

\begin{definition}\label{fo1}
We denote by $[\cW_\xi]^{(0)}$ the following set
\eq{ \label{elfua}
[\cW_\xi]^{(0)} : = \Big\{ [ \cR_\alpha (\uw)] \: \Big | \: \uw \in \cW_\xi^{(\infty)}, \alpha \in [\alpha_-, \alpha_+]   \Big\}\period
}
\end{definition}

\begin{definition}\label{fo2}
We denote by $H[\cW_\xi]^{(0)}$ the set of operators $H$ of the form
\eq{\label{fo3}
H = H([\uw])\comma
}
for some $[\uw] \in  [\cW_\xi]^{(0)}$.

\end{definition}

\subsubsection{The Flow Operator}

\begin{definition}[Flow Operator]
We define the flow function ${\bf \Phi}:H[\cW_\xi]^{(0)} \times [0, \infty) \to H[\cW_\xi]^{(0)}  $
through the formula
\eq{
{\bf \Phi} (H, t) : = \boH_t \comma
}
where $H = H([\uw])$ for some $[\uw] \in  [\cW_\xi]^{(0)}$ and $\boH_t$ is introduced in 
Definition \ref{esta} (with $\uw$ instead of $\uw^{(0)}$). 
\end{definition}

\begin{theorem}[Semigroup Property]\label{sgp}
For every $t, s \geq 0$ and every $H \in H[\cW_\xi]^{(0)}$
\eq{\label{gp}
{\bf \Phi}(H, s+t) = {\bf \Phi}(\boH_s, t)\period
}
\end{theorem}
\noindent \emph{Proof:} The result is a consequence of Theorem \ref{T} and \eqref{aca}.

\QED

\subsubsection{The Flow Operator on the Function Spaces}

\begin{definition}[Flow Operator]
We define the flow function $ \Phi: [\cW_\xi]^{(0)} \times [0, \infty) \to [\cW_\xi]^{(0)}  $
through the formula
\eq{
\Phi ([\uw], t) : = [\uw_t] \comma
}
\end{definition}
where $[\uw_t]$ is introduced in Definition \ref{wchica} (we take $ [\uw^{(0)}]= [\uw] $).

\begin{theorem}[Semigroup Property] \label{itisp}
For every $t, s \geq 0$, and every $  [\uw] \in [\cW_\xi]^{(0)}$,
\eq{\label{gpf}
 \Phi([\uw], s+t) =  \Phi([\uw_s], t)\period
}
\end{theorem}
\noindent \emph{Proof:} The result is a consequence of Theorem \ref{sgp}.
 
\QED

\end{document}